\documentclass[aps,prx,twocolumn,longbibliography, notitlepage]{revtex4-1}
\usepackage{times}
\usepackage{amsmath}
\usepackage{amsfonts}
\usepackage{amssymb}
\usepackage{dsfont}
\usepackage{graphicx}
\usepackage{physics}
\usepackage{tikz}
\usepackage{hyperref}
\usetikzlibrary{arrows.meta}
\usetikzlibrary{decorations.markings}
\tikzset{middlearrow/.style={
        decoration={markings,
            mark= at position 0.6 with {\arrow{#1}} ,
        },
        postaction={decorate}
    }
}

\usepackage{epsfig}
\usepackage{verbatim}
\usepackage{bm}
\usepackage{mathrsfs}
\usepackage{yfonts}

\usepackage[utf8]{inputenc}
\usepackage[english]{babel}

\newtheorem{theorem}{Theorem}

\usepackage{pst-all}
\usepackage{leftidx}

\def\H{\mathcal{H}}
\def\Z{\mathbb{Z}}

\newcommand{\be}{\begin{equation}}
\newcommand{\ee}{\end{equation}}
\newcommand{\bc}{\begin{center}}
\newcommand{\ec}{\end{center}}

\newcommand{\non}{\nonumber}

\definecolor{dualblue}{RGB}{3,101,192}

\makeatletter
\def\l@subsubsection#1#2{}
\makeatother

\begin{document}

\title{Instantaneous braids and Dehn twists in topologically ordered states}
\author{Guanyu Zhu}
\author{Ali Lavasani}
\author{Maissam Barkeshli}
\affiliation{Department of Physics, Condensed Matter Theory Center, University of Maryland, College Park, Maryland 20742, USA}
\affiliation{Joint Quantum Institute, University of Maryland, College Park, Maryland 20742, USA}

\begin{abstract}
A defining feature of topologically ordered states of matter is the existence of locally indistinguishable
states on spaces with non-trivial topology. These degenerate states form a representation of the mapping class
group (MCG) of the space, which is generated by braids of defects or anyons, and by Dehn twists along non-contractible
cycles. These operations can be viewed as fault-tolerant logical gates in the context of topological quantum error
correcting codes and topological quantum computation. Here we show that braids and Dehn twists can in general 
be implemented by a constant depth quantum circuit, with a depth that is \it independent of code distance $d$ and system size.\rm \ 
The circuit consists of a constant depth \it local \rm quantum circuit (LQC) implementing a local geometry deformation of the 
quantum state, followed by a permutation on (relabelling of) the qubits.  The permutation requires permuting qubits that 
are separated by a distance of order $d$; it can be implemented by collective classical motion of
mobile qubits or as a constant depth circuit using long-range SWAP operations (with a range set by $d$) on immobile
qubits. We further show that (i) applying a given braid or Dehn twist $k$ times can be achieved with $\mathcal{O}(\log k)$ time overhead,
independent of code distance and system size, which implies an exponential speedup for certain logical gate sequences by trading
space for time, and (ii) an \it arbitrary \rm element of the MCG can be implemented by a constant depth (independent of $d$) LQC 
followed by a permutation, where in this case the range of interactions of the LQC grows with the number of
generators in the presentation of the group element. Applying these results to certain non-Abelian quantum error correcting codes 
demonstrates how universal logical gate sets can be implemented on
encoded qubits using only constant depth unitary circuits. 
\end{abstract}

\maketitle

\tableofcontents

\section{Introduction}

A profound property of topologically ordered states of matter is the possibility of topologically degenerate ground
states, which arise when the system exists on a topologically non-trivial space \cite{wen04,nayak2008,zhwang2010}. 
The degeneracy is protected by the fact that the states are indistinguishable by any local operators, up to exponentially
small corrections in system size.

This local indistinguishability of topological states is the key feature underlying quantum error correction
and the possibility of creating a fault-tolerant quantum memory \cite{kitaev2003,nayak2008,zhwang2010}. Many well-known
quantum error correcting codes (QECCs), such as Shor's 9-qubit code, the Steane code, and the Reed-Muller code, 
can all be interpreted in terms of the ground state subspace of a topologically ordered state defined on a cellulation of
a topologically non-trivial manifold \cite{freedman2001,campbell2017}. More generally, a large class of QECCs, 
known as topological QECCs, are associated with a particular class of topologically ordered states of matter. 
These include the surface / toric codes and their generalizations: the Kitaev quantum double models \cite{kitaev2003} and 
Turaev-Viro-Levin-Wen models \cite{turaev1992,barrett1996,walker2006,levin2005,Koenig:2010do}. Topological QECCs play 
an important role in the theory of quantum error correction, as they provide the only approach to decrease the 
logical error rate arbitrarily while maintaining local interactions among 
the microscopic degrees of freedom. 

In general, topologically ordered states can be realized in two distinct ways. In the ``passive'' approach, they can be
realized in equilibrium as ground states of an appropriate many-body Hamiltonian. The topological protection derives
from $T / \Delta \ll 1$, where $T$ is the temperature and $\Delta$ is the energy gap. In the ``active'' approach to quantum error
correction (QEC), the topologically ordered states arise as eigenstates of commuting local operators \cite{fowler2012,Bonesteel:2012fl}.
The state can be maintained actively by continuously measuring these local operators.
In the language of QECCs, the topologically degenerate ground state subspace is known as the \it code subspace\rm, and the minimum
length of a string operator that acts non-trivially in the code subspace is known as the \it code distance,\rm \ $d$.

An important question is to understand how to perform robust, non-trivial operations on the code subspace.
It is well-known that non-trivial operations can be obtained by braiding non-Abelian anyons \cite{nayak2008},
twist defects \cite{barkeshli2014SDG}, and holes with gapped boundaries \cite{fowler2012,cong2016}. Alternatively, 
when topological degeneracies arise in a closed genus $g$ surface, non-trivial operations can be obtained by performing 
Dehn twists \cite{witten1989,turaev1994}. In other words, the code subspace forms a representation of the mapping 
class group (MCG) of the space; the braid group on $n$ strands corresponds to the MCG of a disk with $n$ punctures,
while the MCG of a closed genus $g$ surface is generated by Dehn twists along non-contractible cyles.

In the passive approach, elements of the mapping class group have been proposed to be implemented
through adiabatic evolution with a local Hamiltonian, leading to a non-Abelian Berry phase \cite{wen1990naberry,nayak2008,you2015,barkeshli2016fqh}. 
To be adiabatic, the time to implement such transformations must be large compared with $1/\Delta$, and 
increases at least linearly with the code distance (or system size), $d$. In the active approach, known methods to implement 
braids (of holes, twist defects, or non-Abelian anyons) through unitary circuits also increase at least linearly with the 
code distance $d$. Alternatively, mapping class group elements can be effectively achieved through 
measurement based approaches \cite{bonderson2009,bonderson2013braiding,barkeshli2016mcg,lavasani2018},
which also take a time that diverges as $d \rightarrow \infty$. For example, in active approaches, fault-tolerant
readout of a measurement requires either (i) $d$ rounds of measurements \cite{fowler2012}, or
(ii) a single round of measurement with classical processing time that diverges as $d \rightarrow \infty$, and
an extra factor of $\mathcal{O}(d)$ in space overhead \cite{fowler2012time,hastings}.
Therefore in all approaches proposed to date, there is a fundamental tradeoff between space-time overhead and 
accuracy of a fault-tolerant quantum computation; in the limit where the logical error rate goes to zero, 
$d \rightarrow \infty$, and therefore the time to implement logical gates by braiding or Dehn twists also goes to infinity. 

Ideally, it is of interest to implement logical gates in a time that is independent of the code distance, and without increasing
the asymptotic scaling of the space overhead. A certain class of such logical gates are known as transversal 
logical gates \cite{nielsen_chuang_2010}. Transversal gates consist of non-trivial unitary operations on 
the code subspace that decompose as a tensor product of local unitary transformations that do not 
couple different sites within the same code block. Transversal gates are special cases of local constant depth 
quantum circuits, which are intrinsically fault-tolerant as the code distance $d \rightarrow \infty$ due to the locality of error propagation.
The Eastin-Knill theorem establishes that a universal encoded gate set cannot be implemented 
transversally \cite{Eastin:2009cj}. Related theorems impose strong restrictions on logical 
gates implemented with local, finite depth quantum circuits \cite{Beverland:2016bi, Bravyi:2013dx}.  

Recently, it was discovered that certain elements of the MCG of a generic topological state can be effectively implemented in one shot 
as a transversal gate \cite{Zhu:2017tr}. These elements correspond to certain finite order (torsion) elements of the 
MCG, which correspond to specific combinations of braids or Dehn twists. While these results 
do not violate the Eastin-Knill theorem, they do circumvent the assumptions of Ref.~\onlinecite{Beverland:2016bi} 
for implementing non-trivial logical gates in topological QECCs using constant depth local quantum circuits, and hence the corresponding no-go theorem for non-abelian codes.   

In this paper, we demonstrate that braids and Dehn twists in a wide class of topological states can always be achieved
by a quantum circuit of \it constant depth, independent of the code distance  \rm $d$. Specifically, we demonstrate that
elementary braids and Dehn twists can be implemented by (i) a local constant depth quantum circuit (LQC), followed by (ii) a
permutation on the qubits. The permutation requires qubits that have a separation of order $d$ to be permuted. If the qubits
of the system are mobile, the permutation can be physically implemented by shuttling the qubits around, as has been done
experimentally in ion trap systems \cite{Bowler:2012fr,  Walther:2012iz, Wright:2013df}. If the qubits are immobile, 
the permutation can be achieved in one step by utilizing long-range SWAP operations and ancilla qubits. 

Our results imply that by utilizing non-Abelian topological QECCs together with our braiding and Dehn twist protocols, 
a universal logical gate set can be implemented on encoded qubits through a constant depth unitary quantum circuit,
without increasing the scaling of the space overhead. We note that the implementation of these quantum circuits do not require
any additional classical computational resources nor do they depend on the result of measurement outcomes at intermediate
steps in the computation. Furthermore, our protocols for topological codes with local syndromes require 
$\mathcal{O}(d^2)$ space overhead per logical qubit. Therefore the total space-time cost for implementing a single logical gate is 
$\mathcal{O}(d^2)$ per logical qubit. 


This is, to our knowledge, the first result to demonstrate that universal logical gate sets can be implemented on encoded
qubits with constant depth circuits, and without increasing the scaling of the space overhead. Other proposals for implementing
universal logical gate sets, such as those which utilize magic state distillation or code switching, all require $\mathcal{O}(d^3)$
space-time overhead per logical qubit, per logical gate. This space-time overhead can come from either (1) a 
time overhead that diverges at least linearly with the code distance $d$ and $\mathcal{O}(d^2)$ space overhead per logical qubit, 
or (2) polylog time overhead (including classical computational resources) and $\mathcal{O}(d^3)$ space overhead per 
logical qubit \cite{bravyi2005,fowler2012time,Paetznick:2013fu, Jones:2015uv,
Bravyi:2015ue, Bombin:2015jk,Bombin:2016dq, JochymOConnor:2016ck,Yoder:2016cu,hastings}
\footnote{Note that the proposed code switching protocols have intermediate steps that depend on outcomes of measurements
during the protocol. For the measurements to be fault-tolerant, they must either be performed $d$ times or require a factor of 
$d$ increase in space overhead \cite{fowler2012,fowler2012time,hastings}.}.

Our constant depth circuits maps a local operator with support in a region $\mathcal{R}$  to another local operator
with support in a region $\mathcal{R}'$, such that the area of $\mathcal{R}$ and $\mathcal{R}'$ are related by a constant factor, independent of code distance. Consequently,
our logical gates are naturally topologically protected and can be made fault-tolerant, since they only change the length of an error string by an $\mathcal{O}(1)$  constant factor independent of code distance.   The extra time-overhead for decoding and error correction after applying these constant-depth circuits depend on the detailed properties of the logical circuits.  In the presence of noisy syndrome measurements, in the worst-case scenario we expect $\mathcal{O}(d/\log d)$ rounds of syndrome measurements need to be
performed for each application of the logical gate in order to successfully decode error strings. 

We note that some of the results discussed in this paper with respect to braiding with constant depth circuits 
have also been summarized by us in a short paper \cite{zhu2018}.

\subsection{Summary of results}

Our specific technical results are summarized below. 
Let us consider $N + N_a$ physical qubits arranged on a lattice. We further consider a state
\begin{align}
|\Psi \rangle = |\Phi \rangle \otimes |\Pi\rangle_a.
\end{align}
Here, $|\Pi\rangle_a = \otimes_j |\psi_j \rangle_a$ is an arbitrary product state for the $N_a$ ancilla qubits.
$|\Phi\rangle$ is a topologically ordered state on $N$ qubits on a genus $g$ surface with $p$ punctures, $\Sigma_{g, p}$.
The punctures could correspond to holes with gapped boundaries or anyons. $|\Phi\rangle$
is an arbitrary (Abelian or non-Abelian), non-chiral topologically ordered state.  Such states are always related,
by a constant depth local quantum circuit (alternatively, by adiabatic evolution), to an exact ground state of a commuting
projector Hamiltonian, such as the Kitaev quantum double or Levin-Wen models \cite{kitaev2003,levin2005, Koenig:2010do}. 
Alternatively, such non-chiral topological orders can be described within a path integral state-sum construction, as 
described in Ref. \onlinecite{turaev1992,turaev1994,barrett1996,walker2006,barkeshli2016tr}.

$|\Phi\rangle$ belongs to a representation of $\text{MCG}(\Sigma_{g,p})$, the mapping class group of $\Sigma_{g,p}$.
It is well known that $\text{MCG}(\Sigma_{g,p})$ can be generated by $3g-1$ Dehn twists along the simple curves
$\alpha_i, \beta_i, \gamma_j$, where $i = 1,\cdots,g$ and $j = 1,\cdots, g-1$, together with elementary (half-) braids
between neighboring punctures \cite{farb2011primer}. See Sec.~\ref{sec:concepts} and Fig.~\ref{fig:MCG_illustration} for detailed discussions and  illustration. 

Let us denote $\sigma$ to be a permutation on the $N+N_a$ qubits and $\mathcal{P}_\sigma$ the unitary representation of that permutation.
Furthermore, let $\mathcal{LU}$ denote a local, constant depth quantum circuit. In particular, $\mathcal{LU}$ is local in the sense that 
the range $r$ of interactions is independent of system size and code distance. Similarly, constant depth means that the depth of the
circuit is also independent of system size and code distance. 

We first demonstrate the following result:

\begin{theorem}\label{theorem1}
Let $\kappa \in \text{MCG}(\Sigma_{g,p})$ denote either a Dehn twist along a simple curve $\alpha_i, \beta_i, \gamma_j$, 
where $i = 1,\cdots,g$ and $j = 1,\cdots, g-1$, or an elementary braid between neighboring punctures. 
We let $\mathcal{V}_\kappa$ be the unitary representation of $\kappa$ on the topological ground state subspace (i.e. the code subspace). 
Then,
\begin{align}
\mathcal{V}_{\kappa} \otimes I |\Psi\rangle = (\mathcal{V}_\kappa |\Phi\rangle) \otimes |\Pi\rangle_a
= \mathcal{P}_\sigma \mathcal{LU}_\kappa (|\Phi\rangle \otimes  |\Pi\rangle_a) .
\end{align}
\end{theorem}
$I$ is the identity operator on the ancilla qubits and $\mathcal{LU}_{\kappa}$ is a constant depth local unitary that depends
on $\kappa$. The permutation $\mathcal{P}_\sigma$ (which depends on $\kappa$) can be implemented in constant time by utilizing ancilla qubits.
For example, one first performs a SWAP operation between each qubit and an ancilla qubit, followed by a second
SWAP operation between the ancilla qubit and the target location of the SWAPs. (The second SWAP is actually unncessary, as explained
in more detail in Sec. \ref{sec:discussion}). It is crucial to note that these SWAP operations are \it long-range \rm operations. 
In general the range of the SWAPs is set by the code distance $d$. 

It is useful to note that, depending on the physical implementation, the permutation $\mathcal{P}_\sigma$ can also be
performed by physically moving the location of the qubits in physical space. For example, if the qubits are associated with
ions in an ion-trap quantum computer, the ions can be physically moved to their target locations \cite{Bowler:2012fr,  Walther:2012iz, Wright:2013df, Home:jr, Lekitsch:2015ua, Kaufmann:2017kc}.

A corollary of the above theorem is with respect to the space-time overhead for universal fault-tolerant 
quantum computation. It is well-known that mapping class group elements, such as braiding of anyons, in the Fibonacci topological state 
is universal for topological quantum computation \cite{Freedman_Larsen_wang_2002, zhwang2010, Bonesteel:2005ho}.  
We can thus consider the Turaev-Viro code \cite{Koenig:2010do, Bonesteel:2012fl} (and associated Levin-Wen model \cite{levin2005}) 
based on the Fibonacci fusion category, whose topological order corresponds to two time-reversed copies of the Fibonacci state. 
Applying \textbf{Theorem 1} to such a code thus implies that a universal fault-tolerant gate set can be achieved through 
constant-time braiding of Fibonacci anyons, without changing the asymptotic scaling of the space overhead. 

More specifically, in a two-dimensional topological code with local interactions (alternatively, in an active error correction approach, with 
local syndrome measurements), the space overhead is $\mathcal{O}(d^2)$ per logical qubit. The result of \textbf{Theorem 1} thus 
implies that universal fault-tolerant gate sets can be achieved with time overhead that is independent of code distance $d$, 
while keeping the space overhead at $\mathcal{O}(d^2)$ per logical qubit. 

\begin{theorem}
Let $\kappa \in \text{MCG}(\Sigma_{g,p})$ denote either a Dehn twist along a simple curve $\alpha_i, \beta_i, \gamma_j$, 
where $i = 1,\cdots,g$ and $j = 1,\cdots, g-1$, or an elementary braid between neighboring punctures. 
Furthermore, let $\mathcal{V}_{\kappa^n} = \mathcal{V}_{\kappa}^n$ be the unitary representation of $\kappa^n$ 
on the topological ground state subspace (i.e. the code subspace), where $n$ is an arbitrary integer. 
Then,
\begin{align}\label{theorem2.1}
 \mathcal{V}_{\kappa}^n\otimes I |\Psi\rangle = ( \mathcal{V}_{\kappa}^n |\Phi\rangle) \otimes |\Pi\rangle_a
= \prod_{i=1}^k \mathcal{LU}_{i,\kappa} \mathcal{P}_{\sigma_i} (|\Phi\rangle \otimes |\Pi\rangle_a) .
\end{align}
Here, $k = \mathcal{O}(\log n)$, $\mathcal{P}_{\sigma_i}$ is a qubit permutation, which permutes
qubits over a range of $\mathcal{O}(d)$, and $\mathcal{LU}_{i,\kappa}$ are local, finite depth
quantum circuits, where the range $r$ of the gates and depth are independent of $n$, code distance $d$, and
system size. 
In the case of the $\Z_N$ toric code, we also have:
\begin{align}\label{theorem2.2}
 \mathcal{V}_{\kappa}^n \otimes I |\Psi\rangle = ( \mathcal{V}_{\kappa}^n|\Phi\rangle) \otimes |\Pi\rangle_a
= \mathcal{P}_{\sigma} \mathcal{LU}_{\kappa} (|\Phi\rangle \otimes |\Pi\rangle_a) ,
\end{align}
where now $\mathcal{LU}_\kappa$ is a local quantum circuit with maximum range $r = \mathcal{O}(n)$ and 
fixed depth independent of $n$, code distance $d$ and system size.

\end{theorem}

\begin{theorem}\label{theorem3}
Let $\zeta \in \text{MCG}(\Sigma_{g,p})$ be an arbitrary group element, and $\mathcal{V}_\zeta$ its representation on the quantum state.
$\zeta$ has a presentation in terms of a string of $k$ Dehn twists and
braids, for some integer $k$.
Then,
\begin{align}
\mathcal{V}_\zeta \otimes I |\Psi\rangle = (\mathcal{V}_\zeta |\Phi\rangle)  \otimes |\Pi\rangle_a=
\mathcal{P}_\sigma\, \mathcal{LU}_\zeta |\phi \rangle \otimes |\Pi\rangle_a.
\end{align}
Here, $\mathcal{LU}_\zeta$ is a constant (independent of code distance and system size) depth local quantum circuit. 
The range $r$ of gates in $\mathcal{LU}_\zeta$ increases with $k$, such that $r= \mathcal{O}(c^k)$ 
where $c$ is a constant independent of $\zeta$ and code distance and system size.
\end{theorem}

We note that in all the above theorems, the order of $\mathcal{LU}$ and $\mathcal{P}_\sigma$ can in principle be switched 
(with the concrete circuits being modified), and does not affect the final results.

We provide proofs of these statements by explicit construction for ground states of exactly solvable commuting 
projector models. As noted above, any non-chiral topologically ordered state can be transformed 
to the ground state of an exactly solvable commuting projector model by a finite depth local quantum circuit.

An important byproduct of our analysis is to demonstrate how non-Abelian anyons and holes can be
moved by distances of order the code distance by a constant depth local circuit followed by a permutation on the 
qubits. This overturns the general belief that moving non-Abelian anyons by a distance $\ell$ always requires a quantum circuit 
of depth $\propto \ell$. While this belief is correct when restricted to purely local interactions, we see that the use of long-range
permutations allows us to implement the motion by a constant depth circuit. Interestingly, however, our protocols 
can only move anyons by a distance $\ell$ that is a constant factor of the mininum separation between anyons. 
We therefore arrive at an interesting version of Zeno's paradox in the context of non-Abelian topological quantum 
order: the time it takes to create two non-Abelian anyons out of the vacuum and separate them a distance of order $\ell$
requires $\mathcal{O}(\log \ell)$ steps (ignoring the presence of any other anyons). However if the anyons are already
a distance $\ell$ apart, they can be moved by a distance of order $\ell$ in a time that is independent of $\ell$. 

\subsection{Structure of the paper}

This paper is structured as follows. We begin in Sec.~\ref{sec:concepts} by providing a brief review of the mapping class group
of surfaces, Dehn twists, and their relation to braiding. In Sec.~\ref{sec:ZNtc}, we focus on the case of $\mathbb{Z}_N$ topological
order, proving \textbf{Theorems 1} and \textbf{2} by explicitly constructing the quantum circuits with the desired properties. 
In Sec.~\ref{sec:non-abelian}, we then generalize this discussion to encompass arbitrary non-chiral topological orders, which include
both arbitrary Abelian and non-Abelian topological orders.  We prove \textbf{Theorem 3} separately in Sec.~\ref{sec:theorem3}. 
We further show the fault-tolerance aspects of our schemes in Sec.~\ref{sec:fault_tolerance}, and conclude our 
paper with a discussion of the central results in Sec.~\ref{sec:discussion}.

\section{Basic concepts:  Dehn twists, braids and mapping class groups}\label{sec:concepts}

\begin{figure}
\includegraphics[width=1 \columnwidth]{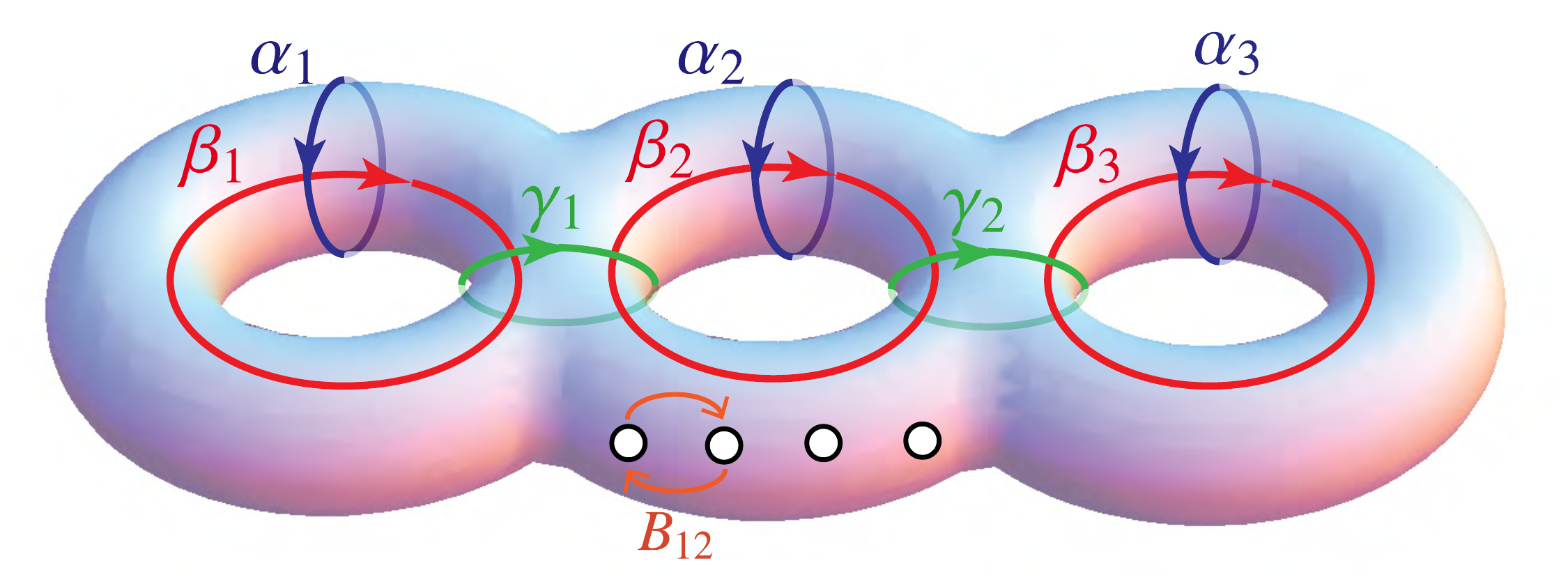}
\caption{
Non-contractible cycles and braid operations on a genus $g = 3$
surface with $p = 4$ punctures.
}
\label{fig:MCG_illustration}
\end{figure}

\begin{figure}
\includegraphics[width=1 \columnwidth]{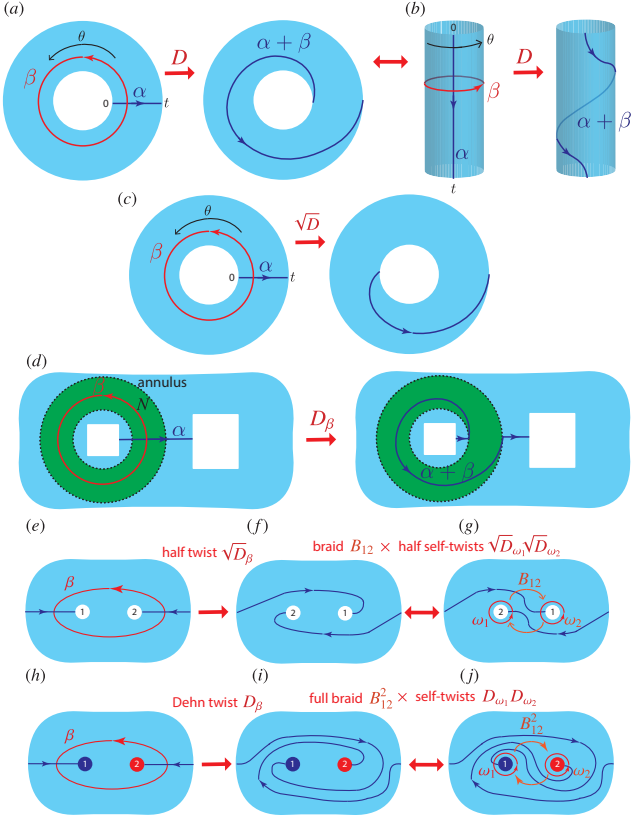}
\caption{Definition of Dehn twists, half twists and braids, as well as their relations.}
\label{fig:twists_definition}
\end{figure}

\begin{figure*}
  \includegraphics[width=1.7\columnwidth]{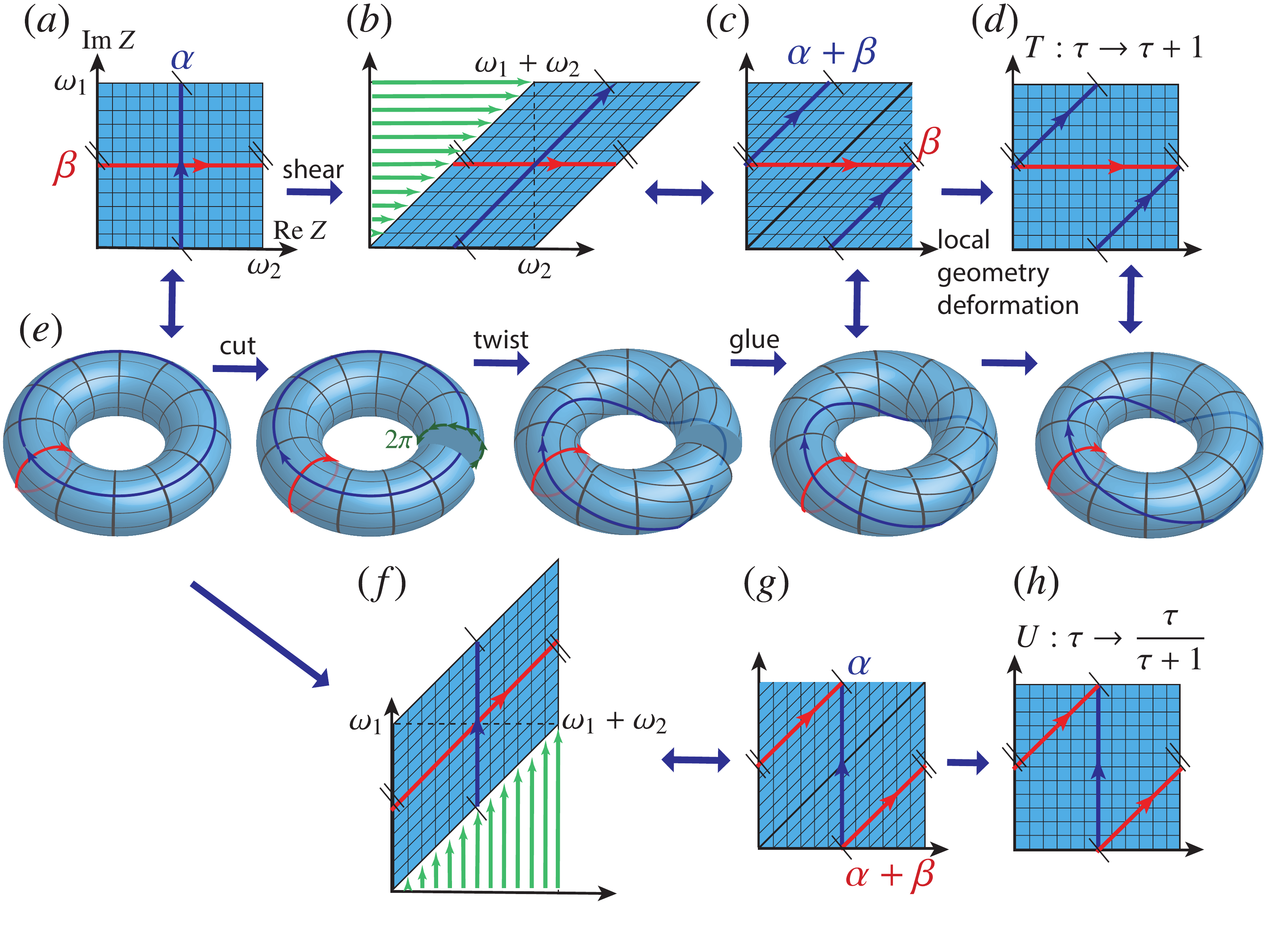}
  \caption{The mathematical definition and implementation of Dehn twists $T$ and $U$ on a torus via Dehn surgery. The coordinate
web represents the metric of the manifold in the continuous case or underlying lattice structure in the discrete case. The shear
deformation (green arrows) in (b) implements a Dehn twist and is equivalent to Dehn surgery in (e): cut, twist, and glue.
The local geometric deformation returns the metric/lattice to the original configuration.  }
\label{fig:MCG_definition}
\end{figure*}

\subsection{Mapping class group and its representation}

We start with the definition of the mapping class group of a surface, which includes all the central concepts discussed in this paper
such as Dehn twists and braids.  Consider a surface of genus $g$ with $p$ punctures, denoted $\Sigma_{g,p}$. The mapping class
group of $\Sigma_{g,p}$, denoted MCG($\Sigma_{g,p}$), is defined to be the group of
orientation-preserving diffeomorphisms of $\Sigma_{g,p}$  modulo those which can be continuously connected to identity:
\be
\text{MCG}(\Sigma_{g,p}) = \text{Diff}^+(\Sigma_{g,p}, \partial \Sigma_{g,p}) / \text{Diff}_0(\Sigma_{g,p}, \partial \Sigma_{g,p}),
\ee
with the diffeomorphisms being restricted to the identity on the boundary $\partial \Sigma_{g,p}$ \cite{farb2011primer}.
Here $\text{Diff}_0(\Sigma_{g,p}, \partial \Sigma_{g,p})$ is the subgroup of $\text{Diff}^+(\Sigma_{g,p}, \partial \Sigma_{g,p})$
which consists of elements that are isotopic (continuously connected) to the identity.  As a consequence of this definition,
any element of the mapping class group, $[\zeta] \in \text{MCG}(\Sigma_{g,p})$, is an equivalence class of a diffeomorphism of manifold
$\Sigma_{g,p}$ which maps the manifold back to itself, i.e., $\zeta : \Sigma_{g,p} \longmapsto \Sigma_{g,p}$.

As illustrated in Fig.~\ref{fig:MCG_illustration}, $\text{MCG}(\Sigma_{g,p})$ can be generated by $3g-1$ Dehn twists along
the non-contractible cycles denoted $\alpha$, $\beta$, and $\gamma$ (see Fig.~\ref{fig:MCG_illustration}),
together with braids between neighboring punctures,  $B_{1,2}$, $B_{2,3}$, ..., and $B_{p-1,p}$. In the following subsections, we will
provide more background on Dehn twists and their relation to braids.

In the context of topological states or codes  supported on a manifold $\Sigma$, the unitary representations
of MCG$(\Sigma)$ are topologically protected (i.e. fault-tolerant) unitary transformations acting on the ground-state subspace
or equivalently the code space $\H_{\Sigma}$. We denote the representation of a mapping class group element $\zeta$
by the unitary operator $\mathcal{V}_\zeta$, which performs an \textit{automorphism} that maps the code space
back to itself, i.e., $\mathcal{V}_\zeta : \H_{\Sigma} \longmapsto \H_{\Sigma}$. Therefore, $\mathcal{V}_\zeta$ is an
element of the automorphism group of the code space, $\mathcal{V}_\zeta \in \text{Aut}(\H_{\Sigma})$, and is hence a logical gate.

\subsection{Dehn twists}

Here we review a specific type of self-diffeomorphism called a Dehn twist. We first consider an annulus
$A $$=$$ S^1$$\times$$[0,1]$, which can be embedded in the $(\theta, r)$-plane
as shown in Fig.~\ref{fig:twists_definition}(a). We define the twist map $D:A \longmapsto A $,   by the following formula
\be\label{twist_annulus_definition}
D \ (\theta, t) = (\theta + 2\pi t, t).
\ee
Note that $D$ is an orientation-preserving diffeomorphism  fixing $\partial A$ pointwise, and hence satisfies the definition of a mapping class. Thus, we call $D$ (and the class of maps up to an additional diffeomorphism continuously connected to the identity) a left \textit{Dehn twist}. The right Dehn twist is its inverse, i.e.,
\be
D^{-1} \ (\theta, t) = (\theta - 2\pi t, t).
\ee

In order to visualize the transformation on the surface, we show two directed non-contractible lines/loops on the surface. The first one is a (blue) line connecting the inner and outer boundary, denoted by $\alpha$, where the arrow represents the direction.   The second one is a (red) non-contractible loop circulating the inner boundary, denoted by $\beta$.   We can see the left Dehn twist performs the following map:
\be
D:(\alpha, \beta ) \longmapsto (\alpha + \beta, \beta),
\ee
where $\alpha + \beta$ is the twisted line shown in the right panel in Fig.~\ref{fig:twists_definition}(a). The above map can also be used as an alternative definition of the Dehn twist.  One can also see that the left Dehn twist is equivalent to a continuous counter-clockwise $2\pi$-rotation of the outer boundary  or a clockwise $2\pi$-rotation of the inner boundary. Similarly, the right Dehn twist performs the map:
\be
D^{-1}:(\alpha, \beta ) \longmapsto (\alpha - \beta, \beta),\ee
where the $\alpha -\beta$ is twisted in the opposite direction.
Since the annulus is equivalent (homotopic) to a cylinder, we hence have also defined the Dehn twist on a cylinder, as shown in Fig.~\ref{fig:twists_definition}(b).

For subsequent discussions, we also introduce the notion of a \textit{half twist} on an annulus as illustrated
in Fig.~\ref{fig:twists_definition}(c). By itself, the half twist is not an element of the mapping class group because it does not leave
the boundary fixed. The left half twist is defined as
\be
\sqrt{D} \ (\theta, t) = (\theta + \pi+ \pi t, t).
\ee
Note that in our convention, the half twist fixes the outer boundary and makes a clockwise $\pi$-rotation on the inner boundary.
Similarly, the right half twist is defined as
\be
\sqrt{D}^{-1} \ (\theta, t) = (\theta+ \pi - \pi t, t),
\ee
which makes a counter-clockwise $\pi$-rotation on the inner boundary.

Now consider the generic surface $\Sigma_{g, p}$. One can perform a Dehn twist along the non-contractible loop $\beta$ in $\Sigma_{g, p}$,
as illustrated in Fig.~\ref{fig:twists_definition}(d).  Note that the direction of the $\beta$-loop (indicated by the arrow) determines whether it is a left or right Dehn twist.
Let $N$ be a regular neighborhood [shown as a green belt in Fig.~\ref{fig:twists_definition}(d)] of $\beta$ and $f$ be an orientation-preserving
diffeomorphism that maps the previously defined annulus to such a neighborhood:
$f$$:$$A \longmapsto N$.
We can hence define a self-diffeomorphism $D_\beta : \Sigma_{g, p} \longmapsto $$ \Sigma_{g, p}$ as a Dehn twist about the $\beta$-loop as follows:
\be
D_\beta (\mathbf{x}) =  \left\lbrace \begin{array}{cc}
f D f^{-1}   & \text{if } \mathbf{x} \in N \\
\mathbf{x} & \text{ otherwise}.
\end{array} \right.
\ee
This definition says that $D_\beta$ performs a Dehn twist $D$ on the annulus N and fixes every point outside the annulus $N$.   To visualize the change of the surface, we again use the non-contractible (blue) line $\alpha$ going across the inner and outer boundary of the neighborhood $N$,  and the (red) loop $\beta$ circulating around $N$ for illustration.  The Dehn twist $D_\beta$ performs the following map on these two loops:
\be
D_\beta:\alpha \longmapsto \alpha + \beta, \qquad D_\beta:\beta \longmapsto \beta,
\ee
where the part of the twisted line $\alpha + \beta$ outside the annulus $N$ remains fixed as illustrated by the right panel of Fig.~\ref{fig:twists_definition}(d).
This map can also be used as the alternative definition of $D_\beta$.   Similarly, one can also define the half twist $\sqrt{D}_\beta$ on an arbitrary surface.

The representation of the Dehn twists on the topological ground state subspace (i.e. the code subspace)
are denoted by $\mathcal{D}_\beta$. Note that we have used different fonts to distinguish an MCG element
and its representation. It induces the following transformations on the Wilson line/loop operators:
\be
\mathcal{D}_\beta W_\alpha^a \mathcal{D}^\dag_\beta = W_{\alpha+\beta}^a,  \quad \mathcal{D}_\beta W_\beta^a \mathcal{D}^\dag_\beta = W_{\beta}^a,
\ee
where $W_\alpha^a$ and $W_\beta^a$ are Wilson line/loop on the $\alpha$-line and $\beta$-cycle with topological charge $a$.
The operator $W_{\alpha+\beta}^a$ represents the twisted Wilson loop.

\subsection{Braids and braid group}\label{sec:concepts_braid}

Although braiding is often discussed in its own context (especially in physics), it is actually just a special type
of mapping class.  A particular example is the case of the braid group on $p$ strands, which is equivalent to the mapping class
group of a disk with $p$ punctures: $\text{B}_p = \text{MCG}(D^2_{p})$, where here $D^2_p$ denotes a disk with $p$ punctures.

Braids between punctures can also be expressed in terms of Dehn twists, as follows. As shown in Fig.~\ref{fig:twists_definition}(e-g),
a right half twist along the $\beta$-loop enclosing two punctures [Fig.~\ref{fig:twists_definition}(e,f)] is equivalent
to braiding the two punctures, with additional half self-twists around both punctures [Fig.~\ref{fig:twists_definition}(g)]. That is,
\begin{align}
B_{1,2} = \sqrt{D}_\beta \sqrt{D^{-1}}_{\omega_1} \sqrt{D^{-1}}_{\omega_2} ,
\end{align}
where $\sqrt{D}_\beta$, $\sqrt{D}_{\omega_1}$, $\sqrt{D}_{\omega_2}$ represent the half-twists around $\beta$, $\omega_1$, and $\omega_2$.

For a full braid, we thus have $B_{1,2}^2 = D_\beta D^{-1}_{\omega_1} D^{-1}_{\omega_2}$, as illustrated in Fig.~\ref{fig:twists_definition}(h-j).

\subsection{Mapping class group, Dehn surgery, and local geometry deformation  on a torus}

We now consider in more detail the case of the mapping class group of a torus, $T^2$. This will help provide
the underlying mathematical intuition that forms the basis of our results.

The points of a torus can be specified by points $z$ in the complex plane, modulo equivalences $z \sim z + \omega_1 \sim z + \omega_2$,
for complex numbers $\omega_1$ and $\omega_2$. The torus is thus parameterized by $(\omega_1, \omega_2)$
or equivalently $(1,\tau)$ as shown in Fig.~\ref{fig:MCG_definition}(a), where the modular parameter is defined
to be $\tau = \omega_1/\omega_2$. The coordinate web indicates the local metric of a continuous manifold,
or represents a lattice in the discrete case.

Arbitrary modular transformations belonging to the MCG of a torus can be achieved by
the following transformation
\be
\left(\begin{matrix}
  \omega_1 \\
  \omega_2
\end{matrix}\right) \mapsto
\left(\begin{matrix}
  \omega'_1 \\
  \omega'_2
\end{matrix}\right)=
\left(\begin{matrix}
 a & b \\
 c & d
\end{matrix}\right)
\left(\begin{matrix}
  \omega_1 \\
  \omega_2
\end{matrix}\right), \quad \ \tau \mapsto\tau'=\frac{a\tau + b}{c\tau + d},
\ee
satisfying $a,b,c,d\in \mathbb{Z}$ and $ad-bc=1$. Therefore, the mapping class group of a torus is
isomorphic to a special linear group, namely MCG$(T^2)=\text{SL}(2, \mathbb{Z})$.
The MCG$(T^2)$ is generated by two transformations  $T:~\tau \mapsto \tau + 1$ and
$U: \tau \mapsto \frac{\tau}{\tau + 1}$. Their matrix representations are
\be
T=\left(\begin{matrix}
 1 & 1 \\
 0 & 1
\end{matrix}\right),
\quad U=\left(\begin{matrix}
 1 & 0 \\
 1 & 1
\end{matrix}\right).
\ee

As shown in Fig.~\ref{fig:MCG_definition}(b), a shear deformation induces a large diffeomorphism of the manifold and
hence maps it back to a torus in Fig.~\ref{fig:MCG_definition}(c), with an additional local metric/lattice deformation
compared to the original torus in Fig.~\ref{fig:MCG_definition}(a) indicated by the slanted coordinate web.  Note that
the original vertical geometric lines get stretched to diagonal lines. The above operation is equivalent to
a Dehn surgery illustrated in Fig.~\ref{fig:MCG_definition}(e), which cuts the torus along the $\beta$-cycle into a
cylinder, twist the cyclinder by $2\pi$ along the $\beta$-cycle (equivalent to the shearing) and then re-glue the
two edges of the cylinder back to a torus.

One can apply an additional \textit{local geometry deformation} to transform the configuration in Fig.~\ref{fig:MCG_definition}(c)
to the one in Fig.~\ref{fig:MCG_definition}(d) with the same metric/lattice structure as the original torus in
Fig.~\ref{fig:MCG_definition}(a).  This local geometry deformation is a diffeomorphism isotopic to an
identity MCG element $\mathbb{I}$, i.e., a trivial mapping class.  In Sec.~\ref{sec:non-abelian},
we will see that this local geometry deformation can be interpreted as a retriangulation of the manifold.

Denoting the two non-contractible cycles of the torus as $\alpha$ (vertical) and $\beta$ (horizontal), the
combination of shear deformation (Dehn surgery) and local metric deformation achieves a
self-diffeomorphism generating the following transformations on these two loops respectively
\be
T: (\alpha,\beta) \longmapsto (\alpha+\beta,\beta).
\ee
Similarly, one apply a combination of shear deformation (Dehn surgery) along the
$\alpha$-loop and a local geometry deformation [Fig.~\ref{fig:MCG_definition}(f-h)] to induce the following transformation on the loops:
\be
U: (\alpha,\beta) \longmapsto (\alpha,\alpha + \beta).
\ee
Therefore, the two generators are Dehn twists along the two cycles, i.e., $T=D_\beta$ and $U=D_\alpha$.

The representation of the above two generators in the topologically ordered ground state subspace (code subspace)
are denoted by $\mathcal{T}=\mathcal{D}_\beta$ and
$\mathcal{U}=\mathcal{D}_\alpha$. They induce the following transformations on the Wilson loop operators:
\be
\mathcal{T} W_\alpha^a \mathcal{T}^\dag = W_{\alpha+\beta}^a,  \quad \mathcal{U} W_\beta^a \mathcal{U}^\dag = W_{\alpha+\beta}^a,
\ee
where $W_\alpha^a$ and $W_\beta^a$ are Wilson loop on the $\alpha$-cycle and $\beta$-cycle with topological charge $a$. The operator $W_{\alpha+\beta}^a$ denotes the twisted Wilson loop.

We emphasize that, from the point of view of the mapping class group, the Dehn surgery (shear deformation) already
performs a self-diffeomorphism which maps the topological manifold back to itself, i.e., $\Sigma \mapsto \Sigma$.
However, such a map changes the local geometry, and in the discrete case the lattice structure, i.e., $\Lambda \mapsto \Lambda'$.
For a topological state or code defined on the lattice $\Lambda$, the code space depends on the lattice structure
and can be denoted by $\H_\Lambda$. Therefore, the Dehn surgery (shear deformation) itself changes the Hilbert space,
i.e., $\H_\Lambda \rightarrow \H_{\Lambda'}$. Note that although $\H_\Lambda$ and $\H_{\Lambda'}$ are isomorphic, they
are distinct subspaces of the full Hilbert space of the microscopic (physical) degrees of freedom (qubits). In order to realize a logical gate, which is an automorphism of the
code space, one has to apply the additional trivial mapping class, i.e., the local geometry deformation in order to
map the Hilbert space back to the original code space $\H_{\Lambda}$.  Such a local geometry deformation can be
implemented by a constant depth local quantum circuit as will be discussed in the later sections. As we will see in
Sec.~\ref{sec:non-abelian}, this geometry dependence is related to the fact that the Hilbert space and
wave function of a topological quantum field theory are not topological invariants and depend on
the local geometry, in particular the triangulation.

Finally, we note that in this paper we focus on the situation that the topological states and codes are defined
on discrete lattices.  However, the notion of local geometry deformation also applies to the
continuum case, and thus our results should be generalizable to
topologically ordered states defined in the
continuum.  

\section{Theory for $\mathbb{Z}_N$ toric code}
\label{sec:ZNtc}

\subsection{Local geometry deformation}

We begin by defining the $\mathbb{Z}_N$ toric code and describing local quantum circuits
that can be used to change the lattice geometry. We will subsequently use these
geometrical transforamtions of the lattice structure to help implement our Dehn twist
and braiding protocols.

\subsubsection{$N = 2$}

Let us begin with the $\Z_2$ toric code model with qubits located on the edges of a square lattice.
We consider the case of periodic boundary conditions, so that the space is topologically
a torus. The $\Z_2$ toric code has the following Hamiltonian:
\be
H_{\Z_2}= - \sum_{\nu} X_{\nu, 1} X_{\nu, 2} X_{\nu, 3} X_{\nu, 4} - \sum_{p} Z_{p, 1} Z_{p, 2} Z_{p, 3} Z_{p, 4},
\ee
where $\nu$ and $p$ specify the vertices and plaquettes of the lattice, $X$ and $Z$ are Pauli-$X$ and Pauli-$Z$ operators, and
the numbers $1,\cdots,4$ index the four qubits associated with each vertex or plaquette. Violations of the vertex stabilizers
are referred to as $e$ particles, while plaquette violations are referred to as $m$ particles.

Let us denote $|0\rangle$ and $|1\rangle$ to be the states associated with $+1$ and $-1$ eigenvalues for the $Z$ operator, respectively.
By taking the state $|0\rangle$ for each qubit to define the absence of a string and $|1\rangle$ to define the presence
of a string, it is straightforward to see that the ground states of $H_{\Z_2}$ are associated with an equal weight superposition of all possible
closed strings. On a torus, there are four topologically degenerate ground states, depending on whether there are an even or odd number
of closed strings wrapping the two non-contractible cycles. In this basis, the ground state is a superposition of closed $m$-strings. Alternatively,
by working in the $X$-basis, we can view the ground state to be a superposition of closed $e$-strings.

\begin{figure}
  \includegraphics[width=1\columnwidth]{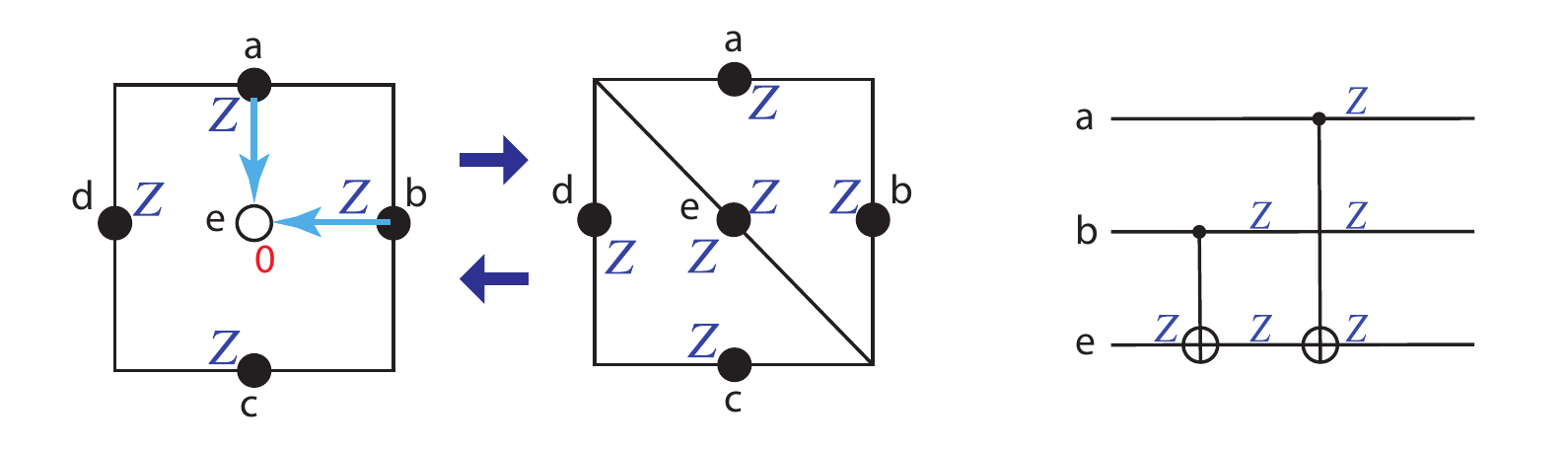}
  \caption{The elementary move of adding (removing) a diagonal edge in the center of the plaquette, which splits (merges) the plaquette(s).
The quantum circuit implementing such a move is composed of two CNOTs (blue arrows) targeting the qubits on the diagonal edge $e$, controlled
by the other two qubits $a$ and $b$ on the same triangular plaquette $\Delta_{abe}$. }
\label{fig:basic-moves}
\end{figure}

The toric code model can in general be defined on any cellulation of a two-dimensional surface; a square lattice is just one particular
simple choice. Given a toric code ground state on one cellulation, it is possible to convert it to a ground state on a different cellulation
with a simple quantum circuit that effectively adds or removes extra qubits. This can be achieved with the basic moves shown in
Fig.~\ref{fig:basic-moves} \cite{Dennis:2002ds,Aguado:2008hn}. We note that these moves have a close connection with notions
of entanglement renormalization \cite{Aguado:2008hn}, as they can be used to rescale the wavefunction to coarser lattices.

\begin{figure*}
  \includegraphics[width=1.5\columnwidth]{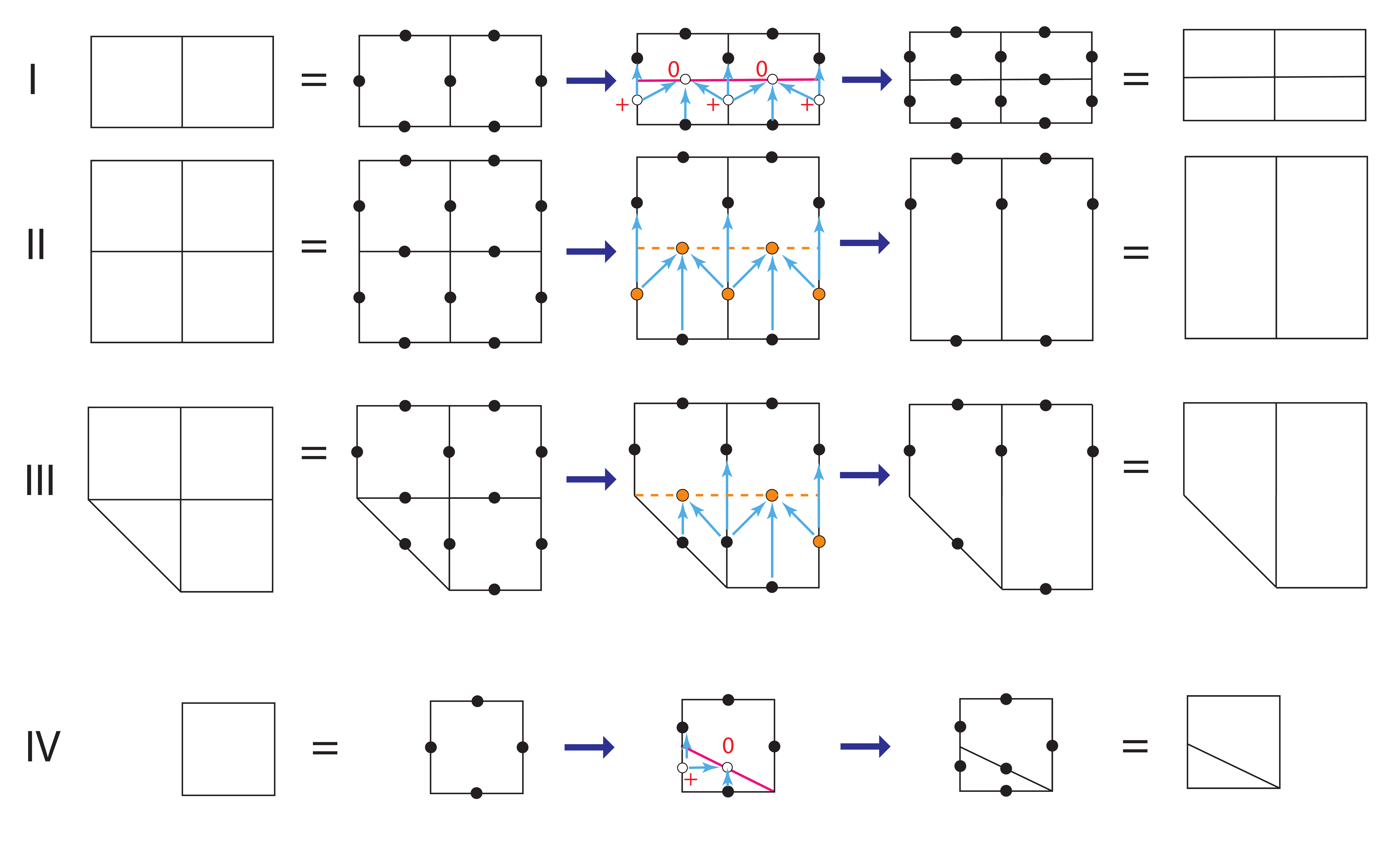}
  \caption{Gadgets of local geometry deformation of the lattice, which will be used for braiding, and Dehn twists on an annulus and high-genus surfaces.
The blue arrows represent the CNOT gates.  The white (empty) circles  represent the added ancilla qubits, which are intialized in
$\ket{0}$ or $\ket{+}$. The yellow (filled) circles represent qubits to be removed from the code.  Pink solid lines represent edges to be
added to the code, while yellow dashed lines represent edges to be removed from the code.}
  \label{gadgets}
\end{figure*}

The basic move in Fig.~\ref{fig:basic-moves} adds an ancilla qubit (white circle in Fig.~\ref{fig:basic-moves}), which effectively
adds an edge $e$ and a triangular plaquette in the center of the square plaquette.
The arrows represent the two-qubit CNOT gate, where the tail and head represent the control and target of the CNOTs. The corresponding
quantum circuit is shown in the right panel. The Pauli operators in the Heisenberg picture are transformed by CNOTs as:
\begin{align}\label{CNOT_relation}
\non \text{CNOT} \ (X\otimes I) \ \text{CNOT} =& X\otimes X, \\
\non \text{CNOT} \ (I\otimes X) \ \text{CNOT} =& I\otimes X, \\
\non \text{CNOT} \ (Z\otimes I) \ \text{CNOT} =& Z\otimes I,\\
\text{CNOT} \ (I\otimes Z) \ \text{CNOT} =& Z\otimes Z.
\end{align}
As indicated by the quantum circuit in Fig. \ref{fig:basic-moves}, a single $Z_e$ operator of the ancilla thus propagates through the two CNOTs into a
3-local stabilizer on the triangular plaquette:
\be
Z_e \longmapsto Z_a Z_b Z_e.
\ee
Since the ancilla qubit is initialized to $\ket{0}_e$, i.e., the eigenstate of $Z_e$ with eigenvalue $+1$, the grown stabilizer
$Z_a Z_b Z_e$ is also fixed at $+1$. According to Eq.~\eqref{CNOT_relation}, the original 4-local stabilizer $Z_a Z_b Z_c Z_d$
is untouched by the quantum circuit, therefore its $+1$ eigenvalue is preserved.    Since $Z_a Z_b Z_c Z_d =(Z_a Z_b Z_e)(Z_c Z_d Z_e)$,
the other triangular stabilizer $Z_c Z_d Z_e$ is automatically fixed at $+1$.
It is also straightforward to verify that the vertex Pauli-$X$ stabilizers will also grow to include the new edge.

This procedure can be reversed to remove (disentangle) a qubit from the toric code ground state by applying the inverse
of this procedure. In this case (but not for the generalization to $\mathbb{Z}_N$) the circuit is its own inverse.  The above elementary move and its reverse is enough for all the local geometric deformation in our protocol of Dehn twist on a torus in Sec.~\ref{sec:Dehn_twist_torus_toric} and high two types of Dehn twists on a high genus surfaces in Sec.~\ref{sec:high_genus_toric}.

For other protocols discussed in the subsequent sections, we will use a number of other simple quantum circuits, which we refer to as gadgets,
in order to implement other local geometric changes of the lattice structure.

\begin{figure*}
  \includegraphics[width=1.8\columnwidth]{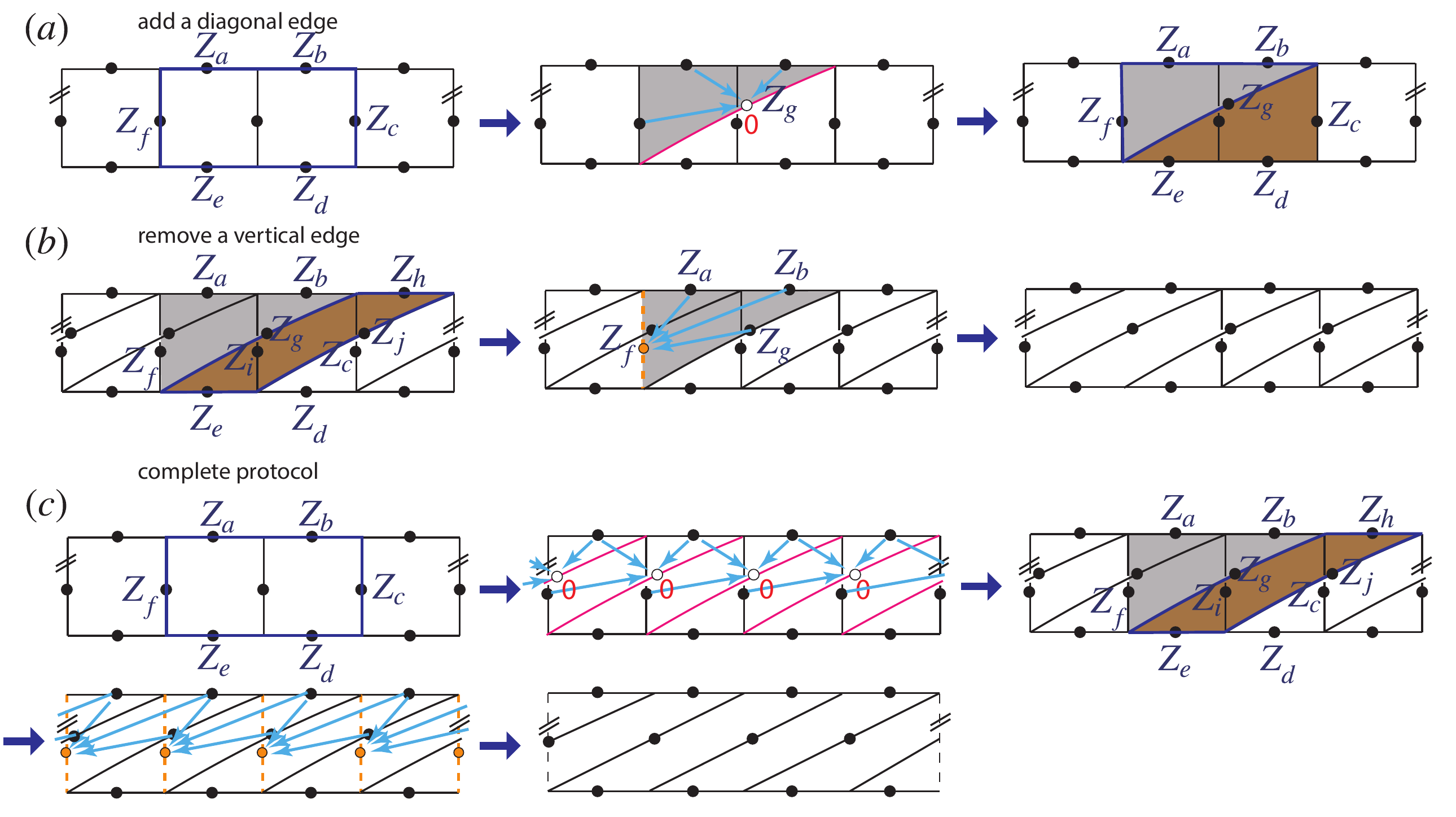}
  \caption{Creating NNN slanted plaquettes through a constant depth circuit. The grey and brown shades represent
stabilizers with triangular or parallelogram shape.}
\label{fig:long_range_plaquette}
\end{figure*}

Consider the moves shown in Fig.~\ref{gadgets}, which will be used for Dehn twists on an annulus in Sec.~\ref{sec:annulus_toric}, and on high-genus surfaces in Sec.~\ref{sec:high_genus_toric}, as well as the braiding protocols in Sec.~\ref{sec:braiding_sections}.   Moves I and II achieve a fine graining (splitting one plaquette into two) and
coarse graining (merging two plaquettes into one), respectively, along the vertical directions.  In move I, the ancilla qubits in
white circles are initialized in the $|0\rangle$ state or the $|+\rangle$ state, depending on whether they are on the horizontal or
vertical edges, as shown.

Move III implements a coarse graining that merges a square and a triangular plaquette together into a trapezoid, while merging
the two neighboring square plaquettes into a single rectangular one.  Finally move IV splits a
square plaquette into a triangular plaquette and a trapezoidal plaquette. Note that all of the required CNOT operations (indicated by blue arrows)
commute with each other and thus can be implemented in any order.

The next set of moves will be used for alternative single-shot protocols for braiding, Dehn twists on an annulus and high-genus surfaces, as well as multiple Dehn twists in a single shot in Sec.~\ref{sec:multiple_Dehn_twists_section}. They create slanted plaquettes with diagonal edges that can be of some arbitrary length, $\ell$. This can
be achieved with a constant depth local circuit with a range $\ell$. The protocol for the simplest case, namely creating a slanted
plaquette with a next-nearest-neighbor (NNN) diagonal edge [which we can label as the vector (2,1)]
on a $\Z_2$ toric code is shown in Fig.~\ref{fig:long_range_plaquette}.
In Fig.~\ref{fig:long_range_plaquette}(a), we apply three CNOTs targeting an ancilla $g$ initialized at $\ket{0}$
(the $+1$-eigenstate of $Z_g$), conditioned by the qubits $a$, $b$, and $f$.   According to Eq.~\eqref{CNOT_relation},
the operator $Z_g$ is transformed as
\be
Z_g \longmapsto Z_a Z_b Z_g Z_f,
\ee
which effectively introduces the NNN diagonal edge overpassing a vertical edge and the triangular plaquette (grey shadow) associated with a
4-body stabilizer $Z_a Z_b Z_g Z_f=+1$ coexisting with all the previous stabilizers,  as shown in Fig.~\ref{fig:long_range_plaquette}(a).
Since the initial configuration in Fig.~\ref{fig:long_range_plaquette}(a) has a 2-plaquette stabilizer $Z_a Z_b Z_c Z_d Z_e Z_f=+1$,  the
stabilizer on the other triangle (brown shadow) divided by the diagonal edge, $Z_g Z_c Z_d Z_e$, is automatically set to $+1$, due to
the decomposition $Z_a Z_b Z_c Z_d Z_e Z_f=(Z_a Z_b Z_f Z_g )(Z_g Z_c Z_d Z_e)$.    One can also easily verify the transformation of the
vertex stabilizers.

One important fact is that the addition of different NNN diagonal edges (2,1) shown in Fig.~\ref{fig:long_range_plaquette}(b) can be done in parallel. In this way, one creates the slanted plaquettes (one shown in brown shadow)
with the NNN diagonal edges overpassing the vertical edges of the original square plaquettes.   To see the stabilizers
corresponding to the slanted plaquettes are fixed at $+1$, we can proceed as follows.

First we note that from the argument above, the elongated triangular stabilizers are all one, e.g. $Z_b Z_h Z_j Z_i=+1$
and  $Z_a Z_b Z_f Z_g=+1$. Next, we consider the trapezoid (union of the region in grey and
brown shadow) which can be considered as a combination of a square plaquette associated with the stabilizer
$Z_a Z_i Z_e Z_f = +1$ and the triangle on the right associated with the stabilizer $Z_b Z_h Z_j Z_i=+1$.
Therefore, the multiplication of these two stabilizers gives rise to the trapezoid stabilizer $Z_a Z_b Z_h Z_j Z_e Z_f =(Z_b Z_h Z_j Z_i) (Z_i Z_a Z_e Z_f)=+1$.
Now, the trapezoid can also be split into the triangle on the left (grey shadow) with stabilizer $Z_a Z_b Z_f Z_g=+1$
and the slanted plaquette with stabilizer $Z_h Z_j Z_e Z_g$, i.e., $Z_a Z_b Z_h Z_j Z_e Z_f = (Z_h Z_j Z_e Z_g)(Z_a Z_b Z_f Z_g) =+1$.
It follows that the stabilizer associated with the slanted plaquette $Z_h Z_j Z_e Z_g$ is also fixed at $+1$.

Now we can get rid of all the verticle edges [which we can label as the vector $(0,1)$] and also the triangular
stabilizers, with the circuit shown in Fig.~\ref{fig:long_range_plaquette}(c), where, for example, the qubit
$f$ is disentangled by the three CNOTs with the controls being the other three sites associated with the triangular
stabilizer (grey shadow).  Similar to the procedure of adding diagonal edges, the removal of the vertical edges can
also be done in parallel as shown in Fig.~\ref{fig:long_range_plaquette}(c).  One hence ends up with a lattice with
slanted plaquettes containing NNN diagonal edges in Fig.~\ref{fig:long_range_plaquette}(c).

This protocol for creating slanted plaquettes spanned by the horizontal edge $(1,0)$ and NNN diagonal edge $(2,1)$
can be generalized to creating more slanted plaquettes spanned by $(1,0)$ and $(n,1)$, where $n$ is an arbitrary integer.
By symmetry, it can also be generalized to creating slanted plaquette spanned by the vertical edge $(0,1)$ and the
diagonal edge $(1, n)$.  Even more generally,  we can create a parallelogram plaquette spanned by the edge
$(m, n)$ and $(m', n')$.

We emphasize that since the plaquettes can all be operated on in parallel, we can convert a lattice with square plaquettes
to any type of slanted plaquettes by a constant depth circuit, independent of the size of the lattice.

\subsubsection{General $N$}

\begin{figure}
 \includegraphics[width=1\columnwidth]{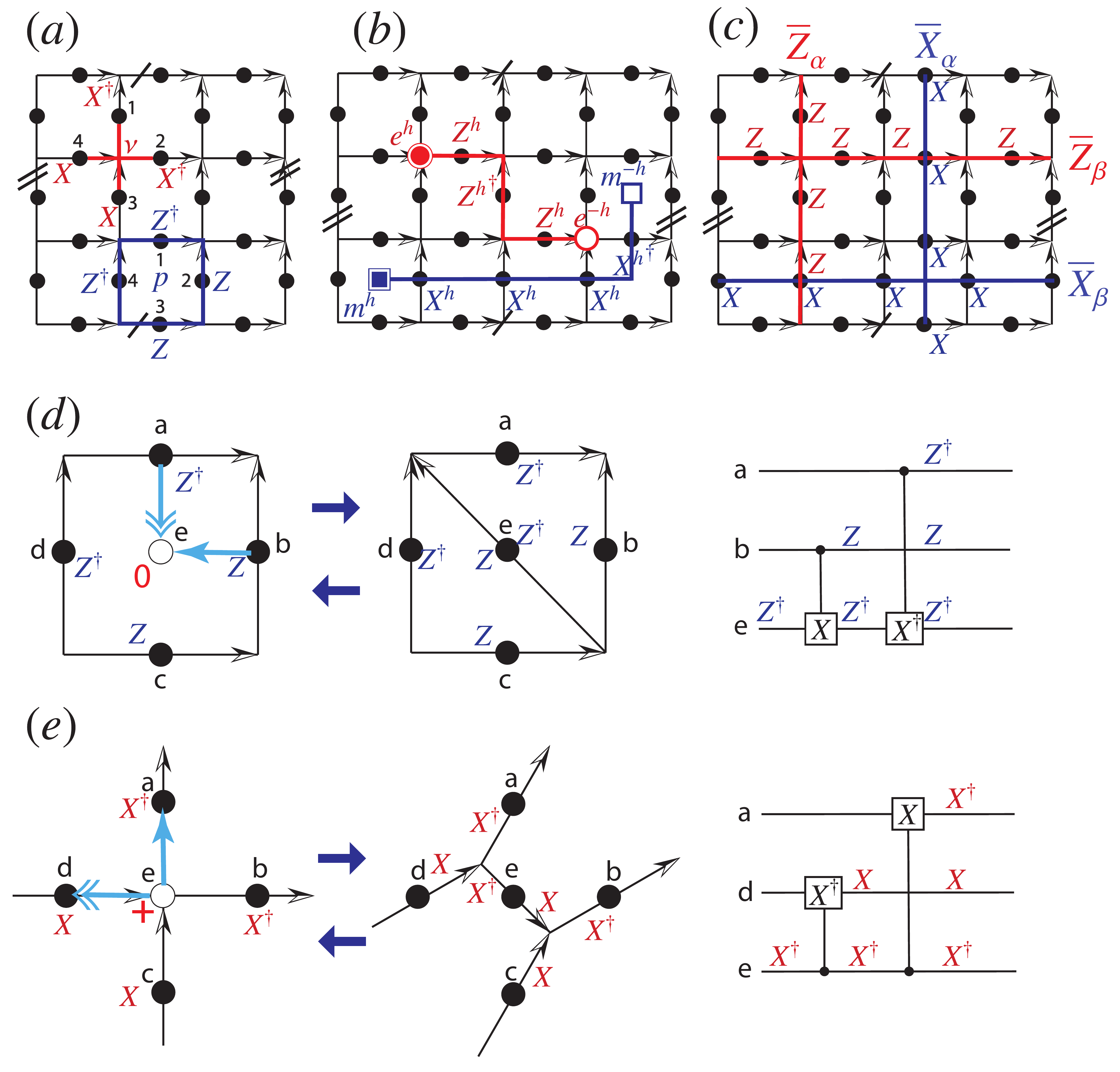}
 \caption{Definition and elementary move for the $\Z_N$ toric code.  The edges of the lattice are directed (with arrows) to represent the Hamiltonian. (a) Two types of stabilizers.  (b) Two types of anyons strings with corresponding anyon pairs in the ends.  (c) Two types of logical string operators along two directions on a torus.   (d) Elementary move for adding an edge. The single and double arrows represent the $CX$ and $CX^\dag$ gates respectively.  (e) Elementary move for removing an edge.   }
 \label{fig:ZN_definition}
\end{figure}

Now we generalize the above results to the $\Z_N$ toric code. On each edge of the lattice we now have a qudit with
$N$ states. The Hamiltonian is
\be
H_{\Z_N}= - \sum_{\nu} X^\dag_{\nu, 1} X^\dag_{\nu, 2} X_{\nu, 3} X_{\nu, 4} - \sum_{p} Z_{p, 1}^\dag Z_{p, 2} Z_{p, 3} Z^\dag_{p, 4},
\ee
where the vertex and plaquette operators are illustrated in Fig.~\ref{fig:ZN_definition}.
Here, the shift operators (generalized Pauli operators) for the $N$-level qudits are defined by
\begin{align}
X =& \sum_{n=0}^{N-1} \ket{(n+1) \ \text{mod} \ N}\bra{n},
\nonumber \\
 X^\dag =& \sum_{n=0}^{N-1} \ket{(n-1) \ \text{mod} \ N}\bra{n},
\nonumber \\
Z =& \sum_{n=0}^{N-1}\omega^n \ketbra{n},
\end{align}
where $\omega = e^{2\pi i/N}$. The shift operators satisfies the Weyl algebra $ZXZ^\dag X^\dag = \omega$.

A useful way of representing the Hamiltonian is to consider the edges to be directed, as the arrows on the edges
indicate in Fig. ~\ref{fig:ZN_definition}.

The topological charges in this model are electric charge $e^h$, magnetic charge $m^h$, and the composites
$e^h m^l$, where $h, l = (0, 1, 2.., N-1)\ \text{mod} \ N$.  As shown in Fig.~\ref{fig:ZN_definition}, anyon
$e^h$ and its antiparticle $e^{-h}$ can be created out of the vacuum by a string operator with $Z^h$ on the
horizontal edges and ${Z^{h}}^\dag$ on the vertical edges.   Similarly, $m^h$ and $m^{-h}$ can be created
by a string involving $X^h$ and ${X^{h}}^\dag$.  The logical qudit shift operators on a torus are
$\overline{X}_\alpha$, $\overline{X}_\beta$, $\overline{Z}_\alpha$ and $\overline{Z}_\beta$ as shown in Fig.~\ref{fig:ZN_definition}(c) up to local deformation.

The generalization of CNOT is the controlled-$X$ gate defined as
\be
CX=\sum_{n=0}^{N-1}\sum_{m=0}^{N-1}\ketbra{n} \otimes \ket{(m+n)\ \text{mod} \ N}\bra{m},
\ee
i.e., the value of the target qudit undertakes a conditional addition of the value of the control qudit.
The shift operators in the Heisenberg picture are transformed by $CX$ or $CX^\dag$ as:
\begin{align}\label{CX_relation}
\non CX \ (X\otimes I) \ CX^\dag =& X\otimes X, \ \ \
 CX \ (I\otimes X) \ CX^\dag = I \otimes X, \\
\non CX^\dag \ (X\otimes I) \ CX =& X\otimes X^\dag, \ \
 CX \ (Z\otimes I) \ CX^\dag = Z\otimes I,\\
CX \ (I\otimes Z) \ CX^\dag =& Z^\dag \otimes Z, \ \ \
CX^\dag \ (I\otimes Z) \ CX = Z \otimes Z.
\end{align}

All of the gadgets described in the previous section for locally changing the geometry of the lattice can be
straightforwardly generalized to the case of the $Z_N$ toric code by replacing CNOT with $CX$ or $CX^\dag$.

For example, consider the splitting of a plaquette shown in Fig.~\ref{fig:ZN_definition}(d).
A single $Z^\dag_e$ operator of the ancilla propagates through the $CX$ (single arrow) and $CX^\dag$ (double arrow)
into a 3-local stabilizer on the triangular plaquette, i.e.,
\be
Z^\dag_e \longmapsto Z^\dag_a Z_b Z^\dag_e,
\ee
according to the Hermitian conjugate of the last line in  Eqs.~\eqref{CX_relation}, namely,
\be\label{CX_relation2}
CX \ (I\otimes Z^\dag) \ CX^\dag = Z \otimes Z^\dag, \quad CX^\dag \ (I\otimes Z^\dag) \ CX = Z^\dag \otimes Z^\dag.
\ee
Note that the tail and head of the arrows represent the control and target of the  $CX$ and $CX^\dag$ gates, as in the $N = 2$ case for CNOT gates.

Since the ancilla qubit is initialized in the state $\ket{0}_e$, i.e., the eigenstate of $Z_e$ and $Z^\dag_e$ with eigenvalue $+1$,
the grown stabilizer $Z^\dag_a Z_b Z^\dag_e$ is also fixed at $+1$. According to Eqs.~\eqref{CX_relation}, the original
4-local plaquette stabilizer $Z^\dag_a Z_b Z_c Z^\dag_d$ is untouched by the quantum circuit, therefore its $+1$ eigenvalue
is preserved. Since $Z^\dag_a Z_b Z_c Z^\dag_d =(Z^\dag_a Z_b Z^\dag_e)(Z_e Z_c Z^\dag_d)$, the other triangle stabilizer
$Z_c Z^\dag_d Z_e$ is automatically fixed at $+1$.

Similarly one can verify that the vertex stabilizer terms are also grown appropriately.

One can also add vertices to the lattice. To split a vertex with four edges into two vertices with three edges,
we follow the procedure shown in Fig.~\ref{fig:ZN_definition}(d). The ancilla qubit $e$
is initialized in the $\ket{+}_e$ state, i.e.  the eigenstate of $X_e$ and $X^\dag_e$ with eigenvalue $+1$.
A single $X^\dag_e$ operator of the ancilla propagates through the
$CX$ (single arrow) and $CX^\dag$ (double arrow) into a 3-local vertex stabilizer, i.e.,
\be
X^\dag_e \longmapsto X^\dag_a X_d X^\dag_e,
\ee
according to the Hermitian conjugate of the relations in  Eqs.~\eqref{CX_relation}, namely,
\be\label{CX_relation3}
CX (X^\dag \otimes I) CX^\dag = X^\dag \otimes X^\dag, \ CX^\dag (X^\dag \otimes I) CX = X^\dag \otimes X.
\ee
Since the ancilla qubit is initialized in the $\ket{+}_e$ state, the eigenvalue of the grown stabilizer $X^\dag_a X_d X^\dag_e$
is also fixed at $+1$. Similar to the case with the plaquette stabilizers, the other 3-term vertex stabilizer
$X^\dag_b X_c X_e$ is also automatically fixed at $+1$.

In the following sections, we describe in detail the protocols for implementing Dehn twists and braids for the $\mathbb{Z}_N$ toric code
state. Most of our results will be presented for the case $N = 2$; the generalization to arbitrary $N$ is straightforward.

\subsection{Dehn twist on a torus}\label{sec:Dehn_twist_torus_toric}

\begin{figure*}
  \includegraphics[width=2\columnwidth]{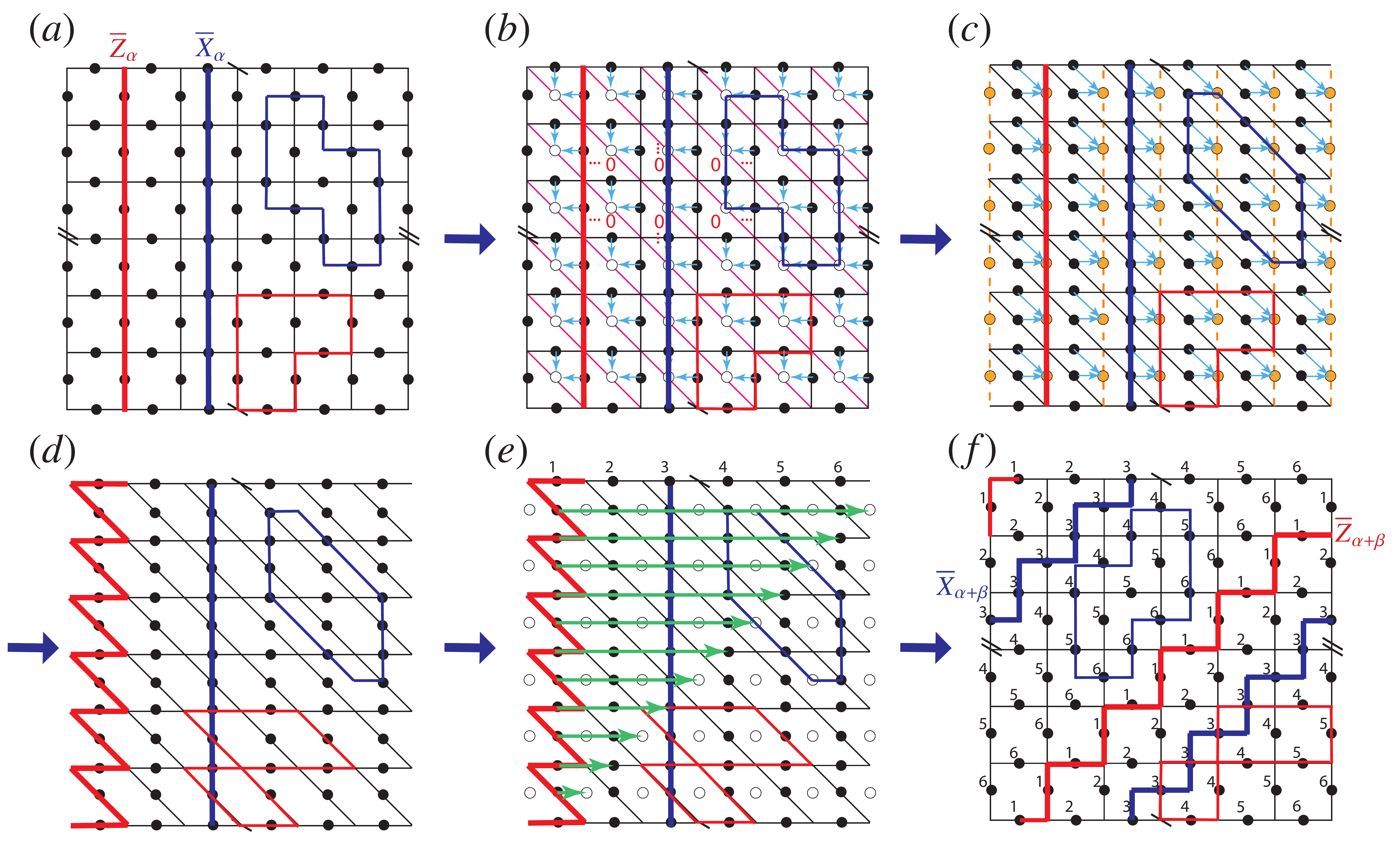}
  \caption{Implementing Dehn twist $\mathcal{D}_\beta$ on a torus for the $\Z_2$ toric code.  The thick lines represent the non-contractible loops of two types $\overline{Z}$ (red) and $\overline{X}$ (blue).  The thin lines represent local contractible loops of two types. The blue arrows in (b) and (c) represent CNOTs, in order to add qubits (white circles, initialized at $\ket{0}$) and edges (pink solid lines), or to remove  qubits (yellow circles) and edges (yellow dashed lines).  The green thick arrows in (e) represents permutation of qubits in a shearing pattern.   }
\label{fig:Dehn-twist}
\end{figure*}

Using the basic moves described in the previous section to entangle / disentangle ancilla qubits in the toric code ground state, we can then implement the protocol
for applying a Dehn twist on the torus. The complete protocol is shown in Fig.~\ref{fig:Dehn-twist}. For illustration purposes,
we show the transformations of both small (contractible) loops and large (non-contractible) loops of $e$ (red)
and $m$ (blue) anyons respectively.  These loops are created by applying Pauli string operators $\overline{Z}=\prod_j Z_j$ and
$\overline{X}=\prod_j X_j$ respectively.

We start with the lattice configuration shown in Fig.~\ref{fig:Dehn-twist}(a). We then incorporate the ancilla qubits (white)
in the center of each plaquette by applying the basic moves, as shown in (b). This can be done with two transversal CNOT
operations, first with the qubits on vertical edges as the control qubits, and next with the qubits on horizontal edges as the control qubits.
Next, as depicted in (c), we apply two transversal CNOT operations targeted on the vertical edges (yellow circles). This causes the qubits on the vertical
edges to become disentangled, thus effectively removing them from participating in the toric code ground state. This
results in the toric code ground state existing on the slanted lattice shown in (d); the ground state is then a superposition of slanted loops.
The sequence of operations from (a) to (d) is therefore a local, finite depth quantum circuit, which we label $\mathcal{LU}_\beta$.
This circuit changes the local geometry of the topological wave function from a condensation of square-shaped loops to a condensation of
slanted loops, which exactly corresponds to the local geometry deformation we introduced before in Sec.~\ref{sec:concepts} (Fig.~\ref{fig:MCG_definition}).

To complete the Dehn twist protocol, we perform a shear deformation by a qubit permutation, $P_\sigma$, shown in (e). The qubits in each row are cyclically permuted
to the right by a number of spacings depicted by the green arrows. This leads to the configuration shown in (f), which we can see is equivalent to the starting
lattice geometry. Therefore the shear recovers the original lattice, and at the same time induces a Dehn twist  of both the large $e$ and $m$ loops along $\beta$, namely
\begin{align}
\mathcal{P}_\sigma \mathcal{LU}_{\beta}: \overline{X}_\alpha &\longrightarrow  \overline{X}_{\alpha+\beta},
\nonumber \\
\mathcal{P}_\sigma \mathcal{LU}_{\beta}: \overline{Z}_\alpha &\longmapsto  \overline{Z}_{\alpha+\beta},
\end{align}
and hence $\mathcal{P}_\sigma \mathcal{LU}_{\beta} =\mathcal{T}=\mathcal{D}_\beta$.  The small
loops are deformed but remain closed under this combined transformation. Therefore the
ground state continues to be an equal weight superposition of closed loops, and the topological ground state (code) subspace
is preserved under this operation.

By implementing the $\mathcal{LU}$ transformation to slant the lattice in the other direction and then a permutation corresponding to a vertical
shear, we can analogously obtain a Dehn twist along $\alpha$.

It is clear that the same operation can also be performed on a cylinder instead of a torus, where we perform a Dehn twist along
the single closed non-contractible cycle of the cylinder.

We note that a different set of protocols for implementing a Dehn twist on a torus in the context of the $\mathbb{Z}_2$ toric code was proposed recently in
Ref.~\onlinecite{Breuckmann:2017hy}. Either one applies a sequence of long-range CNOTs with an $\mathcal{O}(d)$ overhead, or performs only
CNOTs in parallel across the lattice with constant-time overhead. In the latter protocol, the role of the long-range permutations was not discussed,
which is crucial for consideration of multiple Dehn twists.

\subsection{Dehn twist on an annulus}\label{sec:annulus_toric}

In order to set up the rest of the required protocols for arbitrary braids and Dehn twists, we now consider
protocols which implement a Dehn twist along the non-contractible cycle, $\beta$, of an annulus. Again
here we restrict our attention to the case of the $\mathbb{Z}_2$ toric code. We consider
two distinct protocols for implementing the Dehn twist $\mathcal{D}_\beta$ along $\beta$.

\vspace{0.1in}

\subsubsection*{ Protocol 1:  Twist via shearing}
\label{DannulusShear}

\begin{figure*}
\includegraphics[width=1.6\columnwidth]{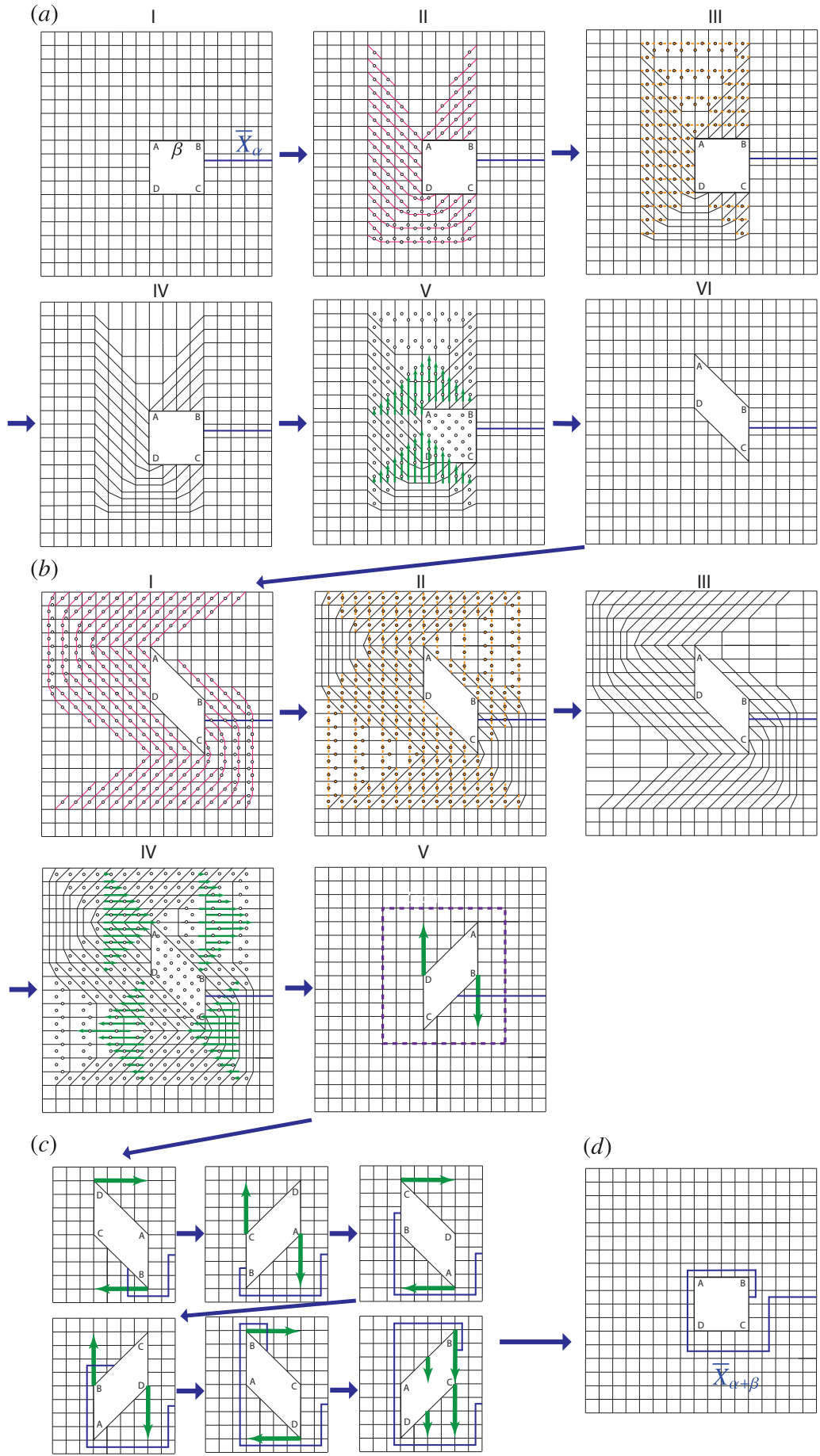}
\caption{Dehn twist along the boundary of the inner defect (the $\beta$-loop) on an annulus via shearing of the defect. An overall $2\pi$-rotation (indicated by the the letters on the 4 corners) is induced by 9 steps of shearing.}
\label{fig:Dehn_twists_annulus}
\end{figure*}

The right Dehn twist $\mathcal{D}_\beta$ on a single annulus can be implemented according to the
protocol shown in Fig.~\ref{fig:Dehn_twists_annulus}.  The essence of this protocol is to implement a
$2\pi$-rotation of the inner boundary by a sequence of operations combining local finite depth circuit
and qubit permutation:
\begin{align}
\mathcal{D}_\beta= \prod_{i=1}^9  \mathcal{P}_{\sigma, i} \mathcal{LU}_i.
\end{align}
Since the $2\pi$-rotation of the square boundary defect can always be decomposed into a sequence of shears, we can just perform
entangling and disentangling gates followed by a qubit permutation to realize each shearing process.

In Fig.~\ref{fig:Dehn_twists_annulus}(a) we illustrate a logical string operator $\overline{X}_\alpha$ along the $\alpha$-line connecting the inner and outer boundary
of the annulus. The first step of shearing the boundary defect is shown in Fig.~\ref{fig:Dehn_twists_annulus}(a)I-VI.  In panel II, we
add (entangle) qubits (white circles) and effectively add edges (red lines) to the code using one step of the elementary CNOT circuits from Fig.~\ref{fig:basic-moves}.
In panel III, we then remove (disentangle) qubits and effectively remove edges (dashed yellow lines), which leads to the deformed lattice
in panel IV.  The lattice in panel IV is a deformed square lattice except the defect region.  In particular,
it is coarse-grained in the region above the defect and fine-grained in the region below, and is sheared on the two sides.  In panel V,
we perform a permutation of qubits, where some of them are moved to the unoccupied ancilla qubits (white circles). After that, we recover the regular square as
shown in panel VI with the original square defect being sheared to a parallelogram.

The second step of shearing the defect is shown in Fig.~\ref{fig:Dehn_twists_annulus}(b)I-V.   Now in panel I and II we
add and remove qubits and edges to get the deformed lattice in panel III,  where the shearing and coarse/fine-graining procedure
is opposite in the top and bottom parts.  Now we again perform qubit permutation in opposite directions in the top and bottom parts as shown in
panel IV, leading to the further sheared parallelogram defect in panel V where the vertex A is permuted to the upper-right corner.  We repeat
this procedure of shearing the parallelogram defect as shown in Fig.~\ref{fig:Dehn_twists_annulus}(c) and finally shear it back to the square defect in
Fig.~\ref{fig:Dehn_twists_annulus}(d). The square boundary defect has been rotated by a full $2\pi$ cycle which leads to a right Dehn twist of the
string $\overline{X}$ or $\overline{Z}$ connecting the inner and outer boundary of the annulus along the $\beta$-loop, i.e.,
\be
\mathcal{D}_\beta: \overline{X}_{\alpha} \longmapsto \overline{X}_{\alpha+\beta},
\quad
\mathcal{D}_\beta: \overline{Z}_{\alpha} \longmapsto \overline{Z}_{\alpha+\beta}
\ee
The transformation of the $\overline{X}$-string is illustrated in Fig.~\ref{fig:Dehn_twists_annulus}(d).

Note that in total we have passed through 8 configurations of the parallelogram defect, and have used 9 composite steps of
local quantum circuit and qubit permutation in total.  Finally we note that if we start with an annulus geometry having a parallelogram
defect in the middle as shown in Fig.~\ref{fig:Dehn_twists_annulus}(a)VI and end up with the same shape, the total
number of composite steps can be reduced to 8.

\vspace{0.1in}

\subsubsection{Protocol 2: Single-shot twist}\label{sec:alternate_beta}

\begin{figure*}
  \includegraphics[width=1.6\columnwidth]{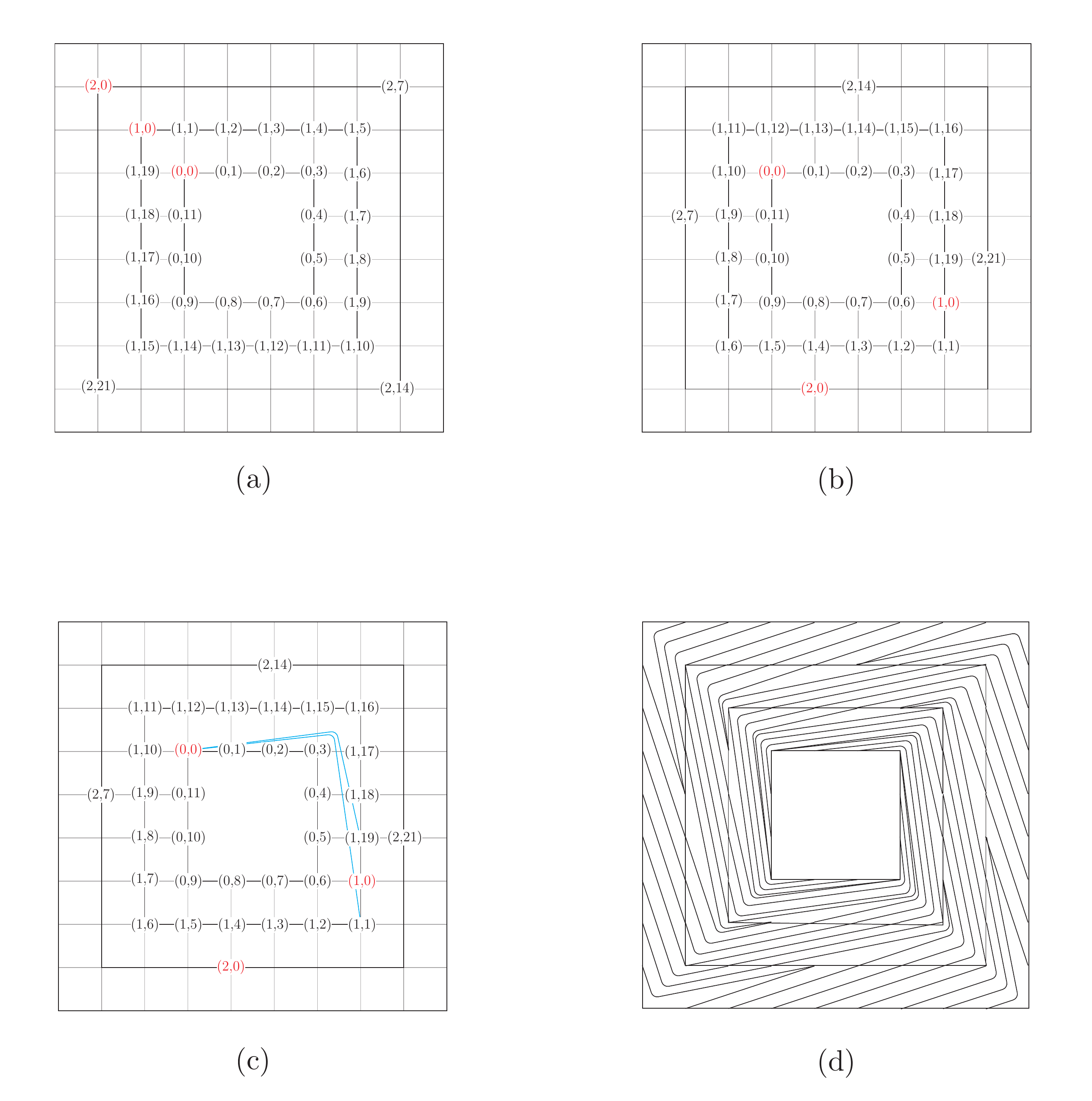}
  \caption{(a) A hole with perimeter $L=12$ in a square lattice. The qubits are on the edges but are not shown for simplicity.
The sites are divided into rings surrounding the hole and are numbered clockwise. We start with the upper left corner, shown
in red numbers, and move clockwise.  (b) We shift the labels on the $n$th ring by $9n$ sites. The labels which originally
corresponded to upper left corners are shown in red. (c) After shifting labels, we reconnect the sites as they were connected originally. For example, since the $(0,0)$ site was connected to $(1,1)$ and $(1,19)$ in panel (a), it should be reconnected by the blue lines shown above. (d) The modified lattice after reconnecting the sites.}
\label{fig:modular_genus_g_beta_long_range}
\end{figure*}

Here we described another procedure for implementing the Dehn twist around the $\beta$-loop on the annulus.
In contrast to the previous protocol, this protocol can be implemented in a single step, so
$\mathcal{D}_\beta = \mathcal{P}_\sigma \mathcal{LU}_\beta$, rather than the $9$ steps used in the previous protocol.
However the price to pay is that this protocol requires the local quantum circuit to have two-qubit gates with a
range $r \approx 9$.

Consider a hole inside a square lattice with perimeter $L$. As an example the case for $L=12$ is shown in
Fig.\ref{fig:modular_genus_g_beta_long_range}(a). Only a portion of the entire lattice has been shown in the
figure. By the following procedure we can perform a Dehn twist around this hole defect.

\begin{enumerate}
  \item
  We start by labeling the sites as shown in Fig.~\ref{fig:modular_genus_g_beta_long_range}(a). For the sake of clarity
only some of the labels are shown. We divide the lattice into rings. The sites that lie on the hole's boundary compose
the $0$th ring ($n=0$). the next ring consists of sites that neighbor the $0$th ring and so on. In Fig.~\ref{fig:modular_genus_g_beta_long_range}(a),
the rings are plotted with thicker lines to be distinguished. We then number the sites in each ring, starting from the site on the
upper left corner ($m=0$, denoted by red color in Fig.~\ref{fig:modular_genus_g_beta_long_range}(a)) and moving clockwise.
Note that the $n$th ring has $L+8n$ sites, so if $(n,m)$ represents the $m$th site on the $n$th ring, $m$ ranges between $0$ to $L+8n-1$.

Given the labels, one can describe the lattice by the set of links between the sites. For example, the $(0,0)$ site is connected to
$(1,1)$ and $(1,19)$ sites from the $n=1$ ring, $(0,1)$ is connected to $(1,2)$ and so on.

  \item
  Now, we shift the \textit{labels} on the $n$th ring by $9n$ sites clockwise along the ring. So, the site that was originally
labeled as $(n,m)$ will now be labeled as $(n,m-9n)$. Fig.~\ref{fig:modular_genus_g_beta_long_range}(b) shows some of
the shifted labels for the $L=12$ case.

  We will do this for $0\le n \le L-1$. Note that the sites on the $L$th ring would have been shifted by $9L$  sites. But since
the $L$th ring has $L+8L=9L$ sites in total, this shift is equivalent to doing nothing. So, For $n\ge L$ we leave the rest of the lattice unchanged.
\item
Now we implement a local, finite depth quantum circuit to reconnect the new labels the same way that the original labels
were connected. So, for example, in the $L=12$ case, the new $(0,0)$ site should be connected to the sites that carry
the labels $(1,1)$ and $(1,19)$ after the shift, as is shown in Fig.~\ref{fig:modular_genus_g_beta_long_range}(c).

Note that although the sites that were connected to each other in the original lattice were nearest neighbors,
after the shift we need to connect sites that are further apart. Nevertheless, the range of those links
remains finite and independent of the code distance. Since the labels on the $n$th and $(n+1)$th ring have been shifted by $9n$ and $9(n+1)$
sites respectively, they have been moved only by $9$ sites \textit{relative} to each other. So, the modified links'
range would be at most $9$.

The whole reconnecting procedure can be done in $2$ steps using the protocol in
Fig.~\ref{fig:long_range_plaquette}. Fig.~\ref{fig:modular_genus_g_beta_long_range}(d) shows
the resulting lattice after changing the links.

  \item
  Finally, we implement a permutation $\mathcal{P}_\sigma$ of qubits such that each site label is moved back to its original position.
\end{enumerate}

\begin{figure*}
  \includegraphics[width=1.8\columnwidth]{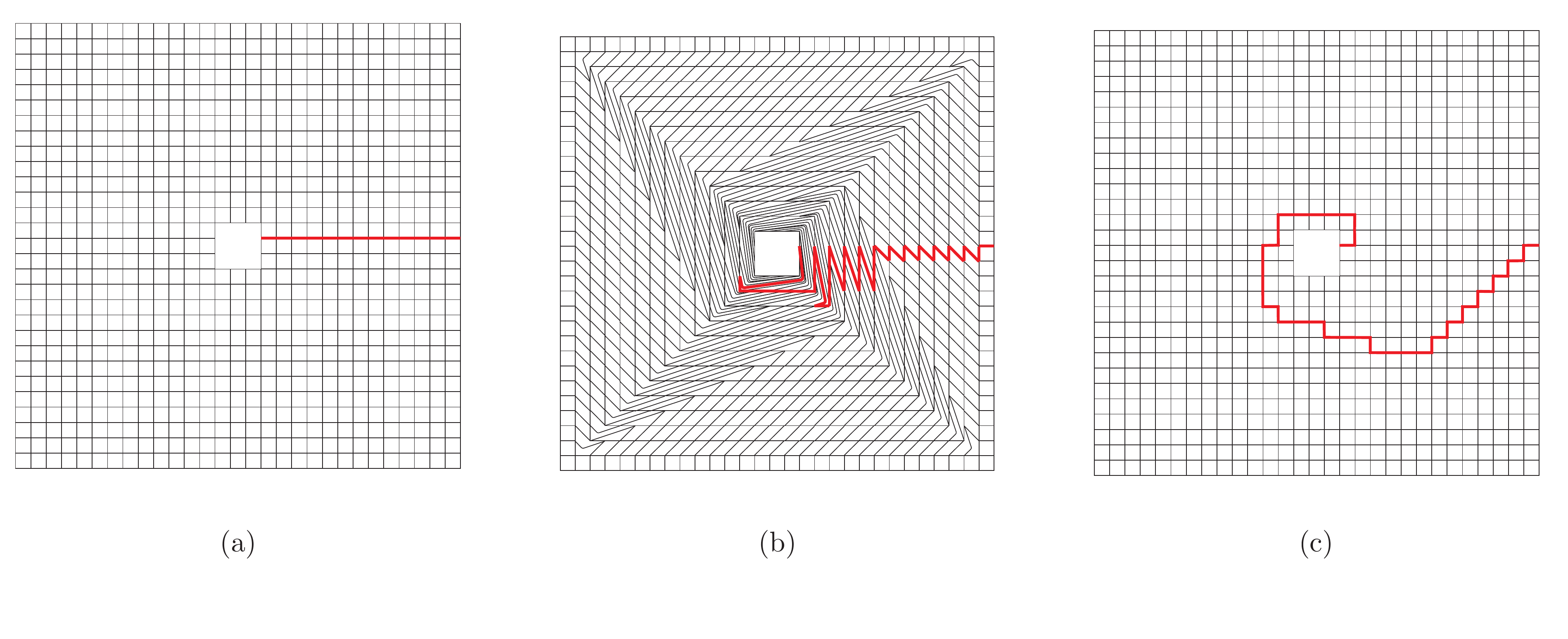}
  \caption{Performing Dehn twist around the $\beta$-loop in one shot (a) A boundary defect with length $L=12$ in a square lattice. A string operator which terminates on the hole's boundary is also shown in red color. (b) How the lattice and the string look like after reconnecting the sites. (c) Following the permutation we recover the original lattice, but now the string operator goes around the hole defect.}
\label{fig:modular_genus_g_beta_long_range_whole}
\end{figure*}

Fig.~\ref{fig:modular_genus_g_beta_long_range_whole} illustrates how a string operator transforms under this procedure.
Fig.~\ref{fig:modular_genus_g_beta_long_range_whole}(a) shows the hole defect and a string that terminates on the
hole's boundary. After reconnecting the sites, the lattice and the string will look as shown in Fig.~\ref{fig:modular_genus_g_beta_long_range_whole}(b).
Finally, after permuting the qubits, we recover the original lattice with the string encircling the hole
as is shown in Fig.~\ref{fig:modular_genus_g_beta_long_range_whole}(c).

\subsection{Braiding}\label{sec:braiding_sections}

\subsubsection{Protocol 1: braiding in multiple steps}\label{sec:braiding_protocol}

\begin{figure}
  \includegraphics[width=1\columnwidth]{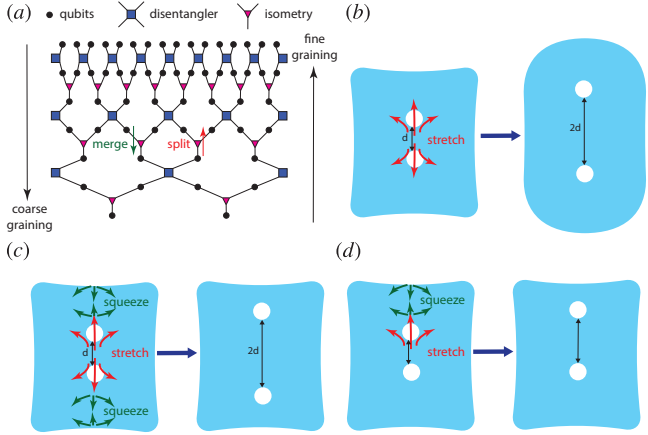}
  \caption{Understanding the essence of braiding via the equivalence of local entanglement renormalization
(its inverse) and manifold stretching (squeezing).  (a) The global MERA circuit performing coarse graining
(downward) or fine graining (upward). (b) Locally stretch the manifold in the region between the two punctures to
enlarge their distance by a factor of 2. (c) Besides stretching, also squeeze the upper and lower region to preserve
the shape of the manifold. (d) Stretch the puncture only in one direction to only move one puncture upward.}
\label{fig:braiding_intuition}
\end{figure}

\begin{figure*}
  \includegraphics[width=2\columnwidth]{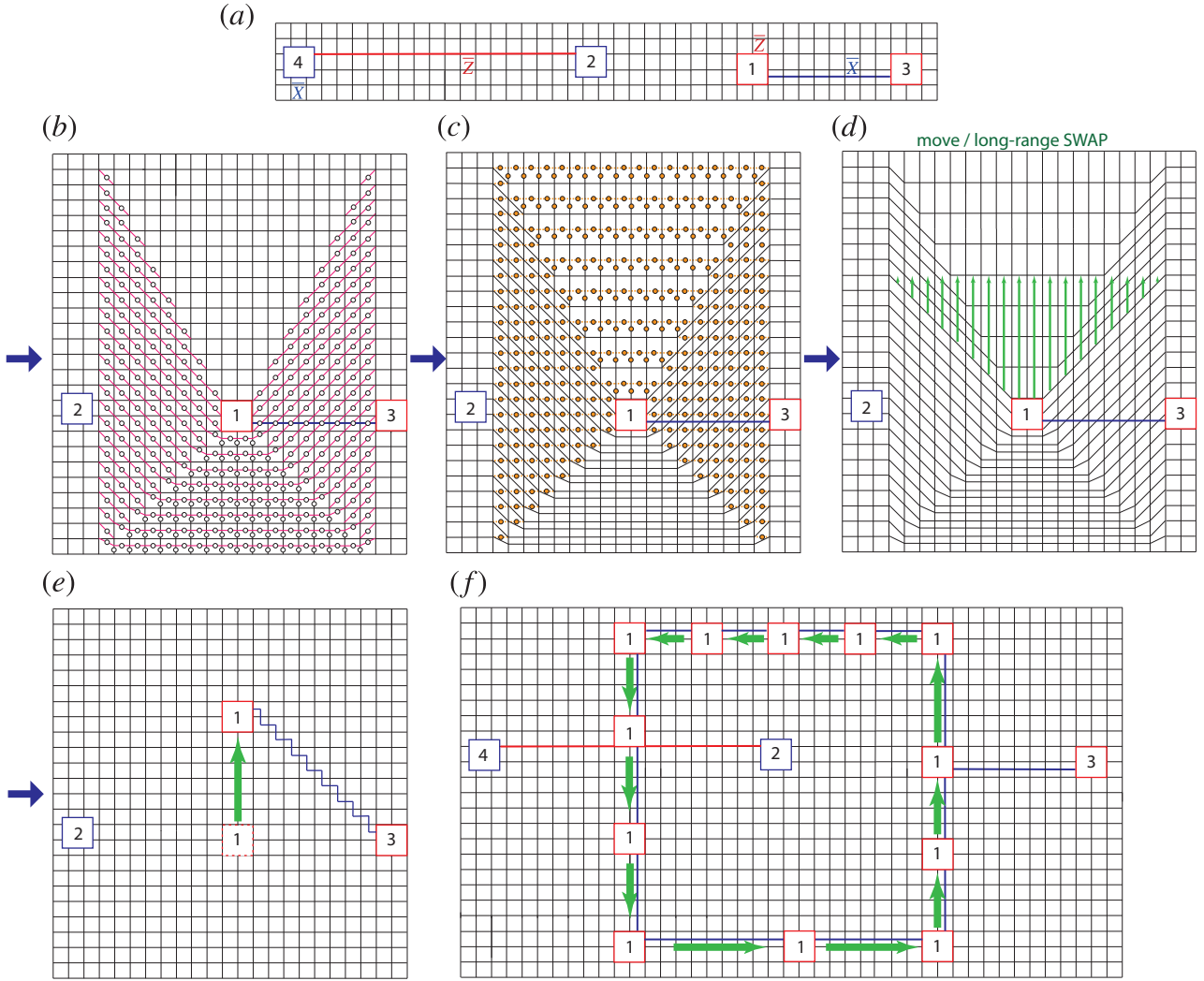}
  \caption{Braiding protocol via single-shot long-distance moving of the boundary defects. (a) Two pairs of boundary defects (X-cut and Z-cut). (b) Adding qubits (white circles) and edges (pink solid lines).  (c) Removing qubits (yellow circles) and edges (yellow dashed lines). (d). Permutation of qubits in a shearing pattern, $\mathcal{P}_\sigma$.  (e)  Effectively move the defect 1 with distance $\mathcal{O}(d)$ after the permutation. (f) A full braiding is implemented by 12 steps of $\mathcal{LU}$ and $\mathcal{P}_\sigma$.}
  \label{fig:braiding_circuits}
\end{figure*}

In this section we demonstrate a protocol for braiding which takes a finite number of steps:
$\mathcal{B}_{12} = \prod_i \mathcal{P}_{\sigma_i} \mathcal{LU}_i$. For concreteness, we will
demonstrate how to implement a full braid between two hole defects in the $\mathbb{Z}_2$ toric code
state. The protocol can be straightforwardly applied to the case of half-braids involving any type of holes,
anyons, or twist defects. Furthermore, the protocol only includes non-trivial operations that act in a given
subregion of the system. Therefore it can be applied to braids on any surface.

An intuitive way to understand the protocol is through the picture of entanglement renormalization and the
multi-scale entanglement renormalization ansatz (MERA).  The essence of entanglement renormalization and
the MERA circuit can be understood as a coarse-graining process that `merges' several qubits together, effectively
removing (disentangling) several qubits, as illustrated in Fig.~\ref{fig:braiding_intuition}(a). In the context of topological
order, one can think of this process as shrinking the manifold which supports the topological state.  The reverse of such a process is `fine-graining' which
`splits' one  qubit into several, effectively adding (entangling) qubits to the code.  One can think of this process as
stretching the manifold.  We consider anyons or defects as punctures in the manifold as illustrated in Fig.~\ref{fig:braiding_intuition}(b) which are distance $d$ apart.
In order to separate the two punctures further, one can perform one layer of the entanglement renormalization circuit (with constant depth)
locally to stretch (fine-graining) the region between the two punctures, which effectively adds qubits into the system.
The distance between the two punctures has effectively been increased by a constant factor that is independent
of the initial separation of the two punctures.  Now the manifold is effectively enlarged due to the addition of qubits.
In order to preserve the shape of the manifold away from the region of the punctures, one can also perform one layer of the
entanglement renormalization circuit locally to squeeze (coarse-grain) the region on the top and bottom sides of the punctures,
as shown in Fig.~\ref{fig:braiding_intuition}(c).  Thus one ends up with the same overall shape of the manifold, with
the two punctures being separated by a factor of 2, i.e., $d\rightarrow 2d$.   One could also stretch the manifold only on one side as illustrated in Fig.~\ref{fig:braiding_intuition}(d), which effectively moves the top puncture upward in this case and keeps the other puncture fixed.

The concrete braiding circuit for the surface code (toric code on a planar geometry) is shown in Fig.~\ref{fig:braiding_circuits}.

In Fig.~\ref{fig:braiding_circuits}(a), we show the setup under consideration, i.e., a pair of Z-cut defects with
smooth boundary (red) and a pair of X-cut defects with rough boundary (blue) in a surface code.  Each pair of
defects form a `double-cut' logical qubit.  The braiding of a Z-cut defect around the another X-cut defect
implements a logical $\overline{\text{CNOT}}$ gate.

In Fig.~\ref{fig:braiding_circuits}(b), one adds qubits (white circles) and edges into the code in the region below defect 1,
which effectively stretches the manifold. One also adds diagonal edges on the two sides which creates a shearing pattern.
Now, in Fig.~\ref{fig:braiding_circuits}(c), one removes half of the qubits (yellow circles) and edges in the region above
defect 1, in order to compensate for the added qubits. Thus we preserve the total number of qubits participating in the topological state.
These operations can be performed on different plaquettes in parallel, so we have a local finite depth quantum circuit that implements
these transformations. After the transformation, one obtains a deformed square lattice,  where the top part is squeezed, the lower part is stretched, and
the side part is sheared.

We then permute the qubits as shown in Fig.~\ref{fig:braiding_circuits}(d), to revert back to the regular square lattice
shown in Fig.~\ref{fig:braiding_circuits}(f). Note that the lattice in (d) and (e) are isomorphic to each other, and the qubit permutation
is exactly the implementation of the isomorphism between the two lattices.    As a consequence, defect 1 effectively moves $\mathcal{O}(d)$ sites
upward, as shown in Fig.~\ref{fig:braiding_circuits}(e). We see that up to the overall permutation of qubits, this transformation occurs
through a finite number of local two-qubit operations, independent of the code distance $d$. This is in sharp contrast to
conventional schemes which need $\mathcal{O}(d)$ steps.

Crucially, the number of sites that a defect can move in our scheme is bounded by the distance to the closest
defect perpendicular to the moving direction. Therefore, to implement a braid, we need to break up the braid
into a finite number of steps, where in each step the defect moves by an amount set by the distance to the closest
defect.

We see that a single full braiding operation can be always performed with a constant number of steps, independent of
system size and code distance. As summarized in  Fig.~\ref{fig:braiding_circuits}(f), we have demonstrated how the
full braid can be achieved with 12 steps:
\begin{align}
\mathcal{B}^2_{1,2}=\prod_{i=1}^{12} \mathcal{P}_{\sigma, i} \mathcal{LU}_i.
\end{align}
We note that there is freedom in our choice of steps to implement this protocol. The precise number of steps can be altered
depending on the exact geometry and finite size effects, but the main point is that the number of steps is finite and independent
of code distance and system size. The precise protocol that we illustrated has a slightly asymmetric nature (4 horizontal steps
in the upper region and 2 horizontal steps in the lower region). This is due to the fact that the maximum distance a puncture
can be moved is bounded by the separation of the defect 1 and 2, various finite size effects, and some choices made
to minimize the number of steps in the protocol.

Note that in this example we considered a full braid because in the $\mathbb{Z}_2$ toric code, that is
the only way to achieve a non-trivial logical operation by braiding hole defects. However one can also consider
half-braids (single exchange) which requires $\approx 6$ steps; a half-braid on two twist defects also
implements a non-trivial logical operation.

\subsubsection{Protocol 2: braiding via half-twist}

\begin{figure*}
  \includegraphics[width=2\columnwidth]{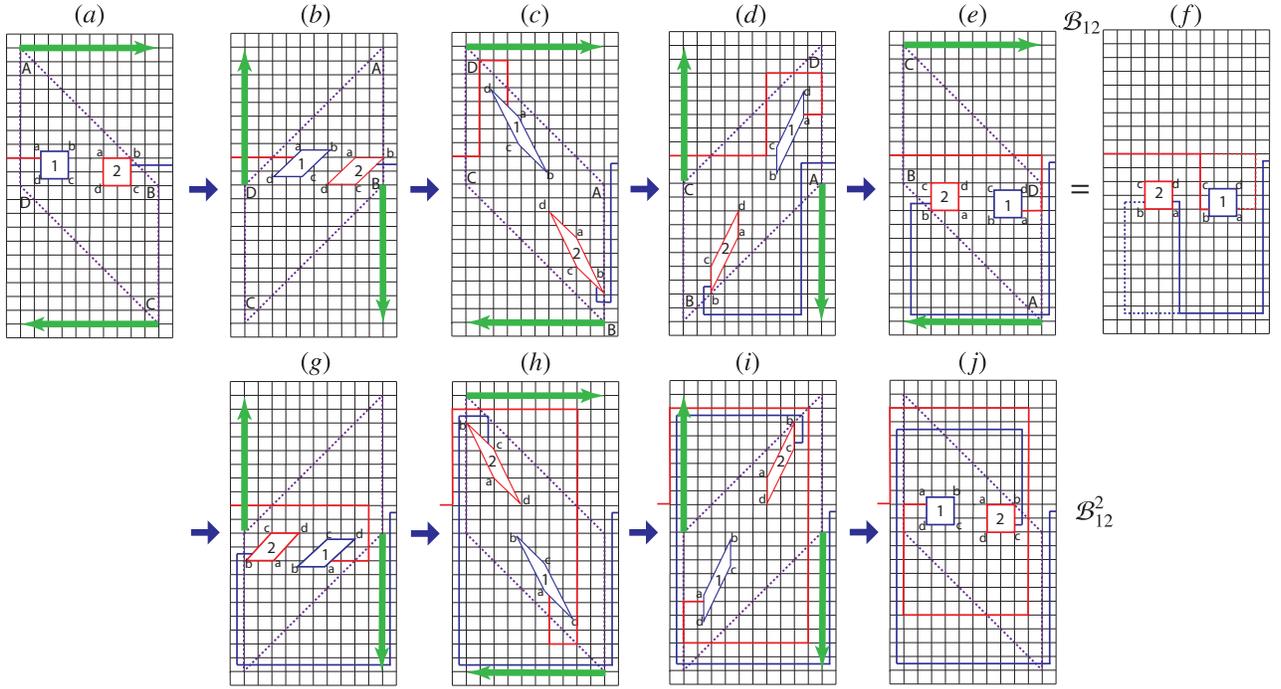}
  \caption{Implementing braiding (full braiding) via half twists (Dehn twists) through multiple steps of shearing operations.
The parallelogram (purple dashed lines) surrounding both punctures is a guide to the eye. The red and blue dashed loops in
(f) indicate gauge transformations, which shows that the half self-twists around each puncture are trivial operations. }
\label{fig:braiding_via_shearing}
\end{figure*}

The previous full braiding protocol in Sec.~\ref{sec:braiding_protocol} needs 12 steps, independent of the code distance.

One can consider a different protocol to further reduce the number of steps, by utilizing the equivalence
between (half) braiding of two punctures and a half twist around the two punctures.  The protocol of the latter can be
readily adapted from the protocol for implementing a Dehn twist on an annulus.

As has been discussed in Sec.~\ref{sec:concepts_braid} Fig.~\ref{fig:twists_definition}(e-g), a half twist along the $\beta$-loop enclosing two punctures is equivalent
to (half) braiding the two punctures, with additional half self-twists around each puncture. That is,
$\sqrt{\mathcal{D}}_\beta=\mathcal{B}_{1,2} \sqrt{\mathcal{D}}_{\omega_1} \sqrt{\mathcal{D}}_{\omega_2}$, where here
$\sqrt{\mathcal{D}}_\beta$, $\sqrt{\mathcal{D}}_{\omega_1}$, $\sqrt{\mathcal{D}}_{\omega_2}$ represent the half-twists around $\beta$, $\omega_1$, and $\omega_2$.

The half-twists $\sqrt{\mathcal{D}}_{\omega_i}$ for $i = 1,2$ can be implemented entirely by a local finite depth quantum circuit.
In general the effect of the half-twists $\sqrt{\mathcal{D}}_{\omega_i}$ around each puncture is simply to ensure that the punctures return
to exactly their original configuration. For example, a hole or an anyon can have some non-trivial geometric shape, such that a half-twist
changes its orientation, so that the half-twists $\sqrt{\mathcal{D}}_{\omega_i}$ are necessary to recover the exact original configuration.
In the case of anyons or twist defects, where the spatial extent is $\mathcal{O}(1)$, independent of the code distance, it is clear
that this can be accomplished entirely by a local finite depth quantum circuit. In the case of holes, even though the linear
size of the hole is $\mathcal{O}(d)$, we have seen how the lattice geometry can be changed everywhere in parallel through
a local finite depth quantum circuit.

Alternatively, we can just implement the half-twists $\sqrt{\mathcal{D}}_{\omega_i}$ using the same protocol for implementing
$\sqrt{\mathcal{D}}_\beta$, except by twisting around each puncture individually.

In the case that the two punctures are not identical, as illustrated in Fig.~\ref{fig:twists_definition}(h-j), we have to perform
a full braid to return the system to the original configuration and perform a logical gate. In this case, we can perform a full
Dehn twist around the $\beta$-loop, which is equivalent to a full braid with additional self Dehn twists around the two punctures,
i.e., $\mathcal{D}_\beta=\mathcal{B}_{1,2} \mathcal{D}_{\omega_1} \mathcal{D}_{\omega_2}$. For anyons, the full twist $\mathcal{D}_{\omega_i}$
simply gives an overall phase (the topological twist of the anyon), and can therefore be ignored; in the case of hole defects, the overall phase
is trivial. Therefore, in both cases the self Dehn twists can be dropped. For twist defects, the branch cut of the twist defect winds around by $2\pi$ locally around the twist defect,
and can be undone by a local finite depth circuit.

We show the protocol for braiding holes in a surface code in Fig.~\ref{fig:braiding_via_shearing}, following the
defect-shearing protocol for the Dehn twist on an annulus in Fig.~\ref{fig:Dehn_twists_annulus}.   We consider
a parallelogram with edges (dashed lines) equivalent to the $\beta$-loop enclosing the two defects, as shown in
Fig.~\ref{fig:braiding_via_shearing}(a). In order to apply a half or full Dehn twist along the $\beta$-loop, we can rotate
the parallelogram by $\pi$ and $2\pi$ respectively, which can be further decomposed into a sequence of shearing
operations as discussed in Fig.~\ref{fig:Dehn_twists_annulus}.  The sequence of shears
spatially changes  the location of the defects and also shears the defects themselves.  After 4 steps of shearing,
we reach the configuration in Fig.~\ref{fig:braiding_via_shearing}(e), with the parallelogram and the two defects
being rotated by $\pi$ (see the labeling of the vertices). By tracking the configuration of reference Wilson loop operators,
we can see that this procedure exactly leads to a half twist, equivalent to braiding and additional half self-twists around
the defects [see Fig.~\ref{fig:twists_definition}(b)], i.e., $\sqrt{\mathcal{D}}_\beta=\mathcal{B}_{1,2} \sqrt{\mathcal{D}}_{\omega_1} \sqrt{\mathcal{D}}_{\omega_2}$.

As we can see from Fig.~\ref{fig:braiding_via_shearing}(f), we can perform a gauge transformation by applying a membrane
of stabilizers to deform the Wilson loop configuration to undo the self-twist, which is a manifestation of the trivial self-twist of the holes.
The result is a configuration that only corresponds to a single braid $\mathcal{B}_{1,2}$ [see Fig.~\ref{fig:twists_definition}(c)].

To get a non-trivial logical gate in the $\mathbb{Z}_2$ surface code, one has to consider the braiding of two different types of hole defects, i.e., those
with smooth and rough boundaries respectively.  Therefore, we need to continue the protocol with an additional 4 steps to achieve a
full braid [see Fig.~\ref{fig:braiding_via_shearing}(g-j)].  In Fig.~\ref{fig:braiding_via_shearing}(j), the two defects come back to the
original configuration in Fig.~\ref{fig:braiding_via_shearing}(a), and hence achieves a full braid $\mathcal{B}^2_{1,2}$.

The protocol described above still uses a finite number of steps, so that $\mathcal{B}_{1,2} = \prod_{i=1}^4 \mathcal{P}_{\sigma_i} \mathcal{LU}_i$,
i.e., 4 steps for braiding. The full braiding needs 8 steps, which is less than the 12-step protocol with cyclic moving of one
defect shown in Fig.~\ref{fig:braiding_circuits}.

We note that by using non-nearest neighbor two-qubit gates, one can implement the Dehn twist protocol described in
Fig.~\ref{fig:modular_genus_g_beta_long_range}, which needs only one step. Therefore, we can have
$\mathcal{B}_{1,2} = \mathcal{P}_\sigma \mathcal{LU}$, at the expense of using a longer-range local quantum circuit.

\begin{figure}
  \includegraphics[width=1\columnwidth]{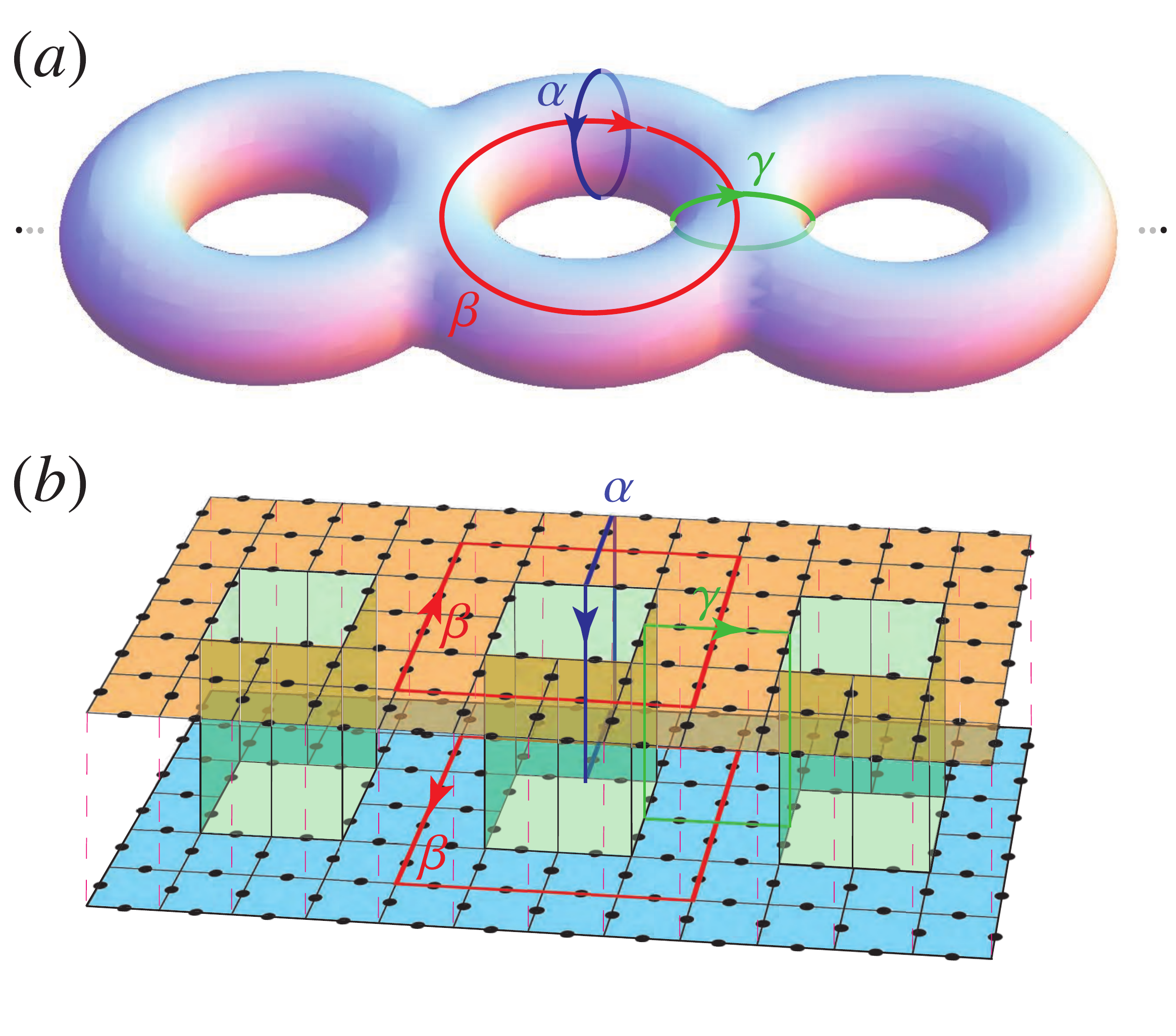}
  \caption{(a) Three types of non-contractible loops on a high genus surface, $\alpha$, $\beta$, and $\gamma$. Dehn twists along these
loops generate the whole mapping class group. (b) The equivalent loops in a bilayer topological system coupled via `wormholes'.
The equivalent $\beta$-loops on the upper and lower layers have opposite orientations. }
  \label{fig:high_genus_loops}
\end{figure}

\begin{figure*}
\includegraphics[width=2\columnwidth]{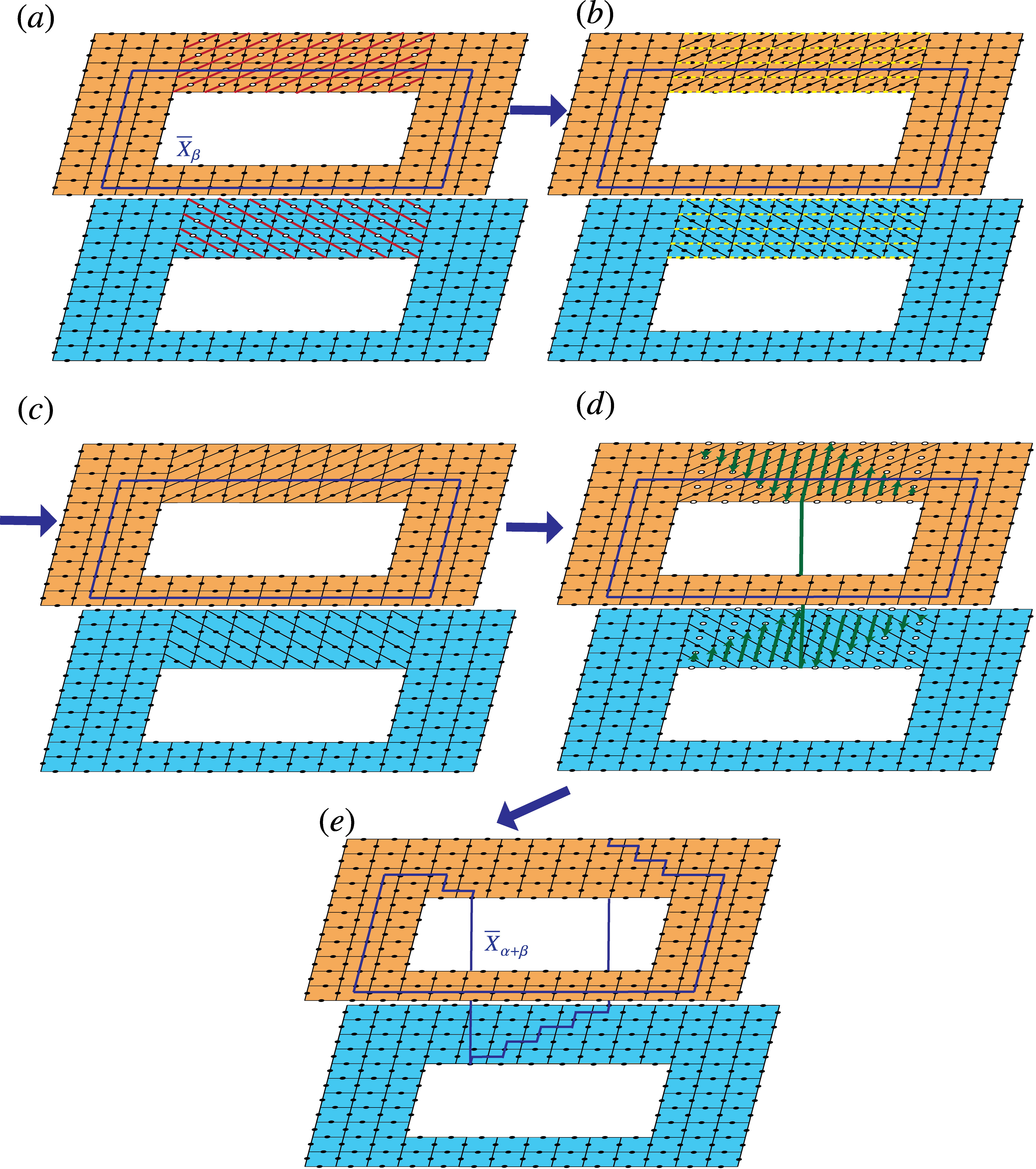}
  \caption{Dehn twist along the $\alpha$-loop.  (a) Adding qubits (white circles) and edges (red solid lines).  (b) Removing
qubits and edges (yellow dashed lines).  (c) Obtain a lattice with a ``gift-wrapping'' (slanted) pattern in the solenoid region.  (d) Shear
permutation (green arrows) of the qubits in the solenoid region. (e) A Dehn twist is achieved and the lattice is mapped to the original geometry in (a).}
\label{fig:modular_genus_g_alpha}
\end{figure*}

\begin{figure*}
\includegraphics[width=2\columnwidth]{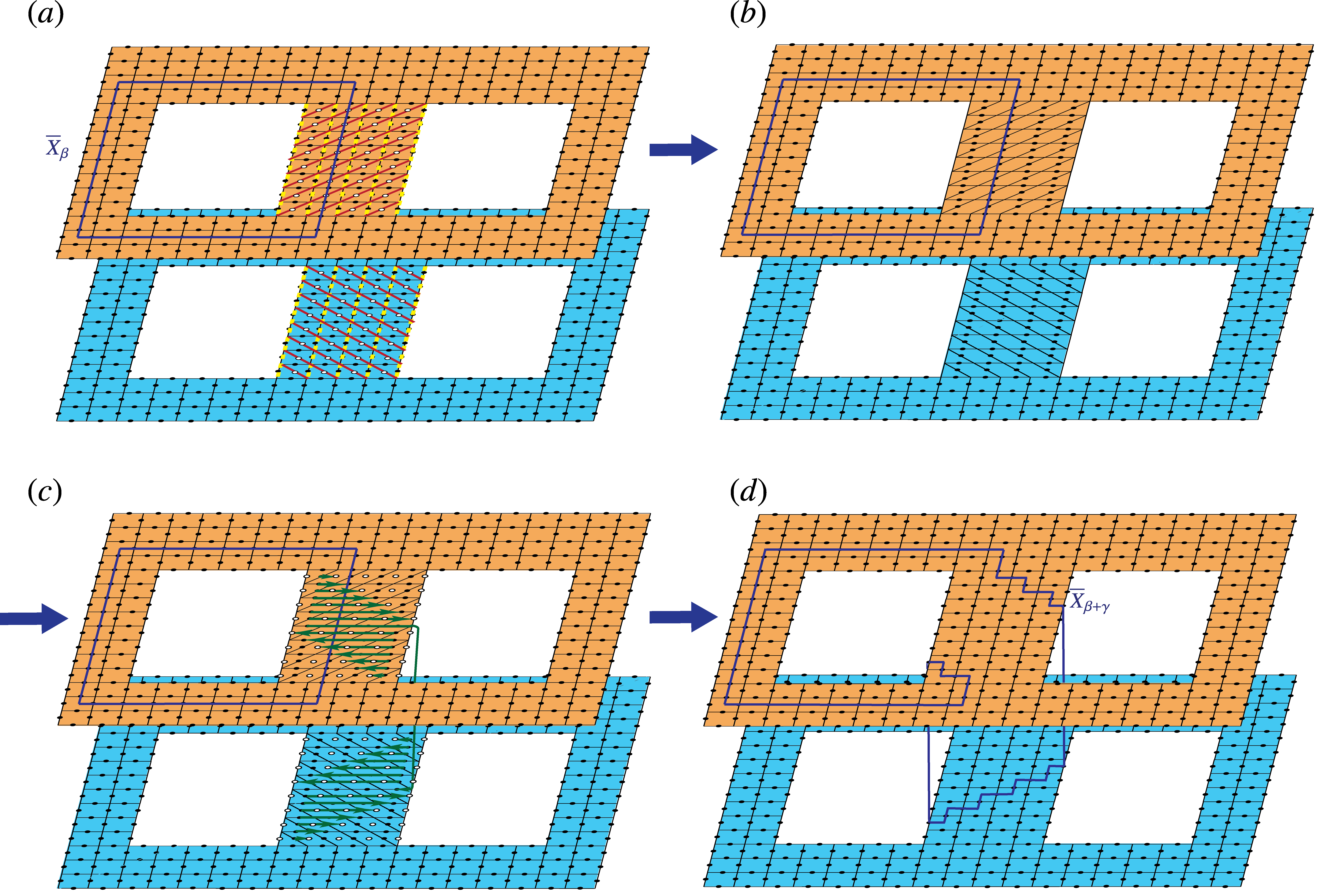}
  \caption{Dehn twist along the $\gamma$-loop.}
\label{fig:modular_genus_g_gamma}
\end{figure*}

\begin{figure*}
\includegraphics[width=1.7\columnwidth]{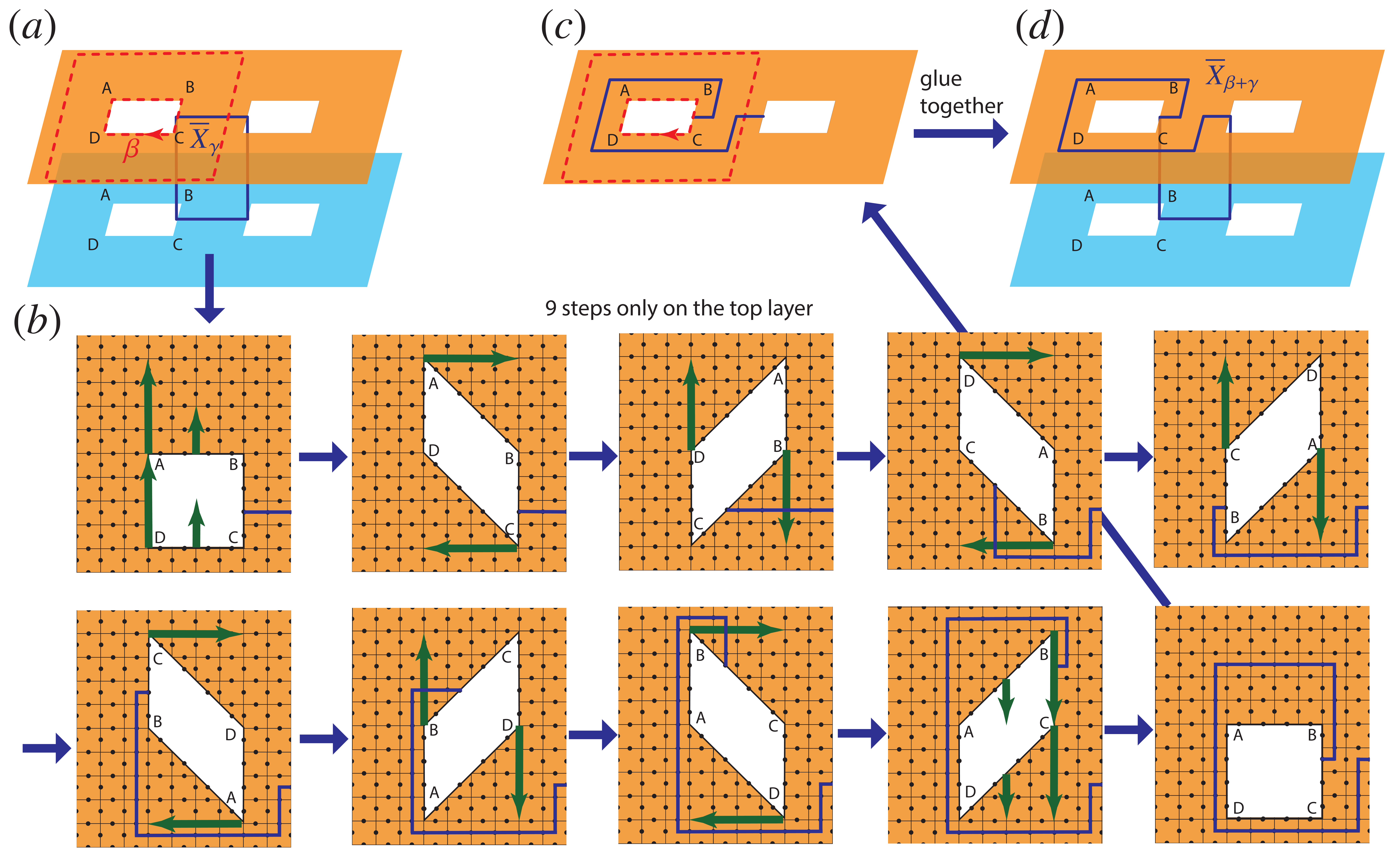}
\caption{Dehn twist along the $\beta$-loop on a high-genus surface via 9 steps of shearing on the top layers.}
\label{fig:modular_genus_g_beta}
\end{figure*}

\begin{figure*}
\includegraphics[width=2\columnwidth]{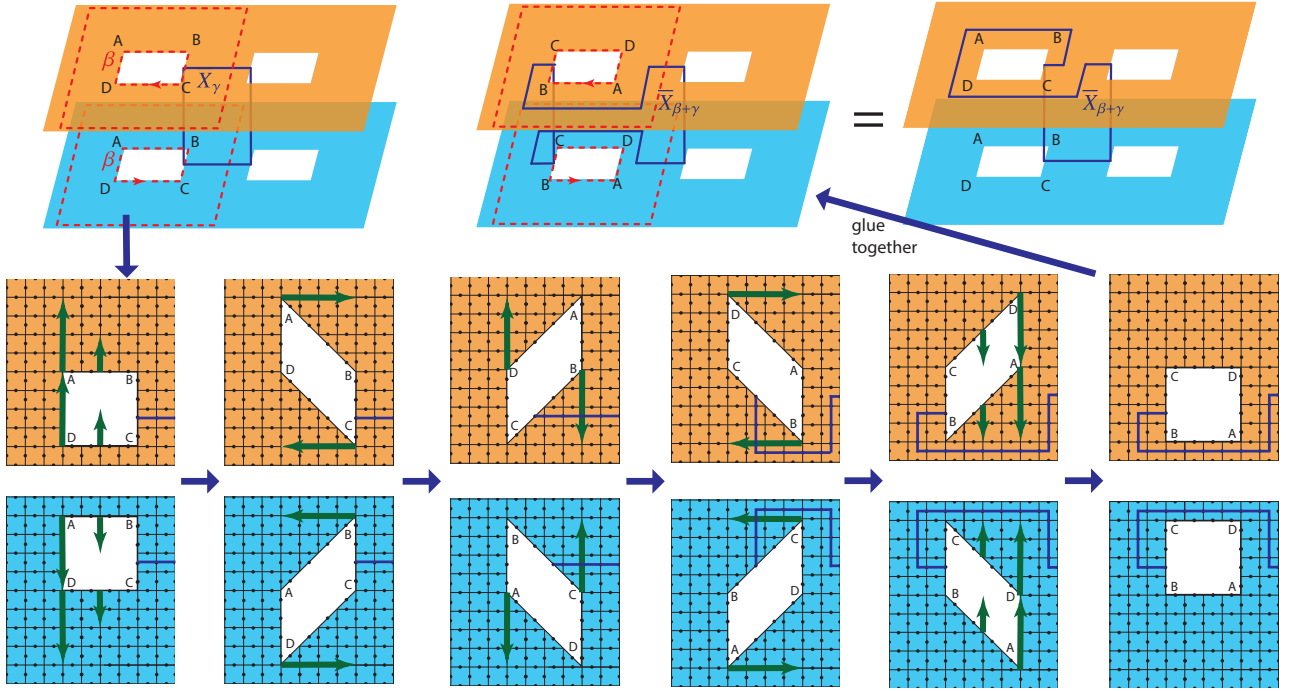}
\caption{Dehn twist along the $\beta$-loop implemented by two half-twists of an annulus on both layers with 5 puncture-shearing steps in total.}
\label{fig:modular_genus_g_beta_alternative}
\end{figure*}

\subsection{Dehn twists on high genus surfaces}\label{sec:high_genus_toric}

Now we consider generating the whole mapping class group of a high genus surface, i.e., MCG$(\Sigma_g)$.
It is well known that MCG$(\Sigma_g)$ can be generated by $3g-1$ Dehn twists \cite{farb2011primer}, which are of
3 types and denoted by $\alpha$, $\beta$ and $\gamma$ respectively, as illustrated in Fig.~\ref{fig:high_genus_loops}(a)
(the arrows indicate the convention of the twist direction in this paper).
As we can see in Fig.~\ref{fig:high_genus_loops}(b), such a high genus surface can be realized in terms of a bilayer
system with holes, where the boundaries of holes in different layers are glued together appropriately.

The protocols we have developed so far for performing Dehn twists on a cylinder and an annulus can easily be adapted to
performing Dehn twists along any of the $\alpha$, $\beta$, and $\gamma$ loops of a high genus surface.
In the following we provide the details for implementing these protocols.

\subsubsection{Dehn twist along $\alpha$- and $\gamma$ loops}

We first show the protocol for implementing a Dehn twist along the $\alpha$-loop in Fig.~\ref{fig:modular_genus_g_alpha}.
As in the situation of Dehn twists on a torus as shown in Fig.~\ref{fig:Dehn-twist}, we perform the local geometry deformation
to get a ``solenoid" region with the slanted plaquette structure as shown in Fig.~\ref{fig:modular_genus_g_alpha}(a-c). Here,
we choose the length of the solenoid such that each diagonal line winds around the solenoid once.  In
Fig.~\ref{fig:modular_genus_g_alpha}(d), we perform the shear through the long-range permutation of the sites,
similar to the situation in Fig.~\ref{fig:Dehn-twist}.  Note that the two ends of the solenoid are fixed and no sites at these ends are permuted.
Therefore, we get the following transformation:
\begin{align}
\mathcal{P}_\sigma \ \mathcal{LU}_{\alpha}: &\overline{X}_\beta \longmapsto  \overline{X}_{\alpha+\beta},
\nonumber \\
\mathcal{P}_\sigma \ \mathcal{LU}_{\alpha}: & \overline{Z}_\beta \longmapsto  \overline{Z}_{\alpha+\beta},
\end{align}
where only the loop $\overline{X}$ is illustrated in the figure.  The transformation is exactly the Dehn twist $\mathcal{D}_\alpha$.

The Dehn twist along the $\gamma$-loop,  illustrated in Fig.~\ref{fig:modular_genus_g_gamma}, is almost identical. Here
the ``solenoid'' region containing the slanted plaquettes is located at the handle between the two pairs of punctures.
We get the following transformation:
\begin{align}
\mathcal{P}_\sigma \ \mathcal{LU}_{\gamma}: &\overline{X}_\gamma \longmapsto  \overline{X}_{\beta+\gamma},
\nonumber \\
\mathcal{P}_\sigma \ \mathcal{LU}_{\gamma}: &\overline{Z}_\gamma \longmapsto  \overline{Z}_{\beta+\gamma},
\end{align}
which is exactly the Dehn twist $\mathcal{D}_\gamma$.

\subsubsection{Dehn twist along $\beta$ loops}

As indicated by Fig.~\ref{fig:high_genus_loops}(b), one can perform the Dehn twist along the $\beta$-loop
in either of the two layers.  As such, it becomes equivalent to performing a Dehn twist on an annulus with
the inner boundary enclosed by the $\beta$-loop. When both layers are viewed from the top as in Fig.~\ref{fig:high_genus_loops}(b),
we see that the directionality of the Dehn twist depends on the layer, as shown.

Now we can directly apply the protocol of the Dehn twist $\mathcal{D}_\beta$ on an annulus to the
Dehn twist $\mathcal{D}_\beta$ on a high-genus surface.

By apply the protocol discussed in Sec. \ref{sec:alternate_beta}, we can apply $\mathcal{D}_\beta$ by a single
step, $\mathcal{D}_\beta = \mathcal{P}_\sigma \mathcal{LU}_\beta$.

Alternatively, we can apply the shearing protocol of Sec. \ref{DannulusShear}. Assuming we choose the $\beta$-loop
located in the upper layer, we can just apply the 9 (for square boundary defect) or 8 (for parallelogram defect)
composite steps of shearing punctures to twist the string in  the upper layer as shown in
Fig.~\ref{fig:modular_genus_g_beta}(a-c).

Note that in the high genus surface, the boundary of the inner hole of the top layer is ``glued'' to the
inner hole of the bottom layer. In the case of passive TQC, during the Dehn twist protocol the Hamiltonian
along this inner boundary should be turned off; alternatively, in the case of active QEC, the stabilizer measurements
along the inner boundary that glue the two layers should be turned off. This is because during the Dehn
twist protocol, the inner boundary of the one layer is twisting relative to the inner boundary of the other layer,
and therefore they cannot be glued together with local interactions until the Dehn twist protocol is completed.

This whole protocol gives the desired Dehn twist $\mathcal{D}_\beta$, which
performs the map $\mathcal{D}_\beta : \overline{X}_{\gamma} \longmapsto \overline{X}_{\beta+\gamma}$ in
the illustrated case in Fig.~\ref{fig:modular_genus_g_beta}(l).

We can further reduce the number of composite steps by a variant of the protocol, shown in
Fig.~\ref{fig:modular_genus_g_beta_alternative}.  Instead of doing the full twist on an
annulus on one of the two layers, we can do half twists on both layers simultaneously, but with
opposite orientation. More concretely, we start with the same square defects on both layers, and
perform the shearing of the defects in opposite directions on both layers, thus effectively performing
a $\pi$ ($-\pi$) rotation of the square defect in the upper (lower) layer.  The half twist is manifested by the fact the vertex
A is permuted to the opposite (lower-right) corner at the end of the protocol.   After gluing the two layers back,
we achieve a Dehn twist along the $\beta$-loop with in total 5 defect-shearing steps, i.e,
$\mathcal{D}_\beta=\prod_{i=1}^5  \mathcal{P}_{\sigma, i} \mathcal{LU}_i$.   Note that the loop configuration of twisted Wilson loops
($\overline{X}$ or $\overline{Z}$) at the end of this protocol can be continuously deformed back to the loop configuration
at the end of the previous protocol, i.e., equivalent to  $\overline{X}_{\beta+\gamma}$ or $\overline{Z}_{\beta+\gamma}$.

\subsection{Multiple Dehn twists in a single shot: Proof of Theorem 2}\label{sec:multiple_Dehn_twists_section}

Here we consider performing $n$ Dehn twists around either the $\alpha$, $\beta$, or $\gamma$ loops.
That is, we consider $\mathcal{D}_{n \omega} = \mathcal{D}_{\omega}^n$, for
$\omega = \alpha_i, \beta_i$, or $\gamma_i$. We will show that
we can perform $\mathcal{D}_{n\omega}$ through a quantum circuit
$\mathcal{P}_{\sigma} \mathcal{LU}_{n\omega}$, where $\mathcal{LU}_{n\omega}$
is a local quantum circuit with finite depth independent of code distance $d$ and system size.

In \textit{Protocol 1} below, we will find that $\mathcal{LU}_{n\omega}$ has a depth that scales
as $\mathcal{O}(\log n)$, but independent of code distance and system
size. In \textit{Protocol 2} below, $\mathcal{LU}_{n\omega}$ has a depth independent of $n$,
but the range of two-qubit gates in $\mathcal{LU}_{n\omega}$ is $r = \mathcal{O}(n)$.
While \textit{Protocol 1} will be generalized to non-abelian codes in Sec.~\ref{sec:non-abelian},
no such generalization exists for \textit{Protocol 2}.

We note that for any topologically ordered phase of matter,
$\mathcal{D}_{\omega}^k = 1$ for some finite $k$ \footnote{This follows from Vafa's theorem \cite{vafa1988},
which states that the topological spins are always rational numbers. }. In the case of the $\mathbb{Z}_N$ toric code,
$k = N$. Therefore, when considering $n$ Dehn twists, we see that $n < k$, while the code distance $d$
can be made arbitrarily large. Below we will always assume we are in the limit $n < k \ll d$.

\subsubsection{Protocol 1: Expanding the lattice}\label{sec:multi_dehn_twist_aspect_ratio_ZN}

\begin{figure}
  \includegraphics[width=1\columnwidth]{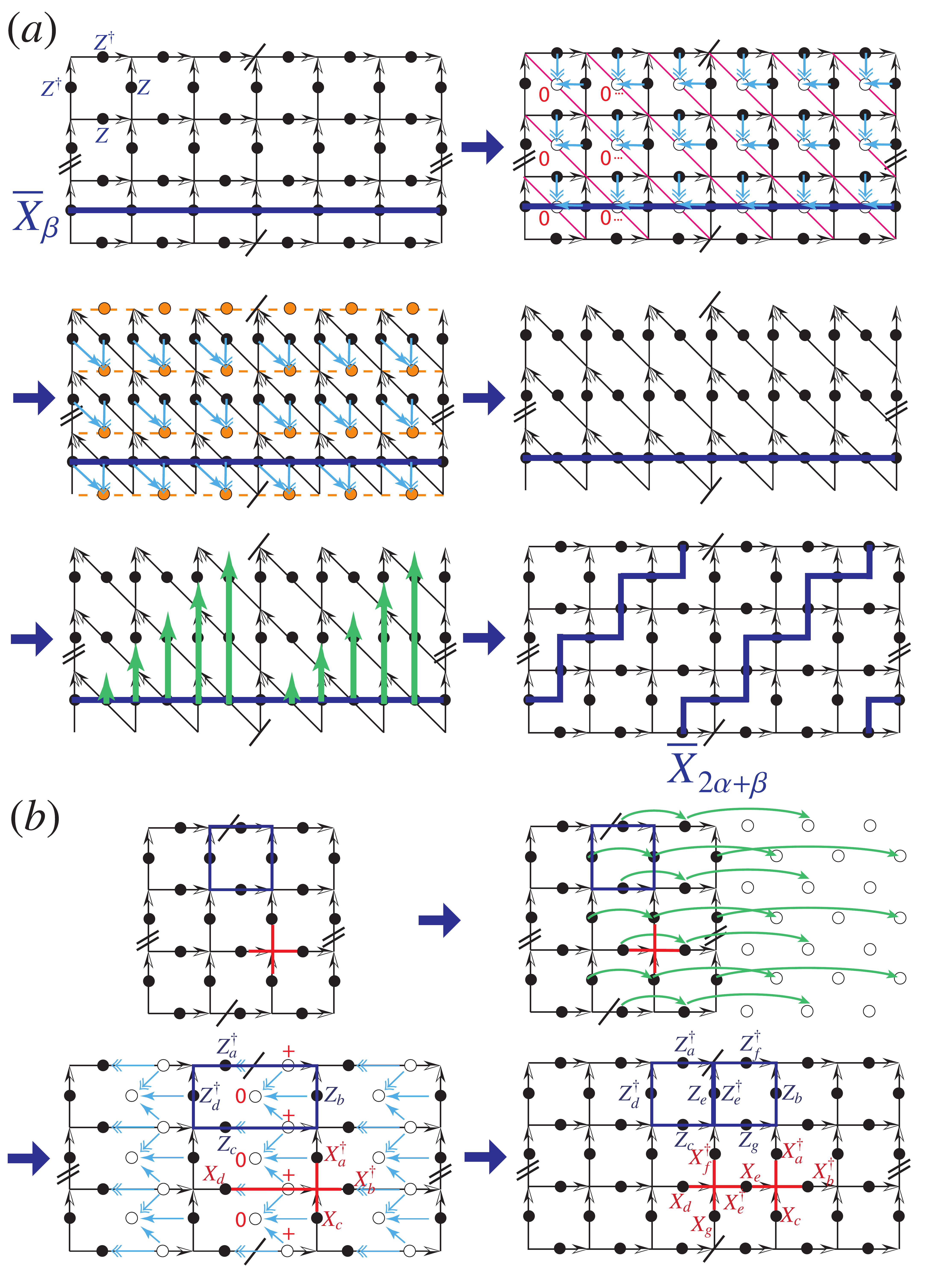}
  \caption{Multiple Dehn twists implemented in a single shot on a $\Z_N$ toric code via changing the aspect ratio of the torus. The single and double arrows represent the $CX$ and $CX^\dag$ gates respectively. The curved green arrows in (b) represent permutations.}
\label{fig:long_solenoid}
\end{figure}

The idea of the first protocol is based on the observation that on an asymmetric torus elongated along one direction, as shown in Fig.~\ref{fig:long_solenoid},
multiple Dehn twists can be applied in parallel along the same cycle. For example, in Fig.~\ref{fig:long_solenoid}
the $\beta$-cycle is twice the length of the $\alpha$-cycle.   The protocol consisting of a finite-depth local quantum circuit followed by long-range qubit permutations,
as illustrated in Fig.~\ref{fig:long_solenoid}, implements a double Dehn twist $D_{2\alpha} = \mathcal{U}^2$ in ``one shot'' (through a constant depth circuit),
leading to the transformation on the illustrated logical string operator
\be
\mathcal{D}_{2\alpha} \overline{X}_\beta \mathcal{D}_{2\alpha}^\dag = \overline{X}_{2\alpha+\beta}.
\ee

In general,  for a fixed code geometry with the $\alpha (\beta)$-cycle $n$ times the length of the $\beta (\alpha)$-cycle,
one can implement the multi-Dehn twist $\mathcal{D}_{n\beta} (\mathcal{D}_{n\alpha})$ in a single shot. This is a remarkable 
result, which demonstrates that by increasing the system size (number of qubits) by $n$ times and with fixed code distance
$d$ (determined by the shorter length of the torus), the time complexity of implementing a particular logical gate sequence
can be decreased by $n$ times, i.e., one can trade space for time. 
Nevertheless, the price to pay is that in such an asymmetric geometry, one can only implement the dual single Dehn 
twist $\mathcal{D}_{\alpha} (\mathcal{D}_{\beta})$ in $n$ shots.

In order to exploit the above observation, we consider the flexibility to adjust the aspect ratio of the torus during the computation,
using entanglement renormalization. As shown in Fig.~\ref{fig:long_solenoid}(b), to be able to implement a double Dehn twist in one shot,
we aim to increase the length of the torus along the $\beta$-direction by a factor of two.  We consider ancilla qubits to the right side of
the system. We next perform a qubit permutation, $\mathcal{P}_{\sigma_1}$ to achieve an effective dilation of the system by increasing the horizontal size of
each plaquette by a factor of 2. Now in order to also increase the number of qubits by a factor of two, we add/entangle the ancilla qubits
(initialized at $\ket{0}$ or $\ket{+}$ ) by the elementary moves composed of $CX$ and $CX^\dag$ gates.
According to Eq.~\eqref{CX_relation2}, the ancilla initialized at $\ket{0}$ is the eigenstate of $Z^\dag_e$, and transformed by the two $CX$ and one $CX^\dag$ as
\be
Z^\dag_e \longmapsto Z^\dag_e Z^\dag_f Z_b Z_g,
\ee
which introduces a new plaquette stabilizer fixed to be $+1$.
Meanwhile, the original plaquette stabilizer involving 4 qubits is transformed by the two $CX^\dag$ as
\be
Z^\dag_a Z_b Z_c Z^\dag_d \longmapsto Z^\dag_a Z_b Z_c Z^\dag_d Z^\dag_f Z_g,
\ee
according Eq.~\eqref{CX_relation} and \eqref{CX_relation2}, which corresponds to a large stabilizer (fixed to be +1) involving two plaquettes
and 7 qubits. This large stabilizer can be decomposed into two stabilizers as $Z^\dag_a Z_b Z_c Z^\dag_d Z^\dag_f Z_g =(Z^\dag_a  Z_c Z^\dag_d Z_e)(Z^\dag_e Z^\dag_f Z_b Z_g)$,
with the operator on $e$ being cancelled. This makes sure the other new plaquette operator $(Z^\dag_a  Z_c Z^\dag_d Z_e)$ is
automatically fixed at $+1$. Therefore, we see that the original plaquette stabilizer is split into two plaquette stabilizers.
Similarly, we have the following transformation for the $X$-stabilizers according to Eq.~\eqref{CX_relation} and \eqref{CX_relation3}, i.e.,
\begin{align}
\non X^\dag_e &\longmapsto X^\dag_e X_g X_d X^\dag_f  \\
X^\dag_a X^\dag_b X_c X_d &\longmapsto X^\dag_a X^\dag_b X_c X_d X^\dag_f X_g ,
\end{align}
which effectively splits the original vertex stabilizers into two. This can be verified by the decomposition $ X^\dag_a X^\dag_b X_c X_g X_d X^\dag_f
$$=$$ (X^\dag_a X^\dag_b X_c X_e)(X^\dag_e X_g X_d X^\dag_f)$.

The above procedure increases the number of plaquette and vertices by a factor of 2.  The whole elongating process is
performed in one shot by a combination of a qubit permutation and a local finite depth quantum circuit. In general,
in order to increase the length of the torus in a particular direction by a factor of $n$, one needs $\log_2 n$ shots of
the above transformations, which is a well-known fact for state-preparation with MERA \cite{Vidal:2007va}.
This protocol provides an exponential improvement over implementing the Dehn twist sequentially $n$ times.

It is straightforward to see that the above result can be extended to any of the $\alpha$, $\beta$, and $\gamma$ loops of a
high genus surface, or to braids. This proves the first statement [Eq.~\eqref{theorem2.1}] of  \textbf{Theorem 2}  in the context of the $\Z_N$ toric code.

\subsubsection{Protocol 2: Increasing the interaction range}\label{sec:increase_range}

In the previous scheme, we fixed the interaction range (nearest-neighbor) of the local unitaries ($\mathcal{LU}$) and
applied multiple Dehn twists in parallel by changing the aspect ratio of the torus. Here we demonstrate a second protocol,
where we apply a single step of $\mathcal{P} \mathcal{LU}$. By increasing the range of $\mathcal{LU}$ to be $\mathcal{O}(n)$,
this allows us to apply $n$ Dehn twists, $\mathcal{D}_{n\omega}$ in a single shot. This protocol shows how the interaction
range can be turned into computational power.

\begin{figure}[hbt]
  \includegraphics[width=1\columnwidth]{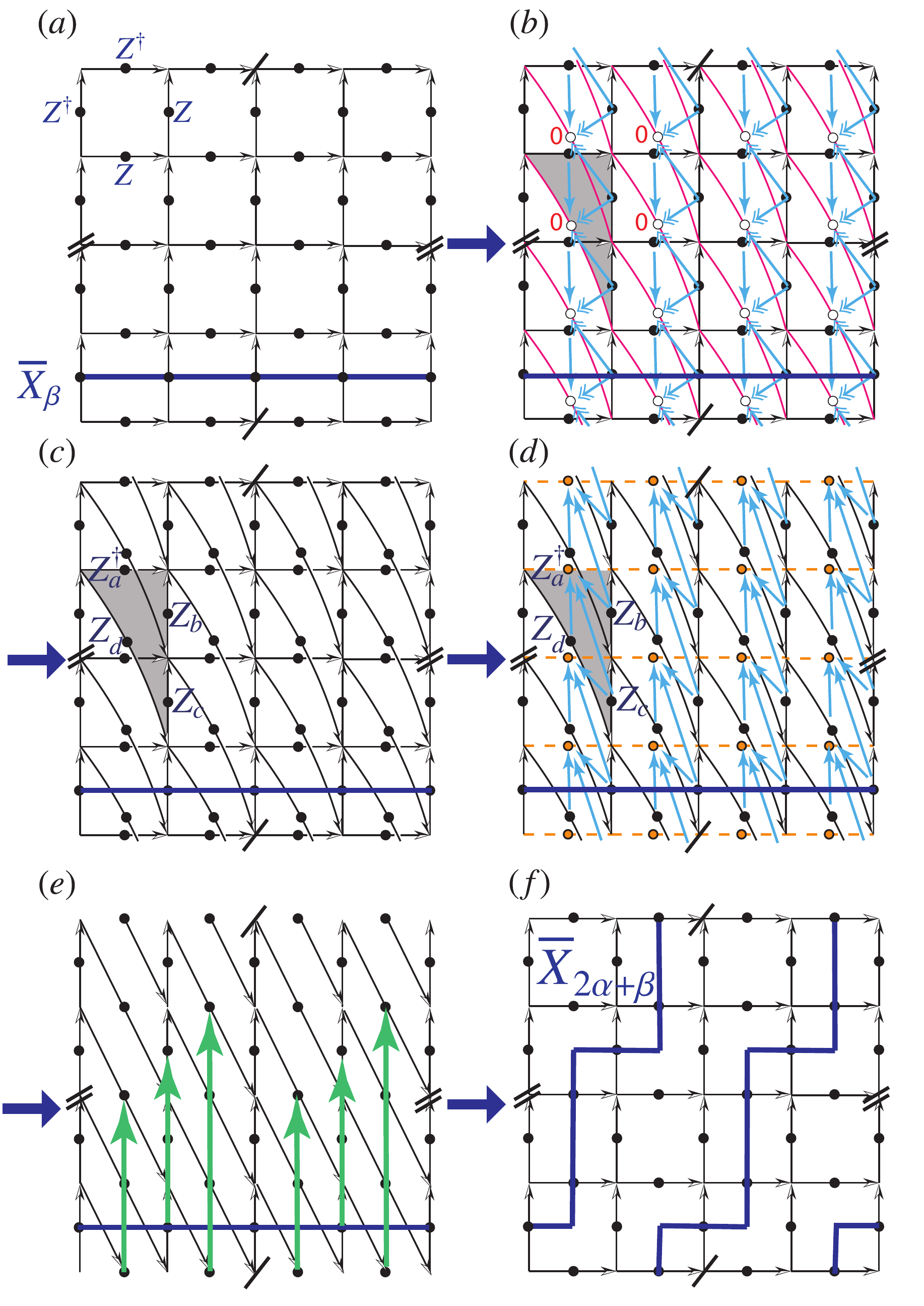}
  \caption{Implementing multiple Dehn twists in a single shot via increasing the range of interaction on a $\Z_N$ toric code. In panel (b), (c) and (d), the diagonal edges overpass the horizontal edges, i.e., equivalent to a bridge structure.}
\label{fig:long_range_solenoid}
\end{figure}

We consider a square lattice in Fig.~\ref{fig:long_range_solenoid}(a). Then we add the NNN diagonal edge $(1,-2)$
and the triangular stabilizer plaquette (shadow) by applying a $CX$ and two $CX^\dag$ gates conditioned by qubits
$a, b$ and $c$, and targeting qubit $d$. $d$ is initialized in state $\ket{0}$. We apply this to other triangular stabilizer plaquettes
in parallel as well, as illustrated in Fig.~\ref{fig:long_range_solenoid}(b, c).   According to the last two identities in
Eq.~\eqref{CX_relation}, the entangling gates on the shadowed plaquette induce the following transformation:
\be
Z_d \longmapsto Z^\dag_a Z_b Z_c Z_d.
\ee
Since $\ket{0}$ is the $+1$-eigenstate of $Z_d$, we fix the triangular stabilizer $Z^\dag_a Z_b Z_c Z_d$ at $+1$.
A similar result holds for all the other added plaquettes.

We now remove all the horizontal edges $(1,0)$ with the $CX$ gates shown in
Fig.~\ref{fig:long_range_solenoid}(d). According to Eq.~\eqref{CX_relation2},
the entangling gates on the shadowed plaquette induce the following transformation:
\be
  Z^\dag_a Z_b Z_c Z_d \longmapsto Z^\dag_a,
\ee
which disentangles the qubit $a$ (yellow circle). We then reach the double slanted lattice with NNN
diagonal edges in Fig.~\ref{fig:long_range_solenoid}(e), which is in contrast to the slanted lattice
with NN diagonal edges in Fig.~\ref{fig:long_solenoid}(a).

Now we apply a qubit permutation $P_\sigma$ shown by the green arrows in
Fig.~\ref{fig:long_range_solenoid}(e). This maps the state back to the original lattice in
Fig.~\ref{fig:long_solenoid}(a) with a double Dehn twist,
leading to the following transformation on the illustrated logical string operator
$\mathcal{D}_{2\alpha} \overline{X}_\beta \mathcal{D}_{2\alpha}^\dag = \overline{X}_{2\alpha+\beta}.$

We see that the maximal range of the finite depth local quantum circuit $\mathcal{LU}$ has an increased
range relative to the case of a single Dehn twist, as now there are two-qubit gates involving a qubit and its
next-nearest diagonal neighbor. One can straightforwardly generalize the above protocol to
apply $n$ Dehn twists in a single shot, with the maximal range $r \sim \mathcal{O}(n)$ in the $\mathcal{LU}$ circuit.

It is straightforward also to adapt this protocol to the case of Dehn twists about any of the $\alpha, \beta, \gamma$ loops of
a high genus surface, or to braids. This proves the second statement [Eq.~\eqref{theorem2.2}] of $\textbf{Theorem 2}$ in the context of $\Z_N$ toric code.

\section{Theory for general topological codes}\label{sec:non-abelian}

In this section we generalize the discussions presented in Sec.~\ref{sec:ZNtc} to the case of arbitrary
non-chiral topologically ordered states. In particular, this analysis applies to both general abelian and
non-abelian topological states. When applied to certain non-abelian codes, such as the Fibonacci
surface code \cite{Bonesteel:2012fl}, our results imply that a universal, fault-tolerant set of logical gates can be
performed with constant time overhead.

The class of topologically ordered states that we consider will be referred to as Turaev-Viro
codes, which are based on Turaev-Viro-Barrett-Westbury (TVBW) topological quantum field theories \cite{turaev1992,turaev1994,barrett1996}.
Ref.~\onlinecite{barkeshli2016tr} contains a recent review aimed at physicists, and contains the conventions that we
follow. These states are associated with exactly solvable models,
such as the Levin-Wen model \cite{levin2005}, which can realize all topologically ordered states that admit gapped boundaries.
Topologically ordered states that can be obtained in this way are referred to as ``non-chiral'' topological
states. Chiral topological states, such as fractional quantum Hall (FQH) states, have topologically
protected gapless edge modes and cannot be obtained from such a construction; thus they are not included in
our analysis.

Ref. \onlinecite{Koenig:2010do, Bonesteel:2012fl} discussed utilizing these TVBW TQFTs as topological QECCs for quantum computation.
As such, the code space corresponds to the ground-state subspace of
the exactly-solvable Levin-Wen Hamiltonians (and their generalizations).

We note that some of the results of this section, in particular those of Sec. \ref{localgeodef} and \ref{tvDiskBraid} are
also summarized in \cite{zhu2018}.

\subsection{Turaev-Viro codes}

The TVBW TQFT associates to a closed surface $\Sigma$ a finite-dimensional Hilbert space $\mathcal{H}_{\Sigma}$.
In the context of QECC, this space can be viewed as the code subspace of a Turaev-Viro code.
We use $\Lambda$ to denote a triangulation of $\Sigma$, together with a local ordering of the vertices of the triangulation.
This local ordering is referred to as a branching structure, and implies that each edge of $\Lambda$ is directed.
We further use $\hat{\Lambda}$ to denote the dual cellulation associated with $\Lambda$, which also defines a directed
graph. For concreteness, we will first consider cases where $\Lambda$ and $\hat{\Lambda}$ define triangular and honeycomb lattices,
respectively.

\begin{figure}
\includegraphics[width=1\columnwidth]{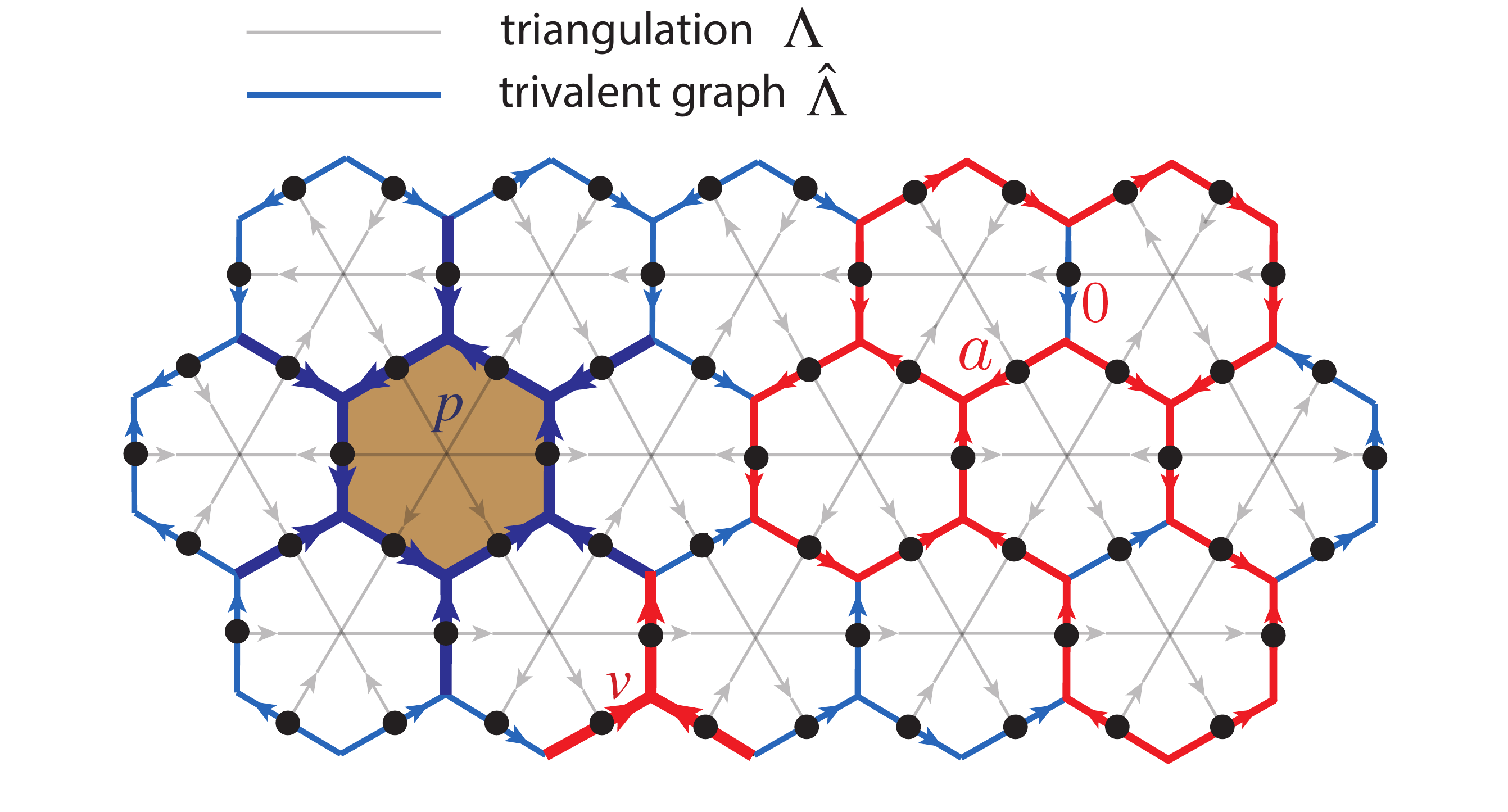}
   \caption{Definition of the Levin-Wen Hamiltonian and Turaev-Viro codes on a triangulated manifold (light grey lines indicate the triangulation $\Lambda$) and the corresponding trivalent graph $\hat{\Lambda}$ (blue lines).  The arrows on the lines specify the branching structure.  The thin  red lines represent the string nets.  The thick blue and red lines illustrates the plaquette and vertex projectors respectively.}
\label{fig:string_net_definition}
\end{figure}

The Turaev-Viro codes (alternatively, the TVBW TQFTs) take as input a unitary fusion category $\mathcal{C}$. The data of $\mathcal{C}$
are specified by the following. $\mathcal{C}$ contains a set of $N$ ``simple objects,''  $\{0,1,2,\cdots, N-1\}$.
Any triplet $(a,b,c)$ of simple objects define a vector space $V_{ab}^c$. The dimension of this vector space
defines the fusion rules $N_{ab}^c$:
\begin{equation*}
N_{ab}^c=\text{dim } V_{ab}^c,
\end{equation*}
where $N_{ab}^c$ is a nonnegative integer.
The fusion rules $N_{ab}^c$ can be summarized through the formal relation
\begin{align}
a \times b = \sum_c N_{ab}^c c .
\end{align}

Given a vector space
$V^{abc}_d \equiv \bigoplus_e V_{ab}^e \otimes V_{ec}^d \simeq \bigoplus_f V_{bc}^f \otimes V_{af}^d$,
$\mathcal{C}$ defines a unitary map $F^{abc}_{d}$:
\begin{align}
F^{abc}_d: \bigoplus_e V_{ab}^e \otimes V_{ec}^d \mapsto \bigoplus_f V_{bc}^f \otimes V_{af}^d.
\end{align}
In components, the $F$-symbols are written as $[F^{abc}_{d}]_{(e,\alpha,\beta),(f,\mu,\nu)}$, where
$\alpha = 0,\cdots, N_{ab}^e - 1$, $\beta = 0,\cdots, N_{ec}^d - 1$, $\mu = 0,\cdots, N_{bc}^f - 1$, and
$\nu = 0,\cdots, N_{af}^d - 1$. When all $N_{ab}^c = 0,1$, the $F$-symbols can be written in components as
$F^{abc}_{def}$. The $F$-symbols satisfy a set of non-trivial self-consistency equations known as the pentagon equations.

In a unitary fusion category, the topological charge conjugate $\bar{a}$ is determined by the unique
label $\bar{a}$ that satisfies $N_{a\bar{a}}^0 = 1$. Furthermore, the identity object $0$ fuses trivially
with all other objects: $N_{a0}^b = N_{0a}^b = \delta_{ab}$.

Below for simplicity we will restrict to cases where $N_{ab}^c = 0,1$, although this restriction
is not necessary for the validity of our results.

The TVBW TQFT provides an explicit wavefunction as follows. Each edge of $\Lambda$ (equivalently, of $\hat{\Lambda}$)
is associated with a local $N$-dimensional Hilbert space (qudit), where the states are labelled by the simple objects
$\{0,1,2,\cdots, N-1\}$. The wavefunction amplitude for a particular state on $\Lambda$ can be explicitly determined
by evaluating a discrete path integral (state sum) over a triangulated 3-manifold $M$, whose boundary $\partial M = \Sigma$.
The triangulation (and corresponding branching structure) of $M$ restricts to $\Lambda$ on $\partial M$. We will not
review the state sum here; we refer the reader to Ref.~\onlinecite{turaev1992,turaev1994,barrett1996,walker2006, Koenig:2010do, barkeshli2016tr} for various
presentations of the state sum.

An important property of the wavefunctions is that, for all states with non-zero amplitude, vertices of the dual graph satisfy the fusion rules, and as a result:
\begin{align}
\label{vertexRel}
\Psi\left(~
\begin{tikzpicture}[baseline={([yshift=-.5ex]current  bounding  box.center)}]
\draw[dualblue,thick,middlearrow={stealth}] (0.8,0) -- (0,0);
\draw[dualblue,thick,middlearrow={stealth}] (-0.3,0.5) -- (0,0);
\draw[dualblue,thick,middlearrow={stealth reversed}] (-0.3,-0.5) -- (0,0);
\draw (0.4,0.15) node {$a$};
\draw (-0.35,0.75) node {$b$};
\draw (-0.35,-0.75) node {$c$};
\end{tikzpicture}
\right)
= 0 \text{\;\;\;if\;\;\;} N_{ab}^c = 0 .
\end{align}
If the qudit on a particular edge is in the state $|a\rangle$, we say that there is a string of type $a$ passing
through that edge. The wavefunction can then be viewed as a superposition of closed string-net
configurations consistent with these string fusion rules \cite{levin2005} .

The original Turaev-Viro construction, together with the corresponding Levin-Wen Hamiltonian, assume a certain
tetrahedral symmetry that imposes many relations among the $F$-symbols; as such, not any unitary fusion category
can be taken as input. The extension due to Barrett and Westbury relaxes the tetrahedral symmetry, at the cost of needing
more careful consideration of the branching structure; the Barrett-Westbury generalization therefore applies
to arbitrary unitary fusion categories \footnote{Actually the Barrett-Westbury construction applies to general `spherical' fusion
categories, but from the perspective of topological quantum states of matter, the stricter condition of
unitarity is required of the fusion category.}. Our protocols in the following sections do not assume any tetrahedral symmetry,
and thus apply to the full Barrett-Westbury generalization.

In the case where the $F$-symbols satisfy additional relations due to tetrahedral symmetry (which we will not summarize here), the wavefunctions of the TVBW TQFT are exact ground states of a commuting projector Hamiltonian known as the Levin-Wen Hamiltonian
\cite{levin2005} \footnote{In the case of Abelian topological states, Ref. \cite{Lin:2014us} provided a generalization of the Levin-Wen models to
relax the assumption of tetrahedral symmetry. }.
The Hamiltonian is
\be\label{Levin-Wen}
H_{\hat{\Lambda}}= - \sum_v Q_v -\sum_p B_p,
\ee
where $v$ and $p$ label the vertices and plaquettes of $\hat{\Lambda}$. The 3-body vertex
projection operator $Q_v$ depends only on the three edges incident to the vertex $v$, and is defined by
\begin{align}\label{branching_rules}
Q_v
\begin{tikzpicture}[baseline={([yshift=-.5ex]current  bounding  box.center)}]
\draw[thick]  (-0.8, -0.5) -- (-0.8, 0.5);
\draw[dualblue,thick,middlearrow={stealth reversed}] (0,0) --  (0.8,0);
\draw[dualblue,thick,middlearrow={stealth reversed}]  (0,0) -- (-0.3,0.5)   ;
\draw[dualblue,thick,middlearrow={stealth reversed}] (-0.3,-0.5) -- (0,0)   ;
\draw (0.5,0.15) node {$a$};
\draw (-0.45,0.45) node {$b$};
\draw (-0.45,-0.45) node {$c$};
\draw (0.1,-0.2) node {$v$};
\draw[thick]   (1, -0.5) -- (1.2, 0);
\draw[thick]   (1,  0.5) -- (1.2, 0);
\end{tikzpicture}
= N_{ab}^c
\begin{tikzpicture}[baseline={([yshift=-.5ex]current  bounding  box.center)}]
\draw[thick]  (-0.8, -0.5) -- (-0.8, 0.5);
\draw[dualblue,thick,middlearrow={stealth reversed}] (0,0) --  (0.8,0);
\draw[dualblue,thick,middlearrow={stealth reversed}]  (0,0) -- (-0.3,0.5)   ;
\draw[dualblue,thick,middlearrow={stealth reversed}] (-0.3,-0.5) -- (0,0)   ;
\draw (0.5,0.15) node {$a$};
\draw (-0.45,0.45) node {$b$};
\draw (-0.45,-0.45) node {$c$};
\draw (0.1,-0.2) node {$v$};
\draw[thick]   (1, -0.5) -- (1.2, 0);
\draw[thick]   (1,  0.5) -- (1.2, 0);
\end{tikzpicture}
\end{align}
Recall for simplicity we have restricted to the case $N_{ab}^c = 0,1$.

As an example, the Fibonacci Levin-Wen model has $N = 2$ and therefore each
edge of the trivalent graph contains two types of strings, corresponding to the
two states ($\{\ket{0}, \ket{1}\}$ of a qubit. The fusion rules are specified as
\be
N_{ab}^c=\left\lbrace \begin{array}{cc}
1  & \text{if} \ abc=000,011,101,110,111, \\
0  & \text{otherwise. }
\end{array} \right.
\ee
The corresponding string-net configuration satisfying the fusion rules can be illustrated on the
right side of Fig.~\ref{fig:string_net_definition}, where the edges with the red string correspond
to an occupied site ($\ket{1}$) and the edges without a string correspond to an unoccupied site
($\ket{0}$) (in the Fibonacci case the arrows on the graph and string-net can be ignored).

On a honeycomb lattice, as shown in Fig.~\ref{fig:string_net_definition}, the plaquette operator $B_p$
is a 12-body operator that depends on the 6 qudits on the hexagonal plaquette and also on
the qudits on the 6 legs connecting to the hexagon. The operator can be written as
$ B_p = \sum_s d_s B_p^s/D^2$, where $d_s = |F^{s\bar{s}s}_{s00}|^{-1}$ is the quantum dimension of the string label
$s$, $D=\sum_s \sqrt{d_s^2}$ is the total quantum dimension, and the operator $B^s_p$ is defined via
\begin{align}
\non & B^s_p \
\begin{tikzpicture}[baseline={([yshift=-.5ex]current  bounding  box.center)}]
\draw[thick]  (-1.7, -1) -- (-1.7, 1);    
\draw[dualblue, thick, middlearrow={stealth  reversed}]  (-1,0) --  (-0.5,0.866);  
\draw[dualblue, thick, middlearrow={stealth}]  (-1,0) -- (-0.5,-0.866);
\draw[dualblue, thick, middlearrow={stealth  reversed}] (-0.5,0.866) -- (0.5,0.866);
\draw[dualblue, thick, middlearrow={stealth}](-0.5,-0.866) -- (0.5,-0.866);
\draw[dualblue, thick, middlearrow={stealth  reversed}] (0.5,0.866) -- (1,0);
\draw[dualblue, thick, middlearrow={stealth}] (0.5,-0.866) -- (1,0);
\draw[dualblue, thick, middlearrow={stealth  reversed}] (-1,0) -- (-1.4,0) ;   
\draw[dualblue, thick, middlearrow={stealth  reversed}] (1,0) --  (1.4,0);
\draw[dualblue, thick, middlearrow={stealth  reversed}] (-0.5,0.866)--(-0.7, 1.212);
\draw[dualblue, thick, middlearrow={stealth  reversed}] (-0.5, -0.866)--(-0.7, -1.212);
\draw[dualblue, thick, middlearrow={stealth  reversed}] (0.5,0.866) -- (0.7,1.212);
\draw[dualblue, thick, middlearrow={stealth  reversed}] (0.5,-0.866) -- (0.7,-1.212);
\draw (-1.53,0) node {$a$}; 
\draw (1.53,0) node {$d$};
\draw (-0.8, 1.312) node {b};
\draw (-0.8, -1.312) node {f};
\draw (0.8, 1.312) node {c};
\draw (0.8, -1.312) node {e};
\draw (-0.95,0.463) node {g};
\draw (-0.95,-0.463) node {l};
\draw (0.95,0.463) node {i};
\draw (0.95,-0.463) node {j};
\draw (0,1.066) node {h};
\draw (0,-1.066) node {k};
\draw[thick]   (1.5, -1) -- (1.8, 0);  
\draw[thick]   (1.5,  1) -- (1.8, 0);
\end{tikzpicture} \\
 =& \sum_{g'h'i'j'k'l'} B^{s, g'h'i'j'k'l'}_{p, ghijkl}(abcdef)
 \begin{tikzpicture}[baseline={([yshift=-.5ex]current  bounding  box.center)}]
\draw[thick]  (-1.7, -1) -- (-1.7, 1);    
\draw[dualblue, thick, middlearrow={stealth  reversed}]  (-1,0) --  (-0.5,0.866);  
\draw[dualblue, thick, middlearrow={stealth}]  (-1,0) -- (-0.5,-0.866);
\draw[dualblue, thick, middlearrow={stealth  reversed}] (-0.5,0.866) -- (0.5,0.866);
\draw[dualblue, thick, middlearrow={stealth}](-0.5,-0.866) -- (0.5,-0.866);
\draw[dualblue, thick, middlearrow={stealth  reversed}] (0.5,0.866) -- (1,0);
\draw[dualblue, thick, middlearrow={stealth}] (0.5,-0.866) -- (1,0);
\draw[dualblue, thick, middlearrow={stealth  reversed}] (-1,0) -- (-1.4,0) ;   
\draw[dualblue, thick, middlearrow={stealth  reversed}] (1,0) --  (1.4,0);
\draw[dualblue, thick, middlearrow={stealth  reversed}] (-0.5,0.866)--(-0.7, 1.212);
\draw[dualblue, thick, middlearrow={stealth  reversed}] (-0.5, -0.866)--(-0.7, -1.212);
\draw[dualblue, thick, middlearrow={stealth  reversed}] (0.5,0.866) -- (0.7,1.212);
\draw[dualblue, thick, middlearrow={stealth  reversed}] (0.5,-0.866) -- (0.7,-1.212);
\draw (-1.53,0) node {$a$}; 
\draw (1.53,0) node {$d$};
\draw (-0.8, 1.312) node {b};
\draw (-0.8, -1.312) node {f};
\draw (0.8, 1.312) node {c};
\draw (0.8, -1.312) node {e};
\draw (-0.95,0.463) node {g'};
\draw (-0.95,-0.463) node {l'};
\draw (0.95,0.463) node {i'};
\draw (0.95,-0.463) node {j'};
\draw (0,1.066) node {h'};
\draw (0,-1.066) node {k'};
\draw[thick]   (1.5, -1) -- (1.8, 0);  
\draw[thick]   (1.5,  1) -- (1.8, 0);
\end{tikzpicture},
\end{align}

where the tensor coefficients are
\begin{align}
\non &B^{s, g'h'i'j'k'l'}_{p, ghijkl}(abcdef) \\
=&F^{b\bar{g}h}_{\bar{s}h'\bar{g}'} F^{c\bar{h}i}_{\bar{s}s\bar{h'}} F^{d\bar{i}j}_{\bar{s}j'\bar{i'}} F^{e\bar{j}k}_{\bar{s}k'\bar{j'}}F^{f\bar{k}l}_{\bar{s}l'\bar{k'}}F^{a\bar{l}g}_{\bar{s}g'\bar{l'}}.
\end{align}

The plaquette term $B_p$ flips the string-net configurations, such as the one shown in
Fig.~\ref{fig:string_net_definition}, to other configurations constrained by the fusion rules.
The ground state of the Hamiltonian Eq.~\eqref{Levin-Wen} in a particular topological sector is an
equal-probability superposition of all the possible string-net configurations connected by the local
action of the plaquette operators.

\begin{figure}
  \includegraphics[width=1\columnwidth]{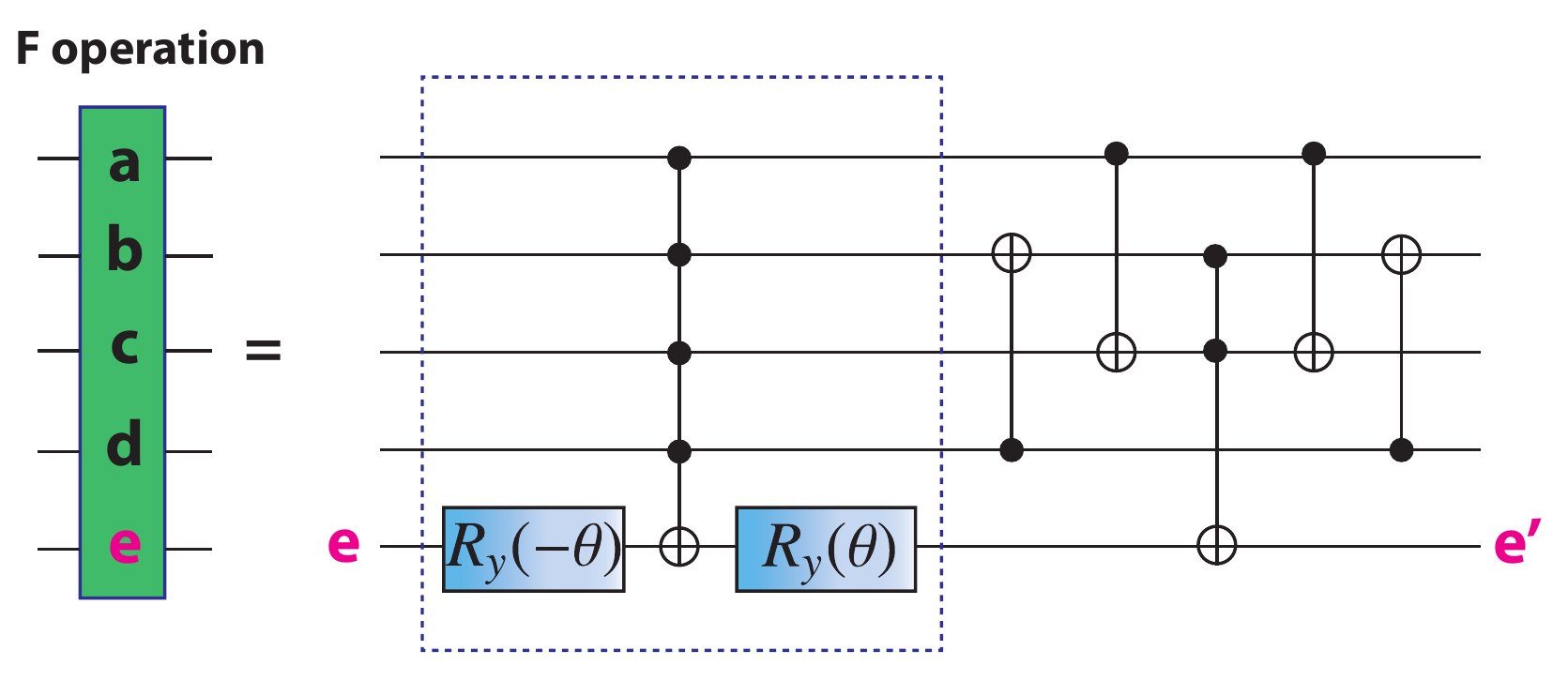}
  \caption{The building block of the quantum computation scheme: the $F$ operation quantum circuit for the Fibonacci surface code.}
\label{fig:F-move_circuit}
\end{figure}

Just as in the case of the $\mathbb{Z}_2$ surface code (toric code), the Turaev-Viro codes can be implemented
through an active error correction approach by repeated measurements of the commuting vertex
and plaquette operators.  Ref.~\onlinecite{Bonesteel:2012fl} presents details of the quantum circuits that can be used to measure these
operators in the context of the Fibonacci surface code. Importantly, the quantum circuits for measuring
these operators need only contain single and two-qubit operations.  Ongoing progress has been made on syndrome extraction, decoding and error correction \cite{Fibonacci_error_correction, Feng:2018, Burton:2017gr, Dauphinais:2017bz}. In particular, the decoder for a phenomenological Fibonacci code has been simulated numerically in Ref.~\onlinecite{Burton:2017gr}.

The basic building block of the whole scheme is the $F$ operation, $F^{abc}_{def}$. The $F$ operation
can be viewed as a controlled-unitary operation, where the external $a,b,c,d$ legs are the
control qudits that determine the resulting unitary $F^{abc}_d$, whose matrix elements
are $[F^{abc}_d]_{ef}$.

In the case of the Fibonacci surface code, the only non-trivial $F$-matrix is:
\be\label{F-matrix}
F^{111}_{1}=\begin{pmatrix}
\phi^{-1} & \phi^{-\frac{1}{2}}  \\
\phi^{-\frac{1}{2}}&  -\phi^{-1}
\end{pmatrix},
\ee
where $\phi=\frac{\sqrt{5}+1}{2}$ is the golden ratio.
All other $F$-symbols are either $1$ or $0$, depending on whether they are consistent with the fusion rules.

A quantum circuit implementing the $F$ operations in the Fibonacci surface code
was presented in Ref.~\onlinecite{Bonesteel:2012fl} and is shown in Fig.~\ref{fig:F-move_circuit}.
The circuit inside the blue dashed box, consisting of a 5-qubit Toffoli gate sandwiched by
 two single-qubit rotations, implements the $F$-matrix in Eq.~\eqref{F-matrix}. Here, $R_y(\pm \theta)=e^{\pm i \theta \sigma_y/2}$
are single-qubit rotations about the y-axis with angle $\theta = \tan^{-1} (\phi^{-\frac{1}{2}})$. All the other maps are taken care of by the rest of the quantum circuit.

\subsection{Local geometry deformation}
\label{localgeodef}

\begin{figure}
\includegraphics[width=1\columnwidth]{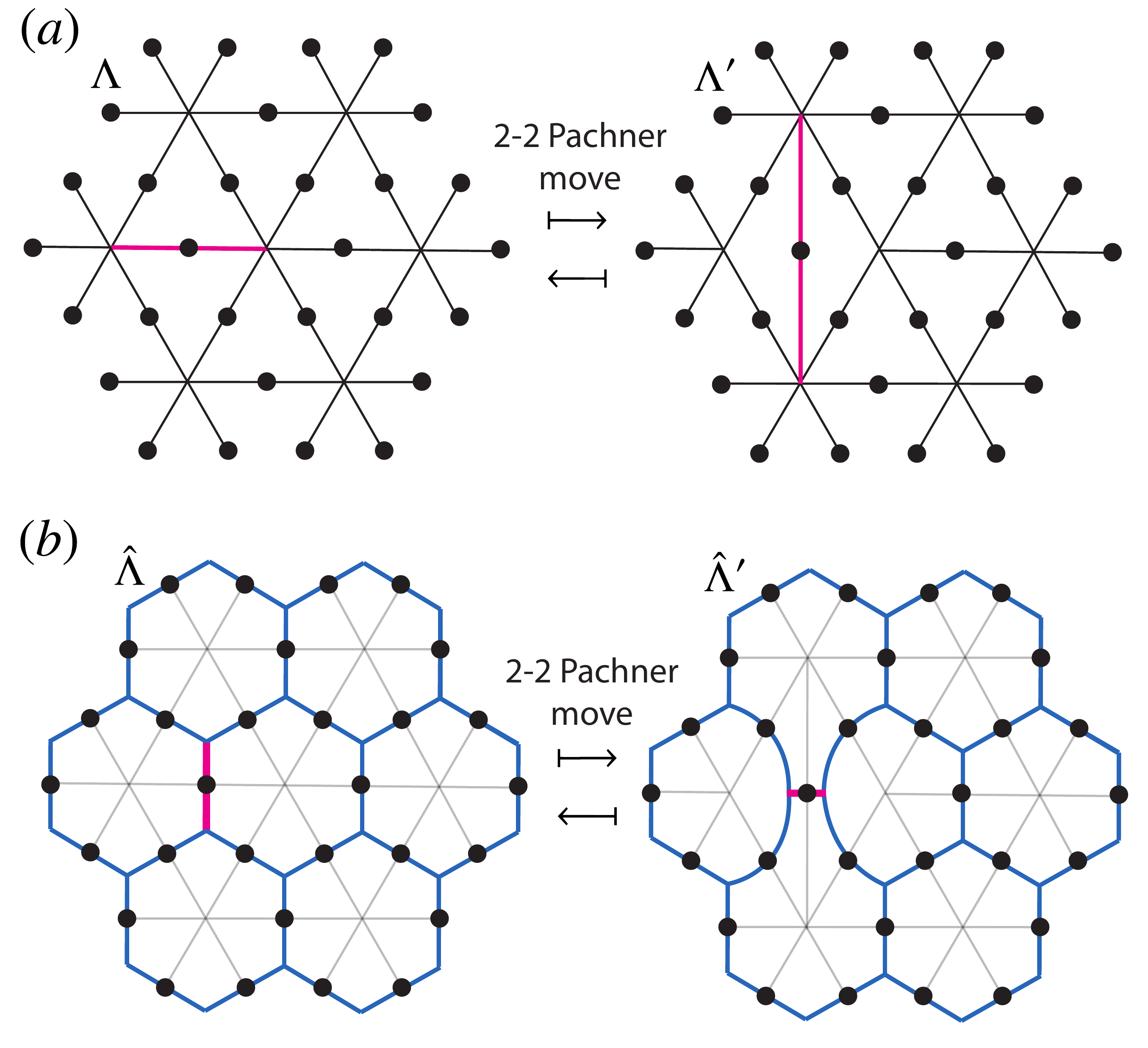}
   \caption{Definition of the 2-2 Pachner move (F-move) on the triangulation grid and the corresponding trivalent graph. The pink edges represent the edges being switched during the moves. }
\label{fig:F-move_definition}
\end{figure}

\begin{figure}
\includegraphics[width=1\columnwidth]{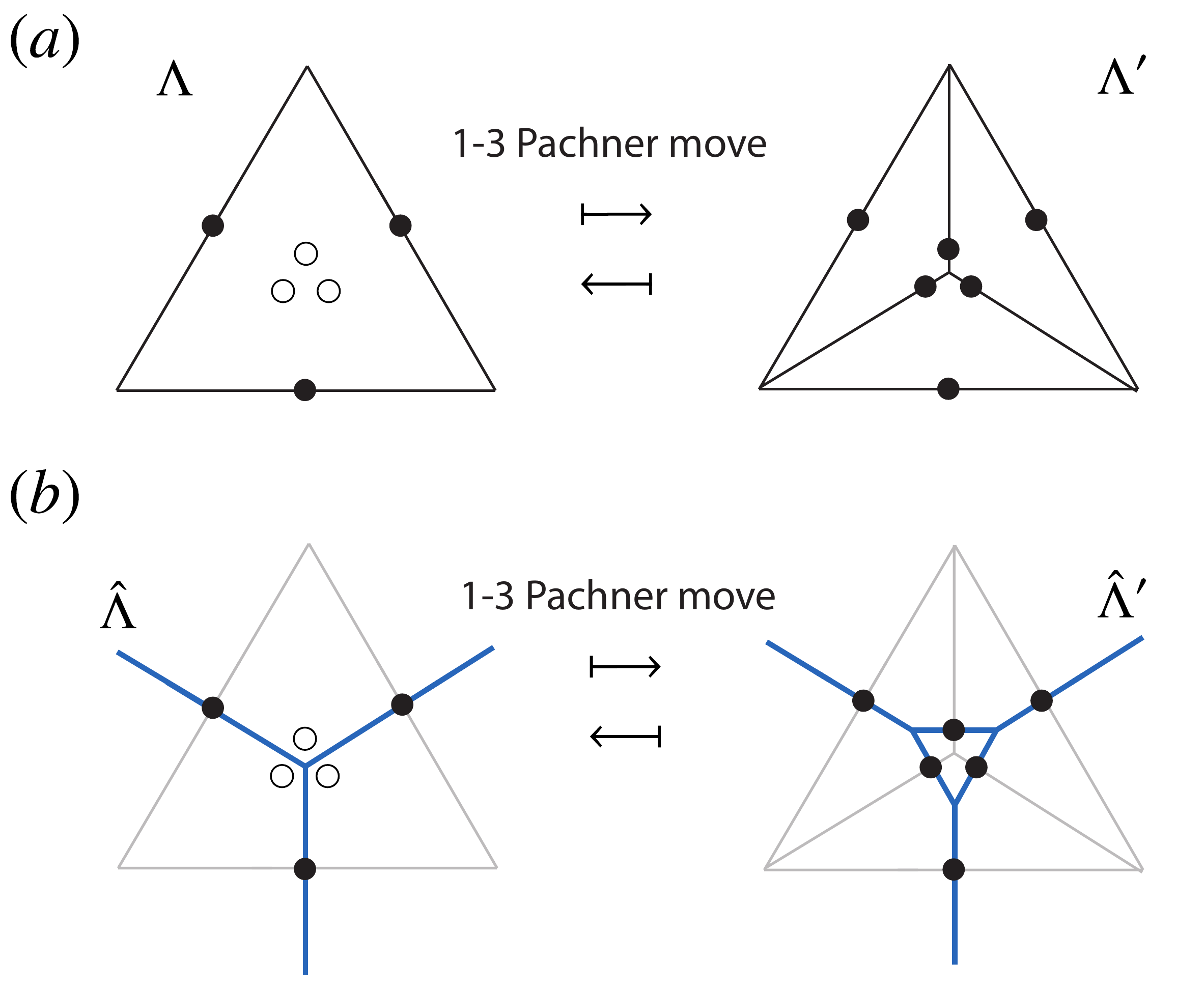}
   \caption{Definition of the 1-3 Pachner move on the triangulation grid and the corresponding trivalent graph.}
\label{fig:pachner_move_definition}
\end{figure}

The key property of the wavefunctions of the TVBW TQFT that forms the basis of our approach
is that wavefunctions associated to different graphs can be related to each other via a series of local moves.
These moves are known as Pachner moves, and correspond to retriangulations of the manifold.
In the (2+1)D path integral state sum, the retriangulations of the 3-manifold can be performed by
2-3 and 1-4 Pachner moves. The 2-3 Pachner move replaces 2 tetrahedra by 3 and vice versa,
while the 1-4 Pachner move replaces 1 tetrahedra by 4 and vice versa. At the 2D surface
$\Sigma = \partial M$ of the 3-manifold, these moves restrict to 2-2 and 1-3 Pachner moves
for retriangulation of a surface, illustrated in Figs. \ref{fig:F-move_definition} and \ref{fig:pachner_move_definition}.

\begin{figure}
  \includegraphics[width=1\columnwidth]{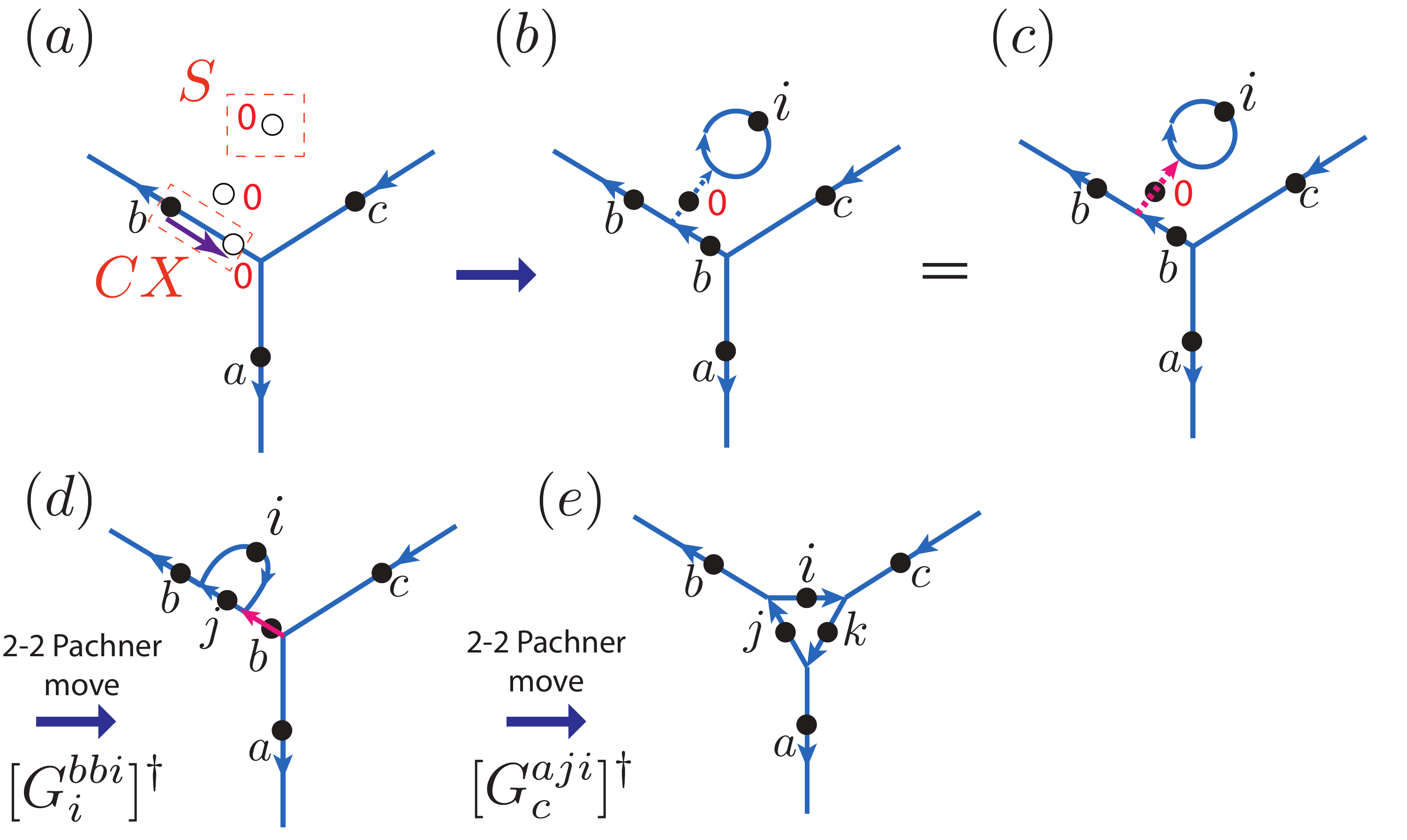}
  \caption{Implementing the 1-3 Pachner move with a unitary cricuit by attaching/detaching a tadpole diagram in
the center of an arbitrary plaquette, followed/preceded by two 2-2 Pachner moves.}
\label{fig:tadpole}
\end{figure}

The path integral state sum of the TVBW TQFT that determines the wavefunction thus also determines how to relate
wavefunctions on different graphs that are related by these local Pachner moves. For example, we have the relations:
\begin{align}
\label{FmoveRel1}
\Psi_{\hat{\Lambda}'}\left(~
\begin{tikzpicture}[baseline={([yshift=-.5ex]current  bounding  box.center)}]
\draw[dualblue,thick,middlearrow={stealth reversed}] (0,0) -- (0.8,0);
\draw[dualblue,thick,middlearrow={stealth reversed}] (-0.3,0.5) -- (0,0);
\draw[dualblue,thick,middlearrow={stealth reversed}] (-0.3,-0.5) -- (0,0);
\draw[dualblue,thick,middlearrow={stealth reversed}] (1.1,0.5) -- (0.8,0);
\draw[dualblue,thick,middlearrow={stealth reversed}] (0.8,0) -- (1.1,-0.5);
\draw (0.4,0.15) node {$e$};
\draw (-0.35,0.75) node {$b$};
\draw (-0.35,-0.75) node {$a$};
\draw (1.15,0.75) node {$c$};
\draw (1.15,-0.75) node {$d$};
\end{tikzpicture}
\right)
 = \sum_f F^{abc}_{def} \
\Psi_{\hat{\Lambda}}\left(~
\begin{tikzpicture}[baseline={([yshift=-.5ex]current  bounding  box.center)}]
\draw[dualblue,thick,middlearrow={stealth}] (0,0) -- (0,0.8);
\draw[dualblue,thick,middlearrow={stealth reversed}] (0,0) -- (0.5,-0.3) ;
\draw[dualblue,thick,middlearrow={stealth reversed}] (-0.5,-0.3) -- (0,0);
\draw[dualblue,thick,middlearrow={stealth reversed}] (0.5, 1.1) -- (0,0.8) ;
\draw[dualblue,thick,middlearrow={stealth reversed}] (-0.5,1.1) -- (0,0.8);
\draw (0.15,0.4) node {$f$};
\draw (-0.6,1.25) node {$b$};
\draw (-0.6,-0.45) node {$a$};
\draw ( 0.6,1.25) node {$c$};
\draw (0.6,-0.45) node {$d$};
\end{tikzpicture}
\right)
\end{align}
\begin{align}
\label{FmoveRel3}
\Psi_{\hat{\Lambda}'}\left(~
\begin{tikzpicture}[baseline={([yshift=-.5ex]current  bounding  box.center)}]
\draw[dualblue,thick,middlearrow={stealth reversed}]  (0.8,0) -- (0,0);
\draw[dualblue,thick,middlearrow={stealth reversed}] (-0.3,0.5) -- (0,0);
\draw[dualblue,thick,middlearrow={stealth reversed}]  (0,0) -- (-0.3,-0.5);
\draw[dualblue,thick,middlearrow={stealth reversed}] (1.1,0.5) -- (0.8,0);
\draw[dualblue,thick,middlearrow={stealth reversed}] (0.8,0) -- (1.1,-0.5);
\draw (0.4,0.15) node {$e$};
\draw (-0.35,0.75) node {$b$};
\draw (-0.35,-0.75) node {$a$};
\draw (1.15,0.75) node {$c$};
\draw (1.15,-0.75) node {$d$};
\end{tikzpicture}
\right)
=\sum_f G^{abc}_{def} \ \Psi_{\hat{\Lambda}}\left(~\begin{tikzpicture}[baseline={([yshift=-.5ex]current  bounding  box.center)}]
\draw[dualblue,thick,middlearrow={stealth}] (0,0) -- (0,0.8);
\draw[dualblue,thick,middlearrow={stealth reversed}] (0,0) -- (0.5,-0.3) ;
\draw[dualblue,thick,middlearrow={stealth reversed}] (0,0) --(-0.5,-0.3) ;
\draw[dualblue,thick,middlearrow={stealth reversed}] (0.5, 1.1) -- (0,0.8) ;
\draw[dualblue,thick,middlearrow={stealth reversed}] (-0.5,1.1) -- (0,0.8);
\draw (0.15,0.4) node {$f$};
\draw (-0.6,1.25) node {$b$};
\draw (-0.6,-0.45) node {$a$};
\draw ( 0.6,1.25) node {$c$};
\draw (0.6,-0.45) node {$d$};
\end{tikzpicture}
\right)
\end{align}
\begin{align}
\label{FmoveRel2}
\Psi_{\hat{\Lambda}'}\left(
\begin{tikzpicture}[scale=0.7, baseline={([yshift=-.5ex]current  bounding  box.center)}]
\draw[dualblue,thick,middlearrow={stealth reversed}] (0,-4/3.4) -- (0,-4*0.4/3.4);
\draw[dualblue,thick,middlearrow={stealth reversed}] (-1, 2/3.4 ) -- (-0.4,0.8/3.4);
\draw[dualblue,thick,middlearrow={stealth}] (1,2/3.4) -- (0.4,0.8/3.4);
\draw[dualblue,thick,middlearrow={stealth reversed}] (0.4,0.8/3.4) -- (-0.4,0.8/3.4);
\draw[dualblue,thick,middlearrow={stealth reversed}] (-0.4,0.8/3.4) -- (0,-1.6/3.4);
\draw[dualblue,thick,middlearrow={stealth}] (0.4,0.8/3.4) -- (0,-1.6/3.4) ;
\draw (-1.15,0.75) node {$b$};
\draw (0,-1.4) node {$a$};
\draw (1.15,0.75) node {$c$};
\draw (0, 0.55) node {$d$};
\draw (-0.46,-0.12) node {$e$};
\draw (0.46,-0.12) node {$f$};
\end{tikzpicture}
\right)
= [F^{abd}_{fce}]^* \sqrt{\frac{d_d d_f}{d_c}} \
\Psi_{\hat{\Lambda}}\left(
\begin{tikzpicture}[scale=0.4, baseline={([yshift=-.5ex]current  bounding  box.center)}]
\draw[dualblue,thick,middlearrow={stealth reversed}] (0,-4/3.4) -- (0,0);
\draw[dualblue,thick,middlearrow={stealth reversed}] (-1, 2/3.4 ) -- (0,0);
\draw[dualblue,thick,middlearrow={stealth}] (1,2/3.4) -- (0,0);
\draw (-1.25,0.8) node {$b$};
\draw (0,-1.7) node {$a$};
\draw (1.25,0.8) node {$c$};
\end{tikzpicture} \right)
\end{align}
We have defined
\begin{align}
\label{Gmove}
G^{abc}_{def} = F^{bed}_{fac}\sqrt{\frac{d_e d_f}{d_a d_c}} ,
\end{align}
where $G^{abc}_d$ is a unitary matrix:
\begin{align}
G^{abc}_{d} [G^{abc}_{d}]^\dagger = \sum_f G^{abc}_{def} (G^{abc}_{de'f})^* = \delta_{ee'}
\end{align}

These equations are to be interpreted as follows. We consider two different graphs, specified by
$\hat{\Lambda}$ and $\hat{\Lambda}'$, whose duals $\Lambda$ and $\Lambda'$ can be viewed as
two different triangulations of the same surface. $\hat{\Lambda}$ and $\hat{\Lambda}'$ only differ
in the local patch that is illustrated. The equations show how to relate the wavefunction amplitudes
for the states associated with the different graphs.

We have chosen to illustrate the wavefunctions using the dual cellulations $\hat{\Lambda}$, but of course
one could also use the triangulation $\Lambda$.

Other similar relations exist as well; we do not list all of them here, but they can be easily derived from the path integral
state sum of the TVBW TQFT by considering the 2-2 and 1-3 Pachner moves with various choices of branching structures.

Since the external legs of the above diagrams are all fixed, $F^{abc}_d$ and $G^{abc}_d$ can be viewed as controlled-unitary
gates, which effectively change the lattice geometry inside the plaquette defined by the edges $a,b,c,d$ of the original
triangulation $\Lambda$ (see Fig. \ref{fig:F-move_definition}). This implies that these local moves can be performed
in parallel over extensive regions of the lattice -- a property which we will exploit in subsequent sections.

The 1-3 Pachner move, shown in Fig.~\ref{fig:pachner_move_definition} and Eq.~\eqref{FmoveRel2}, adds additional edges and thus additional qudits
to the lattice. We can obtain the state on the new lattice from a state on the old lattice by appropriately initializing
new qudits and applying a local unitary circuit shown in Fig.~\ref{fig:tadpole}. We first consider three extra qudits, each initialized
to the $|0\rangle$ state. Next we consider applying a $CX$ operation, which takes $|b\rangle|0\rangle \mapsto |b\rangle |b\rangle$,
to the qudits shown in Fig.~\ref{fig:tadpole}(a). At the same time, we apply $S: |0\rangle $$\mapsto $$ \sum_i \frac{d_i}{D} |i \rangle$
to the top-most qudit, which effectively builds a `tadpole diagram' connected to the original graph through the edge with the remaining
ancilla in the $\ket{0}$ state, as shown in Fig.~\ref{fig:basic-moves}(b).  Note that the original edge labeled by $b$
is split into two edges with the same label $b$.  The equivalence to the tadpole diagram has implicitly used the concept
of a smooth string net (see Refs.~\onlinecite{levin2005} and \onlinecite{Koenig:2010do} for details).
Next, we apply a sequence of two 2-2 Pachner moves by applying the unitary operations $[G^{bbi}_i]^\dag$
and $[G^{aji}_c]^\dag$, as shown in Fig.~\ref{fig:tadpole}(e).

Therefore we see that the 2-2 Pachner moves and the 1-3 Pachner moves, which locally change the
lattice geometry, can be effectively implemented as unitary operations on the many-qudit quantum state.
In what follows, we will demonstrate how to use these moves in parallel across large regions of the lattice,
in order to build up non-trivial larger scale geometry deformations through constant depth circuits. We will then
subsequently use these large scale geometry deformations for braiding and Dehn twists.

For simplicity, from now on we will call all the 2-2 Pachner moves as F-moves, with the implication that
depending on the branching structures they can correspond to different unitaries such as those in
Eqs.~\eqref{FmoveRel1}, \eqref{FmoveRel3}, and others.   All these unitaries will be proportional to the
$F$-symbols with aditional normalization factor such as the case in Eqs.~\eqref{FmoveRel3} and \eqref{Gmove}.

\subsection{Dehn twist on a torus and a cylinder}

\begin{figure*}
\includegraphics[width=2\columnwidth]{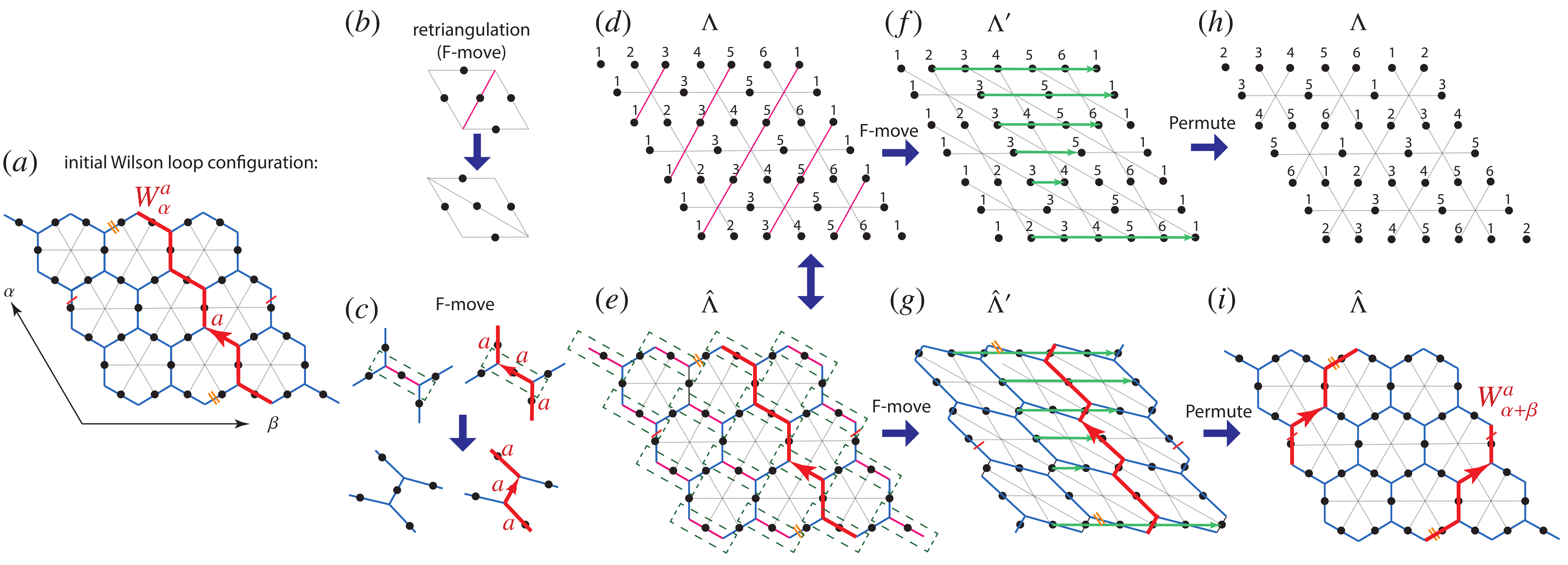}
\caption{Dehn-twist on a torus for the Turev-Viro code. The upper row (b,d,f,g) shows the protocol on the triangulation (light grey), while the lower row (c,e,g,i) illustrates the protocol on the dual trivalent graph (blue). The pink lines indicate the edges to be switched during the F-moves.  The green dashed box specifies the four external legs of the F-moves on the trivalent graph.  The green arrows represent qubit permutations, with the column numbers specified before and after the permutation.}
\label{fig:Dehn-twist-string-nets}
\end{figure*}

Let us now consider the case where $\Sigma = T^2$ is a torus. Thus we consider the case where the honeycomb lattice $\hat{\Lambda}$
and the triangulation $\Lambda$ have periodic boundary conditions, as shown in Fig.~\ref{fig:Dehn-twist-string-nets}(a). The initial lattice and triangulation correspond to a microscopic Hilbert space $\mathcal{H}_\Lambda$. To help demonstrate
the resulting Dehn twist, we follow an initial Wilson loop $W^a_\alpha$, i.e., a non-contractible string with string type $a$ passing
along the $\alpha$-cycle.

The elementary move in this protocol is the retriangulation obtained by the F-move (2-2 Pachner move). As illustrated in Fig.~\ref{fig:Dehn-twist-string-nets}(b),
this corresponds to flipping the edge (pink) between two neighboring triangles, and can be achieved by the controlled-unitary $F$ operation
discussed in the previous section. Fig.~\ref{fig:Dehn-twist-string-nets}(c) illustrates the effect on the dual lattice and on the string operator $W^a_\alpha$.

We mark all the varying edges of the triangulation by pink lines in Fig.~\ref{fig:Dehn-twist-string-nets}(d), with the corresponding moves in the dual graph
indicated by the dashed box and pink edges in Fig.~\ref{fig:Dehn-twist-string-nets}(e).  The new triangulation $\Lambda'$ and honeycomb lattice
$\hat{\Lambda}'$ are shown in Fig.~\ref{fig:Dehn-twist-string-nets}(f,g), corresponding to a different microscopic Hilbert space
$\mathcal{H}_{\Lambda'}$.  Since we can perform all of these moves by controlled-unitary operations in parallel, this retriangulation can be done
via a local quantum circuit with $\mathcal{O}(1)$ depth.

To recover the original triangulation $\Lambda$ and trivalent graph $\hat{\Lambda}$, we now perform a collective permutation of the qudits,
$\mathcal{P}_\sigma$, indicated by the arrows in Fig.~\ref{fig:Dehn-twist-string-nets}(f,g). The qudits in each row are cyclically permuted by a spacing
indicated by the length of the arrow in that row.

After the permutation, the column labels of the qubits are changed into the configuration shown in Fig.~\ref{fig:Dehn-twist-string-nets}(j),
which corresponds to the honeycomb lattice configuration in Fig.~\ref{fig:Dehn-twist-string-nets}(i) where the Wilson loop now also
winds around the $\beta$-cycle.  In the end of the protocol, we thus come back to the original triangulation $\Lambda$ and honeycomb lattice
$\hat{\Lambda}$ with permuted sites, and hence the same microscopic Hilbert space $\mathcal{H}_{\Lambda}$. This is
exactly a self-diffeomorphism and corresponds to a Dehn twist $\mathcal{D}_\beta : W^a_{\alpha} \longmapsto  W^a_{\alpha + \beta}$.
We have thus shown that the Dehn twist $\mathcal{D}_{\beta}$ can be implemented by a constant depth local unitary quantum circuit, followed by a permutation on
the qubits.

We note that in the illustration depicted in Fig. \ref{fig:Dehn-twist-string-nets}, as well as all the illustrations for the following protocols,
we have not drawn the branching structure in order to keep the illustrations simple. It can be verified that all the protocols can be adapted
to the case where the branching structure is taken into account, so that the branching structure after the protocol is the same as before
the protocol.

\subsection{Braiding}

Now we consider constant-depth circuits for implementing braiding in Turaev-Viro codes. We will consider braiding of `punctures',
with the understanding that punctures can refer to anyons (including non-Abelian anyons), holes, or bulk twist defects.
An important ingredient in the braiding protocols includes the ability to move the punctures by a distance of order the
code distance $d$, by a local finite depth quantum circuit, followed by a permutation of qubits.
When these protocols are applied to the case of the non-Abelian anyons of the Fibonacci surface code, they imply the possibility of
a universal fault-tolerant gate set by a constant depth local quantum circuit, together with qubit permutations.

\subsubsection{Braiding on a space with periodic boundary conditions}

\begin{figure*}[hbt]
  \includegraphics[width=2\columnwidth]{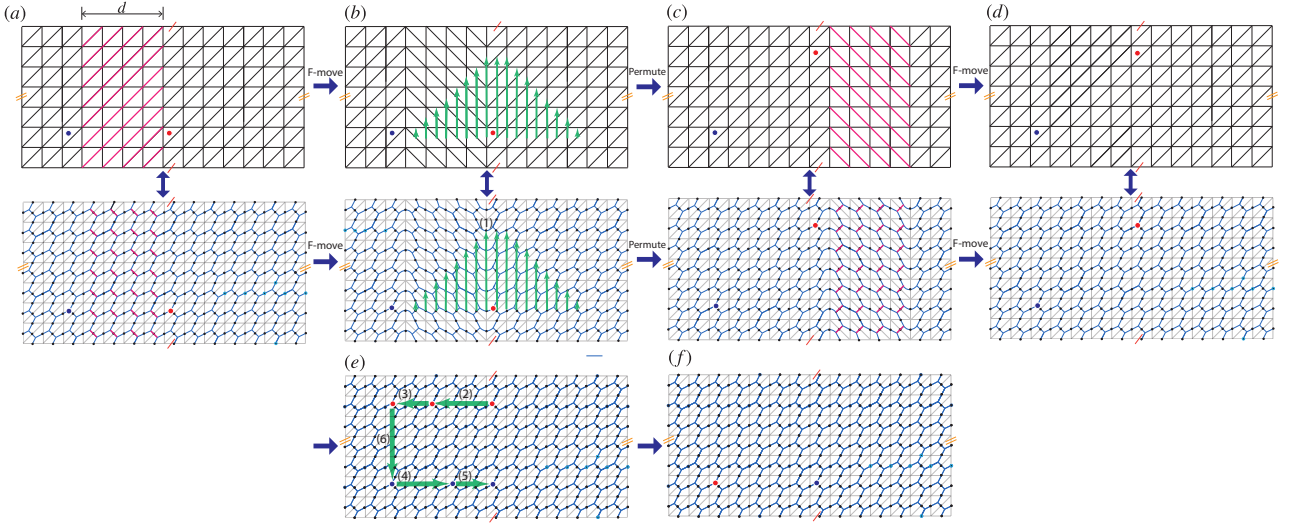}
  \caption{Braiding of two punctures, e.g. non-Abelian anyons (blue and red circles) in Turaev-Viro codes,  with constant depth circuit on a space with periodic boundary conditions (e.g. a genus $g > 0$ surface).
The scheme can also be applied to punctures on a bilayer system with holes and appropriate boundary conditions, which is a way to effectively obtain a single layer
system on a high-genus surface.}
  \label{fig:braiding_string_net_periodic}
\end{figure*}

We first consider the simpler situation where the punctures are located on a space with periodic boundary conditions, such as a
torus (Fig.~\ref{fig:braiding_string_net_periodic}). Physically this can be realized in a planar geometry by considering a bilayer system
with holes and appropriate boundary conditions.

We start the protocol in Fig.~\ref{fig:braiding_string_net_periodic}(a) with two punctures separated by $d$ plaquettes. The code distance $d = 4$
in Fig.~\ref{fig:braiding_string_net_periodic}(a). The upper panel of Fig.~\ref{fig:braiding_string_net_periodic}(a) shows the triangulation $\Lambda$
and the lower panel shows the corresponding trivalent graph $\hat{\Lambda}$. As above, for simplicity we do not illustrate the branching structure.

The first step is to perform a retriangulation inside the cylindrical strip of width $d$ between the two punctures. The retriangulation corresponds
to the F-moves (2-2 Pachner moves) discussed above; the edges to be flipped are indicated in pink thick lines, and the corresponding change of the dual graph
$\hat{\Lambda}$ is indicated in the lower panel. The retriangulation can be performed via the controlled $F$ operations discussed in the
previous sections, and can be performed in parallel over the entire strip. Therefore this corresponds to a local quantum circuit with depth $\mathcal{O}(1)$.

The resulting triangulation and trivalent graph is shown in Fig.~\ref{fig:braiding_string_net_periodic}(b).
We then permute the qudits in a shearing pattern (indicated by the green arrows) to reach the configuration in Fig.~\ref{fig:braiding_string_net_periodic}(c). Specifically,
the qudits in each column are permuted cyclically, by a number of steps indicated by the length of the arrow in each column. This
sequence of moves causes the puncture on the right (red) to be moved vertically by $d$ spacings via a local constant depth quantum circuit, followed by a permutation.
Note that due to the periodic boundary conditions, some qudits on top are permuted to the bottom.

Now the triangulation pattern is modified in the strip of width $d$ to the right of the (red) puncture, compared to the original
configuration in Fig.~\ref{fig:braiding_string_net_periodic}(a).  To recover the original triangulation, we perform another
retriangulation, as shown in Fig.~\ref{fig:braiding_string_net_periodic}(d).

The overall effect of this cycle, therefore, is that we have moved the puncture vertically upward by $d$ spacings, via
a constant depth local quantum circuit, followed by a permutation of qudits, and followed finally by another constant depth
local quantum circuit.

Another 5 steps completes the cycle of braiding, exchanging the red and blue punctures as shown
in Fig.~\ref{fig:braiding_string_net_periodic}(e,f). Therefore we have seen how we can braid two punctures
in $6$ steps, each of which contains a permutation of qudits sandwiched between two constant depth local quantum circuits.

Note that in the above protocol, in each step the punctures can only be moved by a distance that is bounded by the distance to the closest
puncture.

If we have $n$ punctures, we can consider placing the punctures along a line, with a spacing $d$ between each puncture. Then
neighboring punctures along the line can be braided using the protocol described above. Therefore, this method demonstrates how
to perform elementary braids for any number of punctures on a space with periodic boundary conditions.

\subsubsection{Braiding on a disk subregion}
\label{tvDiskBraid}

\begin{figure*}[hbt]
 \includegraphics[width=2\columnwidth]{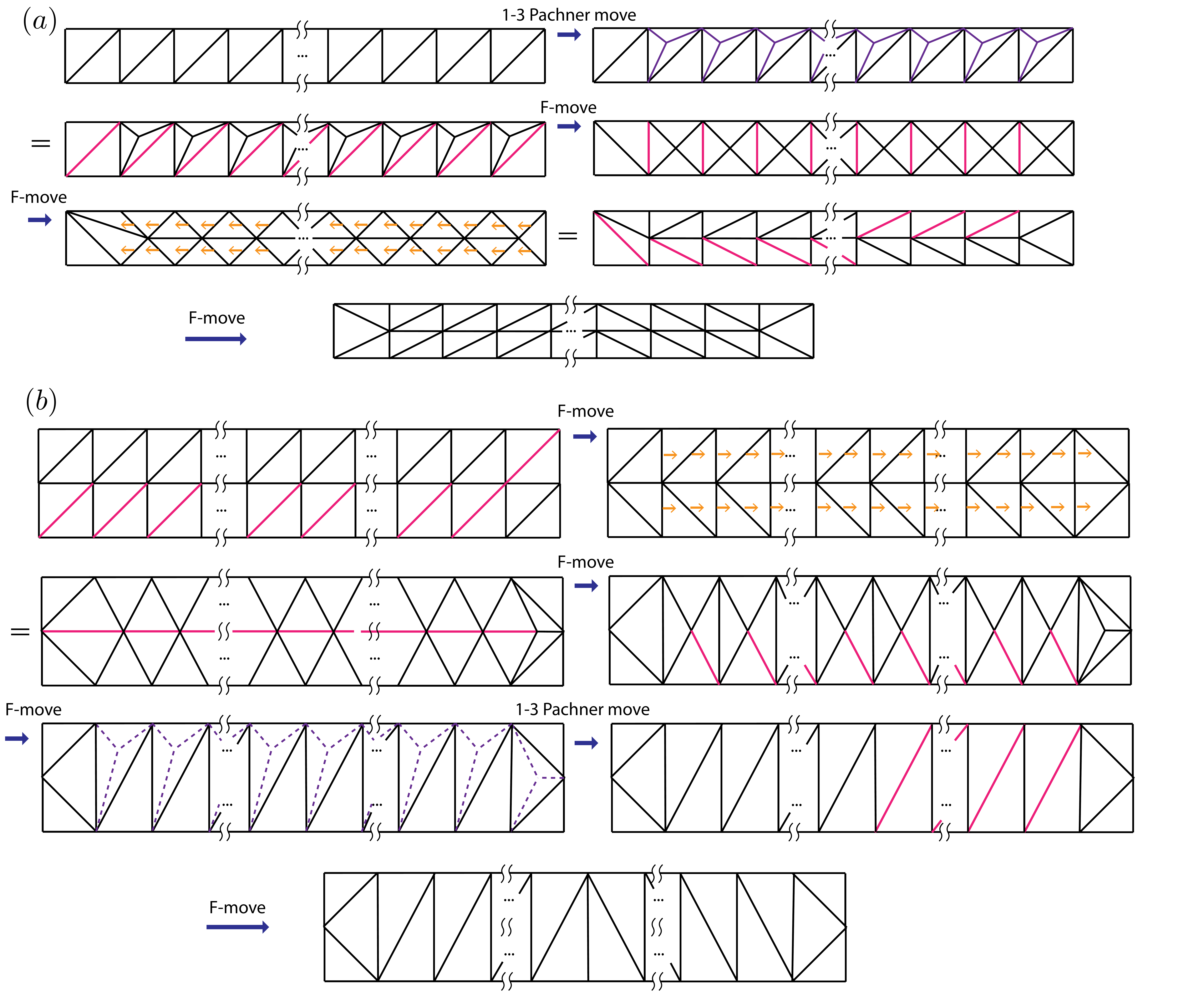}
  \caption{ Gadgets for local geometry deformation in Turaev-Viro codes.  The solid (dashed) purple lines represent added (removed) edges during the 1-3 Pachner moves.  The pink line represent the switched edges in F-moves.  The yellow arrow indicates the equivalence between two triangulations by locally shifting the positions of the edges, which can be physically implemented by local SWAPs.
\label{fig:longGadgets}
}
\end{figure*}

\begin{figure*}[hbt]
  \includegraphics[width=2\columnwidth]{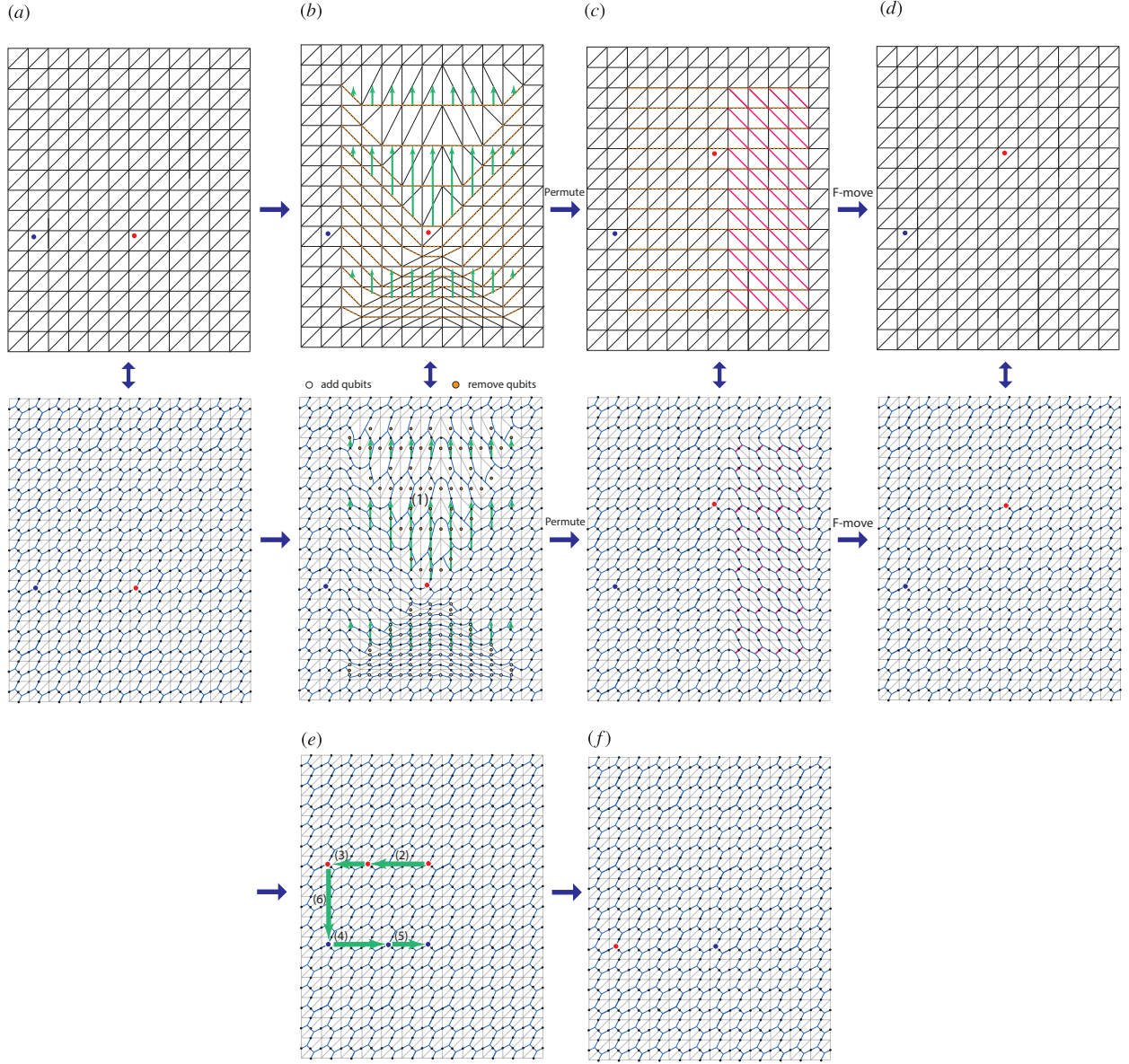}
  \caption{Braiding of two non-abelian anyons in Turaev-Viro codes with constant-depth
    circuit on a disk subregion, keeping the boundary fixed. The yellow dashed lines in (b) and (c) show the equivalent edges before/after the permutation.}
  \label{fig:braiding_string_net}
\end{figure*}

Here we demonstrate a protocol for braiding punctures where the whole protocol acts only within a disk subregion
of the space. As such, it can be applied to braiding punctures on any space, and in a way which does not affect the system
outside of the disk subregion.

In the case of braiding on a disk subregion, our protocol will require merging and splitting
plaquettes in a manner similar to the case of the toric code in Fig.~\ref{fig:braiding_circuits}.
Two important building blocks of our protocol are demonstrated in Fig.~\ref{fig:longGadgets}.
In Fig.~\ref{fig:longGadgets}(a), we consider a single row of arbitrary length. By utilizing
ancilla qudits, we can implement the 1-3 Pachner moves, which increase the number of vertices
of the triangulation. By a finite sequence of F-moves (2-2 Pachner moves) and local SWAPs, we can
effectively split a single row of arbitrary length $L$ into two rows, with a constant (independent of the length)
number of steps (i.e. a constant depth local unitary circuit). In Fig.~\ref{fig:longGadgets}(b), we illustrate how
two rows can be converted into a single row by a finite number of steps.

Note that in both of these protocols, the qudits on the outer boundary of the rows shown are completely unaffected,
acting as control qudits for the unitary operations. This then allows the similar transformations to be applied to
a large number of rows in parallel.

Using the above gadgets for splitting or combining rows, we can now demonstrate our braiding protocol, shown
in Fig.~\ref{fig:braiding_string_net}. In the first step, in the region below the right puncture (red), we split rows
of varying lengths in two rows, while combining rows in the region above the puncture, in a manner illustrated in Fig.~\ref{fig:braiding_string_net}(b).
This can be performed by a local quantum circuit with a constant depth, independent of the spacing between punctures.
Similar to the protocol for toric code (Fig.~\ref{fig:braiding_circuits}), we create a lattice with a shearing pattern on the left and right sides
of the (red) puncture; the regions above the (red) puncture being coarse grained (effectively compressed) while the region below the (red) puncture
is fine grained (effectively stretched).

The dual graph in the lower panel illustrates the positions of the qudits. The white and yellow circles represent
added (entangled) and removed (disentangled) qudits respectively.  Now via long-range permutation of qudits
(indicated by green arrows), we reach the configuration in Fig.~\ref{fig:braiding_string_net}(c) where the (red) anyon
is moved up.  To recover the original triangulation and corresponding trivalent graph, we apply another step of
retriangulation in the strip on the right of the (red) anyon (indicated by the pink thick lines), and hence map back
to the original lattice in Fig.~\ref{fig:braiding_string_net}(d).

The above protocol, which uses a constant-depth local unitary circuit and long-range qudit permutations,
effectively moves one puncture vertically by a distance of the order of the separation between the nearest
puncture, which is on the order of the code distance. To complete a braiding cycle, we apply another 5 shots
of a similar procedure, which then effectively braids the two punctures around each other as illustrated
in Fig.~\ref{fig:braiding_string_net}(e, f).

To summarize, a single braiding operation -- either in a single patch or utilizing periodic boundary conditions --
can be performed in a constant number of steps, independent of the system size and code distance.   Note that this is in
contrast with the previous computation schemes of the Turaev-Viro code presented in Ref.~\onlinecite{Koenig:2010do},
where braiding or Dehn twists are implemented by sequential F-moves with circuit depth of $\mathcal{O}(d)$.

In this case we have demonstrated a 6-step procedure:
\begin{align}
\mathcal{B}_{1,2}=\prod_{i=1}^{6} \mathcal{LU}'_i \mathcal{P}_{\sigma, i} \mathcal{LU}_i.
\end{align}
Note that each step is composed of a  constant-depth local quantum circuit $\mathcal{LU}_i$ corresponding to a
retriangulation of the manifold, a permutation of qubits $P_{\sigma, i}$ over a distance $\mathcal{O}(d)$, and
another local circuit $\mathcal{LU}'_i$ in order to retriangulate the manifold back to the original triangulation.
Here, we choose three smaller steps in each of the $6$ steps of the move because of the symmetry of the
intermediate retriangulation pattern shown in Fig.~\ref{fig:braiding_string_net_periodic}(c) and
Fig.~\ref{fig:braiding_string_net}(c). Alternatively, one can also exchange the last retriangulation with the
permutation and merge the two retriangulations into one, so that
\begin{align}
\mathcal{B}_{1,2}=\prod_{i=1}^{6} \tilde{\mathcal{P}}_{\sigma, i} \widetilde{\mathcal{LU}}_i.
\end{align}
This results in a less symmetric triangulation during intermediate steps of the protocol, and the local circuit $\widetilde{\mathcal{LU}}_i$ will have a somewhat longer (but still finite, independent of $d$) interaction range.

We note that, as discussed in the context of the $\mathbb{Z}_N$ toric code, another approach to braiding is to consider a half Dehn twist
along a loop that surrounds the two punctures. In this case, the protocol to perform a Dehn twist on an annulus can then be easily adapted
to perform a half Dehn twist along the loop surrounding two punctures. In the following section we demonstrate protocols for a full Dehn twist
on an annulus for the Turaev-Viro codes. The adapation of this protocol to a half Dehn twist to perform braiding is straightforward and will thus
not be presented explicitly here. Nevertheless, we note that it can be used to effectively perform the braiding in ``one shot'' by a single
$\mathcal{LU}$ followed by a permutation.

\subsection{Dehn twists on an annulus}

In this section, we demonstrate protocols for implementing a Dehn twist on an annulus in the
Turaev-Viro codes. This operation can then be utilized for braiding (discussed above), or for
performing Dehn twists along the $\beta$ loops of a high genus surface, discussed
in the subsequent section.

\begin{figure*}
  \includegraphics[width=2\columnwidth]{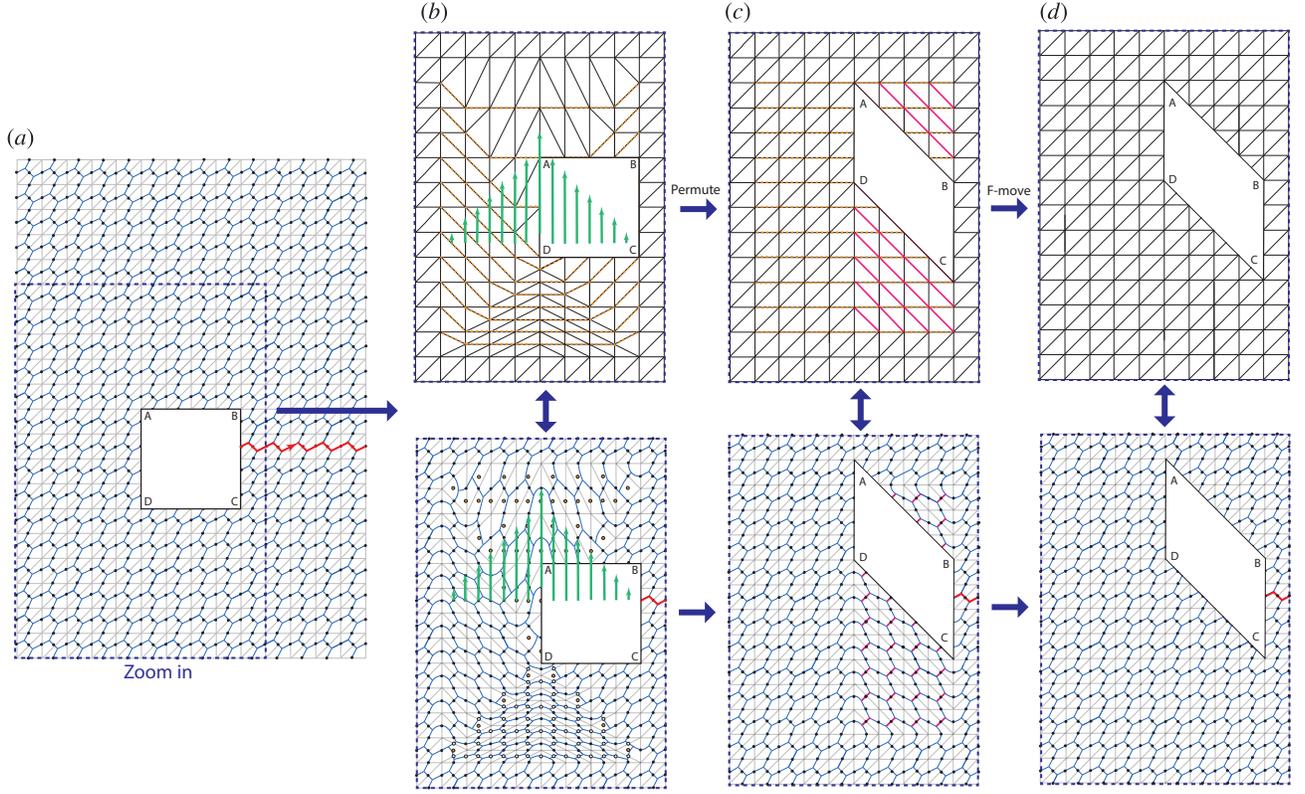}
  \caption{Implementation of Dehn twist $\mathcal{D}_\beta$ in Turaev-Viro codes, on an annulus along the $\beta$-loop encircling the boundary defect via a sequence of shearing operation.  The red line with arrows represents a Wilson-line operators connecting the inner and outer boundaries.}
\label{fig:modular_genus_g_beta_string_nets}
\end{figure*}

\begin{figure*}
  \includegraphics[width=2\columnwidth]{modular_genus_g_beta_string_nets_step2.pdf}
  \caption{(continued) Implementation of Dehn twist $\mathcal{D}_\beta$ in Turaev-Viro codes on an annulus.}
\label{fig:modular_genus_g_beta_string_nets_step2}
\end{figure*}

\begin{figure*}
  \includegraphics[width=2\columnwidth]{modular_genus_g_beta_string_nets_step3.pdf}
  \caption{(continued) Implementation of Dehn twist $\mathcal{D}_\beta$ in Turaev-Viro codes on an annulus.}
\label{fig:modular_genus_g_beta_string_nets_step3}
\end{figure*}

\subsubsection{Protocol 1: Twist via shearing}

We begin with an annulus depicted in Fig.~\ref{fig:modular_genus_g_beta_string_nets}(a), where we have
drawn a red string, $\alpha$, from the inner to the outer boundary to track its motion during the course of the
protocol and to verify when we have completed a Dehn twist. We note that the qudits on the inner
and outer boundaries are unaffected, aside from undergoing the relevant permutations. Therefore our protocol
can be used for any choice of boundary conditions, or even when this annulus is viewed as a piece of a larger surface.

As we have demonstrated in the case of the toric code, we can perform an effective rotation of the inner boundary by $2\pi$ relative to the outer boundary
through a sequence of shear deformations. The steps in this protocol are analogous to those used in the braiding protocol above.
For the first shearing, we start with a constant-depth local quantum circuit $\mathcal{LU}$, which implements a retriangulation to reach the
configuration shown in Fig.~\ref{fig:modular_genus_g_beta_string_nets}(b).  The corresponding geometry deformation on the trivalent graph is
shown in the lower panel. After a permutation $P_{\sigma}$, one reaches the pattern in Fig.~\ref{fig:modular_genus_g_beta_string_nets}(c).
To revert to the original triangulation, we apply another constant-depth local circuit  $\mathcal{LU}'$, to reach the configuration in
Fig.~\ref{fig:modular_genus_g_beta_string_nets}(d). We see that the lattice has returned to the original form, while the inner hole has
been effectively sheared upward.

Now we continue and perform the second shear. We first apply a constant depth local unitary to effectively induce the retriangulation shown in
Fig.~\ref{fig:modular_genus_g_beta_string_nets_step2}(a),  followed by a permutation to reach the configuration in
Fig.~\ref{fig:modular_genus_g_beta_string_nets_step2}(b). A third constant-depth local unitary effectively returns the system to the original
triangulation, with the net effect being a horizontal shear that takes the vertex $A$ to the upper right corner and the vertex $C$ to the lower-left corner.

We continue this basic procedure through another seven shears, as shown in Fig.~\ref{fig:modular_genus_g_beta_string_nets_step3}.
The net effect of the whole procedure is thus to effectively rotates the inner boundary of the annulus by $2\pi$. The system is returned
to its original configuration. By tracking the red string $\alpha$, we see that $\alpha \rightarrow \alpha + \beta$, as shown in
Fig.~\ref{fig:modular_genus_g_beta_string_nets_step3}(h), and thus we have performed a Dehn twist along the $\beta$ curve of the annulus.

In sum, our protocol uses $9$ steps of shearing, each of which is composed of 3 smaller steps: a constant-depth local circuit, followed by a permutation,
and then another constant-depth local circuit. Thus, when acting in the code subspace, we have
\begin{align}
\mathcal{D}_\beta &= \prod_{i=1}^9 \mathcal{LU}_i' \mathcal{P}_{\sigma_i} \mathcal{LU}_i
\nonumber \\
&= \prod_{i=1}^9 \tilde{\mathcal{P}}_{\sigma_i} \widetilde{\mathcal{LU}}_i.
\end{align}
As in the braiding discussion, in the second line we have commuted the last permutation through the last
$\mathcal{LU}$ and combined neighboring local quantum circuits.

\subsubsection{Protocol 2: Single shot twist}

As in the discussion for the toric code in Sec.~\ref{sec:alternate_beta}, in the general Turaev-Viro codes it is also possible to perform the Dehn twist
around the annulus in a `single shot,' in the sense that we need only one constant-depth local quantum circuit
$\mathcal{LU}$ followed by a permutation.  The previous shearing protocol, in contrast, used $9$ such steps. However
in this single shot procedure the local unitary circuit has longer range, although the range is still independent of code distance and system size.

The implementation of this protocol follows almost the same steps described in Sec.~\ref{sec:alternate_beta} for the case of the toric code. However
in contrast to the toric code case, the local unitary circuit required in this case to perform the analog of step 3 in Sec.~\ref{sec:alternate_beta}
is somewhat more involved. To explain the protocol, we first focus on a single ring encircling the inner boundary
and explain how one can utilize F-moves (2-2 Pachner moves) to reconnect the vertices accordingly. After we have understood
the case of a single ring, we can then use the same gadget in parallel across different rings to implement the full protocol.

\begin{figure*}[hbt]
  \centerline{\includegraphics[width=2\columnwidth]{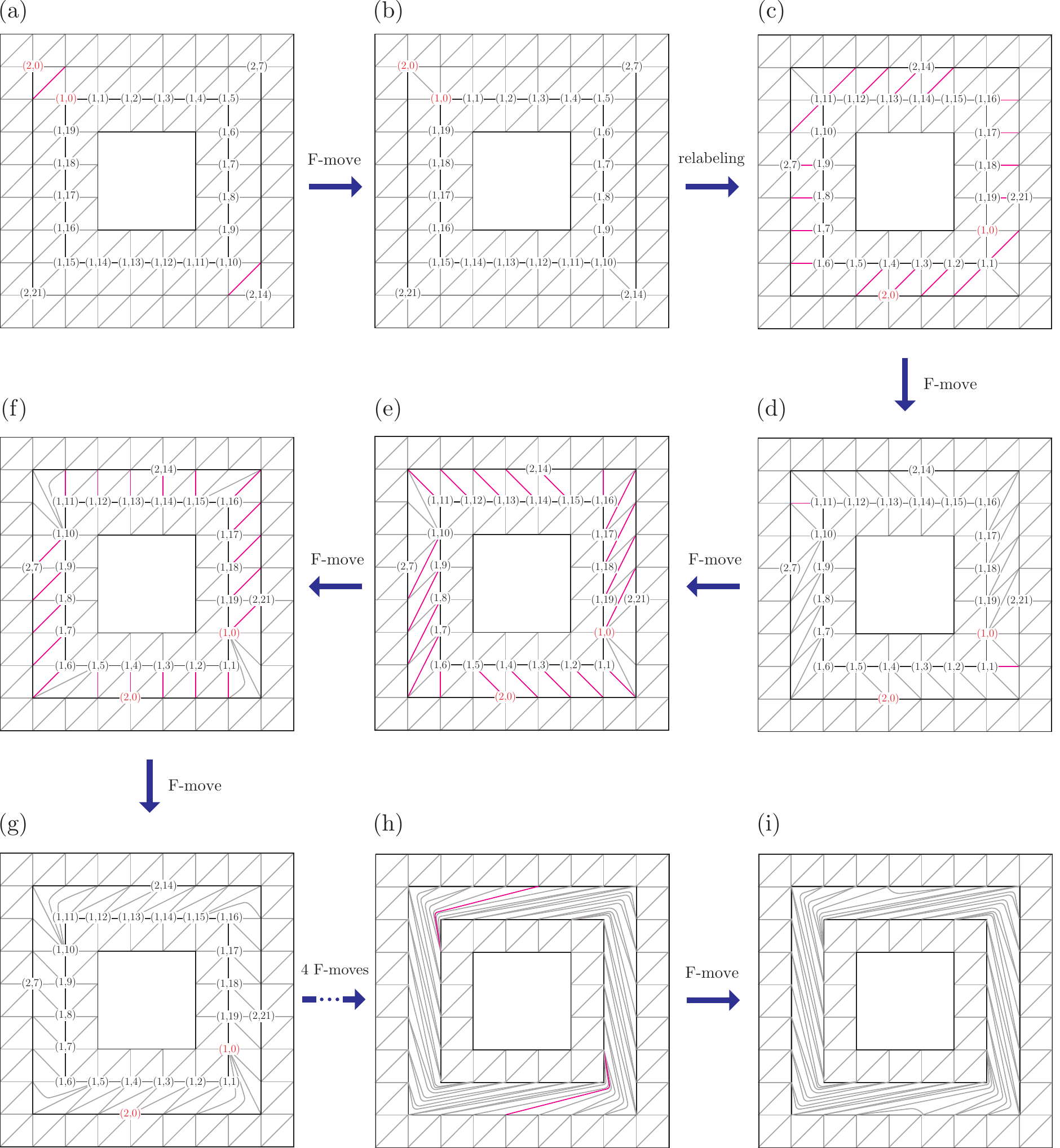}}
  \caption{The gadget used for the alternative protocol of Dehn twist on an annulus in Turaev-Viro codes.}
  \label{fig:NA_Beta_Gadget}
\end{figure*}

We begin by grouping the vertices of the triangulation into rings of increasing size surrounding the inner boundary; we thus
label the vertices as $(n,m)$ where $n$ labels the ring and $m$ labels the vertices along the ring, with $m$ increasing
clockwise starting at the top left [see Fig.~\ref{fig:NA_Beta_Gadget}(a)]. We first perform an F-move (2-2 Pachner move)
as shown in Fig.~\ref{fig:NA_Beta_Gadget}a which facilitates reconnection of vertices.

\begin{figure*}[hbt]
  \centerline{\includegraphics[width=2\columnwidth]{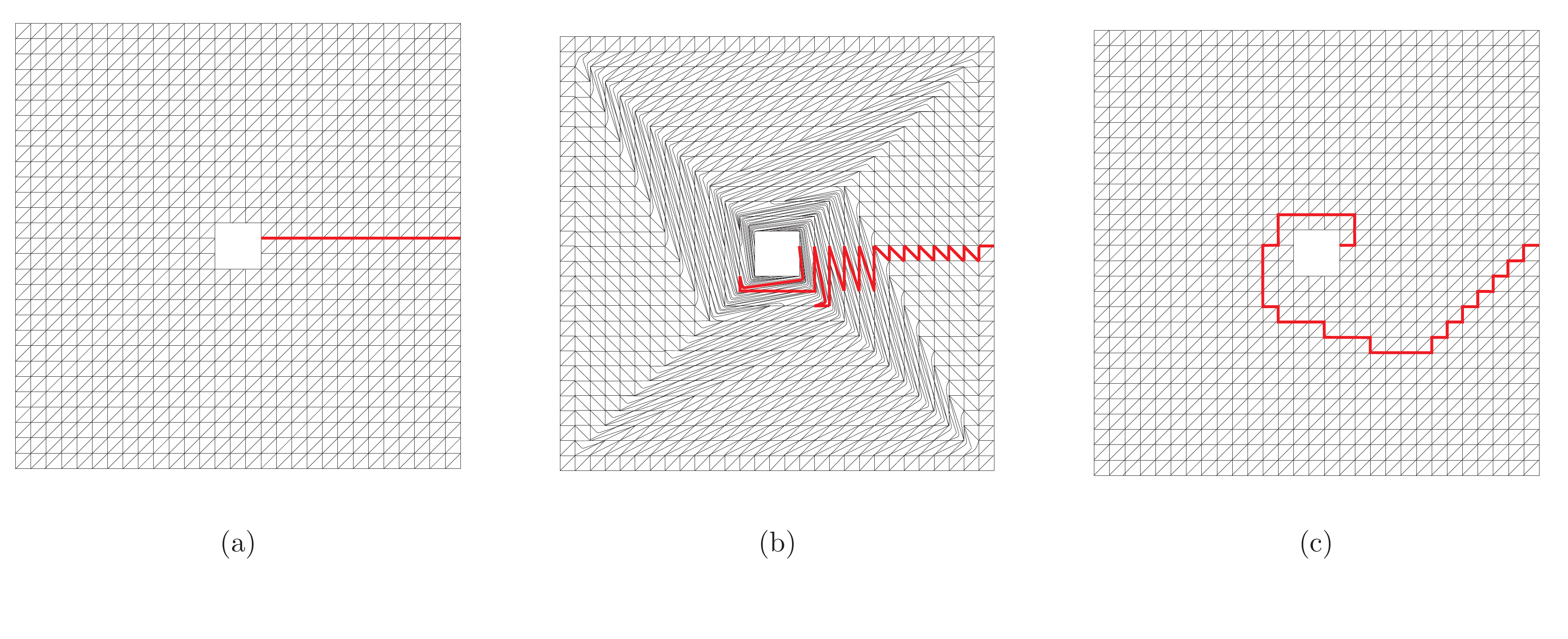}}
  \caption{Alternative protocol of Dehn twist on an annulus for Turaev-Viro codes. }
  \label{fig:NA_Beta}
\end{figure*}

Next, we relabel the vertices by shifting the labels $(n,m) \rightarrow (n,m-9n)$ as shown in Fig.~\ref{fig:NA_Beta_Gadget}(c). Our goal now is to perform a constant depth local unitary,
corresponding to a constant number of F-moves, such that the edges of the final lattice connect vertices with the same labels as at the start
of the protocol. To do this, consider first the two sequences of F-moves shown in Fig.~\ref{fig:NA_Beta_Gadget}(c) and (d). After these F-moves, we see that the number
of edges coming out of each vertex matches that of the original lattice in Fig.~\ref{fig:NA_Beta_Gadget}(b), up to a cyclic permutation of the vertices.

Therefore to get to the desired configuration, we perform a total of $6$ steps, each of which consists of applying F-moves in parallel
around the ring, as shown in Fig.~\ref{fig:NA_Beta_Gadget}(e)-(g). The exact number of steps varies among the rings, starting from
$9$ for the inner most ring and decreasing as one moves further away. After these steps, we see that the graph consisting of the vertices and edges
is the same as it was before the relabelling $(n,m) \rightarrow (n,m-9n)$. Finally we apply the F-moves shown in Fig.~\ref{fig:NA_Beta_Gadget}(h)
to account for the first F-moves done in Fig.~\ref{fig:NA_Beta_Gadget}(a).

Note that in the gadgets described in Fig.~\ref{fig:NA_Beta_Gadget}, the qudits at the boundaries of the rings (i.e on the horizontal and vertical
thick edges in Fig.~\ref{fig:NA_Beta_Gadget}) are used only as control qudits. This allows us to perform this transformation on all of the rings
in parallel. For a hole with $L$ qudits on its boundary, we have to modify $L$ rings to perform the twist. Fig.~\ref{fig:NA_Beta}(b) shows the state of the whole annulus after performing the gadget on all 12 rings in parallel. We see that the configuration in Fig.~\ref{fig:NA_Beta}(b)
can be achieved by a constant-depth local quantum circuit.

Finally, by a permutation we can recover the original lattice (Fig.~\ref{fig:NA_Beta}(c) ). By following a red string $\alpha$ throughout the protocol, we
see that a Dehn twist around the $\beta$ loop has been performed. In sum, we thus see that a Dehn twist around a loop encircling the inner boundary of an annulus
can be achieved by a constant-depth local quantum circuit, $\mathcal{LU}_\beta$, followed by a permutation: $\mathcal{D}_\beta = \mathcal{P}_\sigma \mathcal{LU}_\beta$.

\subsection{Dehn twists on a high-genus surface for Turaev-Viro codes}

\begin{figure*}[hbt]
  \includegraphics[width=2\columnwidth]{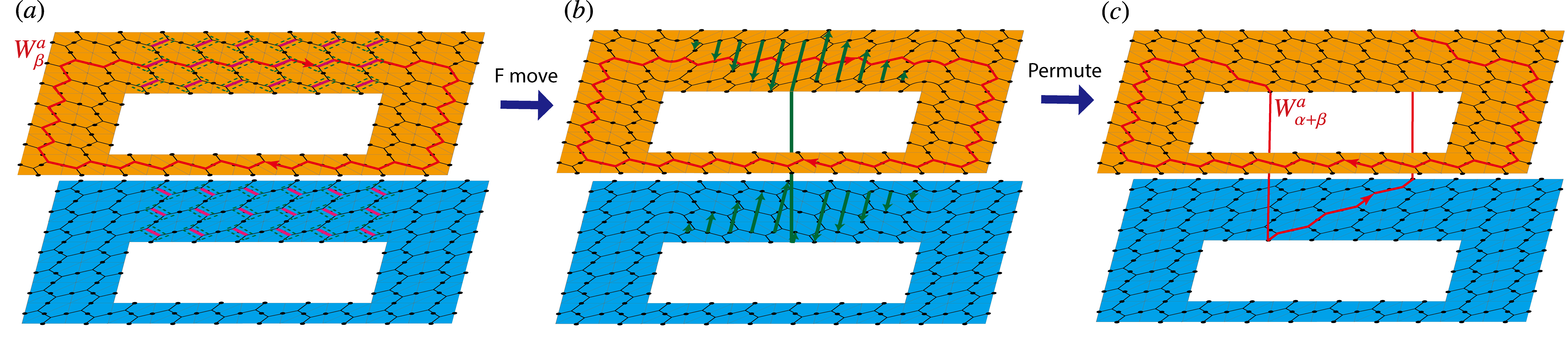}
  \caption{Implementation of Dehn twist along the $\alpha$-loop in Turaev-Viro code. The red line with arrows represent the Wilson loop.  The pink thick lines represent the edges being switched during the F-moves, with the green dashed boxes specifying the four external legs.  The dark green arrows represent qubit permutation with a shearing pattern.}
\label{fig:modular_genus_g_alpha_string_net}
\end{figure*}

\begin{figure*}[hbt]
  \includegraphics[width=2\columnwidth]{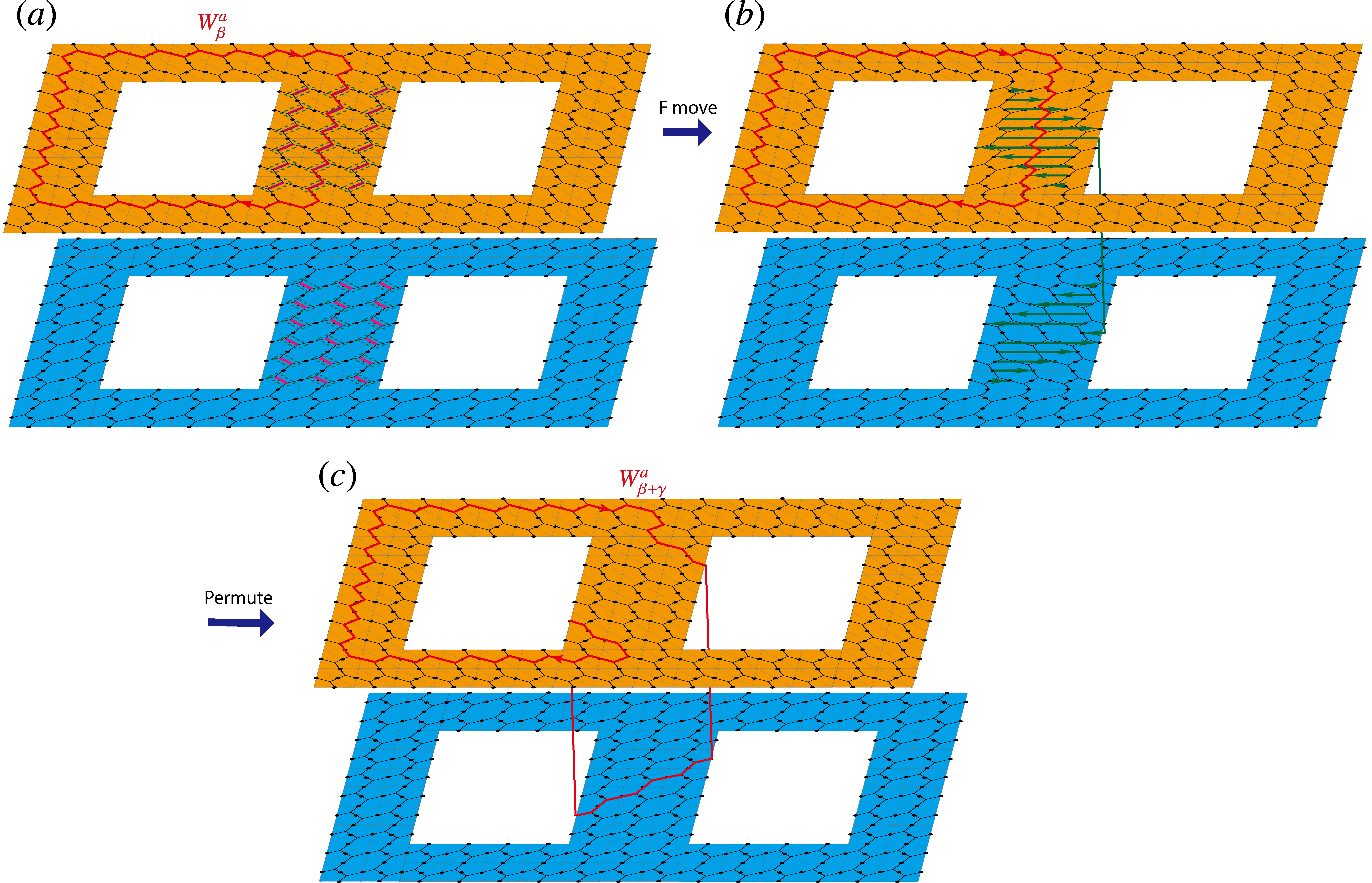}
  \caption{Implementation of Dehn twist along the $\gamma$-loop in Turaev-Viro code.}
\label{fig:modular_genus_g_gamma_string_net}
\end{figure*}

The results of the previous sections can be straightforwardly adapted to the case of braids
and Dehn twists on a genus $g$ surface with $p$ punctures, i.e., the generators of $\text{MCG}(\Sigma_{g,p})$.

To see this explicitly, we can think of a genus $g$ surface in terms of a two layers of a Turaev-Viro code,
connected via `wormholes' along boundaries of disconnected holes, as shown in Fig.~\ref{fig:modular_genus_g_alpha_string_net}
and Fig.~\ref{fig:modular_genus_g_gamma_string_net}. In other words, a genus $g$ surface is equivalent two a bilayer
version of the state, with $g$ disconnected gapped boundaries. Note that we choose the triangulation and trivalent
graph patterns on the two layers being symmetric up to a mirror reflection.

The $\alpha$ and $\gamma$ Dehn twists, shown explicitly in Fig.~\ref{fig:modular_genus_g_alpha_string_net}
and Fig.~\ref{fig:modular_genus_g_gamma_string_net} can be applied by straightforward generalizations of the
Dehn twist protocols on a torus.

The Dehn twist along the $\beta$-loop can be applied by using the protocol for the Dehn twist on an annulus for the top or bottom layer, as illustrated in
Fig.~\ref{fig:modular_genus_g_beta}.

Finally, in the presence of $p$ punctures, the additional generators needed for $\text{MCG}(\Sigma_{g,p})$ are the braids between punctures $i$ and $i+1$, i.e. $\mathcal{B}_{i,i+1}$.
Here we can straightforwardly use the protocols shown in Fig.~\ref{fig:braiding_string_net_periodic}-\ref{fig:braiding_string_net} to implement these braids.

\subsection{Multiple Dehn twists in a single shot: Proof of Theorem 2}

\begin{figure*}
\includegraphics[width=2\columnwidth]{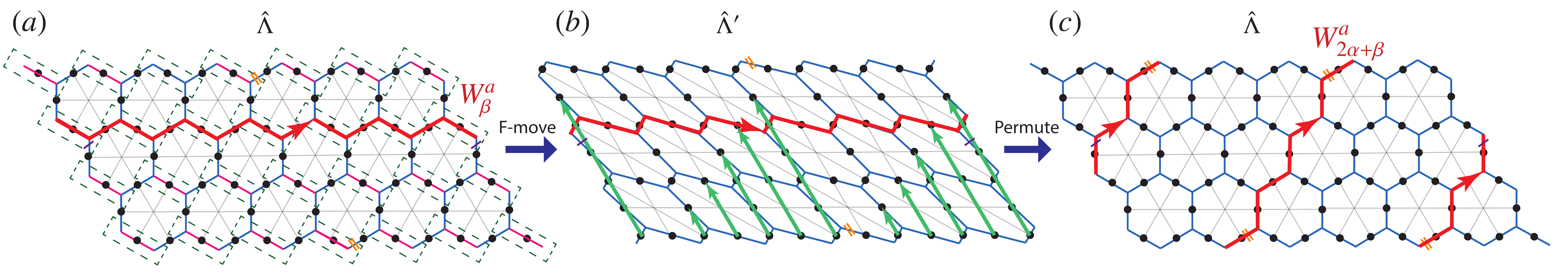}
  \caption{Implementing multiple Dehn twists in Turev-Viro code in a single shot by choosing a particular aspect ratio of the torus.  }
\label{fig:multi_Dehn_twists_string_net}
\end{figure*}

\begin{figure*}[hbt]
  \includegraphics[width=2\columnwidth]{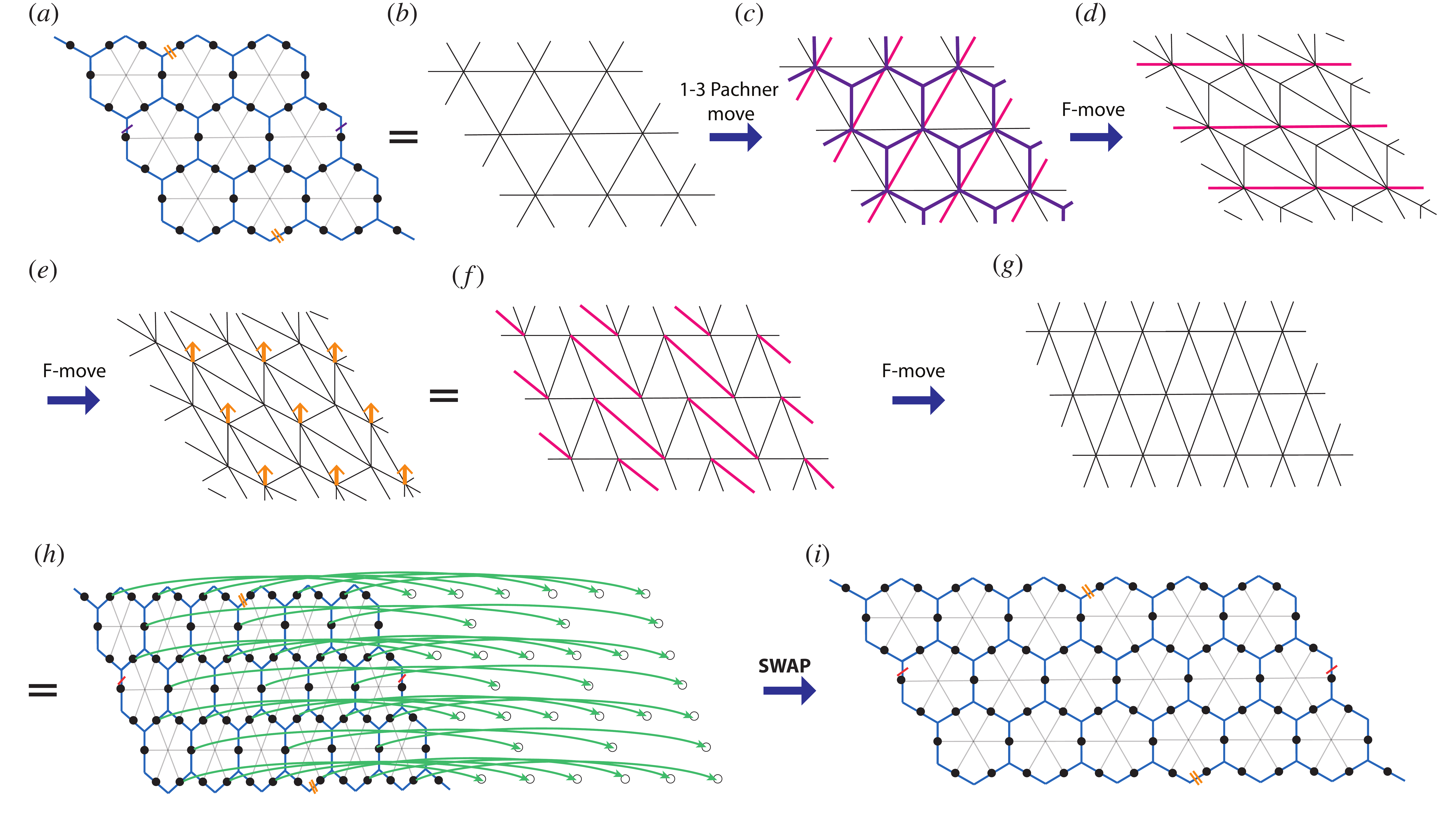}
  \caption{Double the size of the torus along one cycle with by dividing a plaquette into two and long-range permutation. The yellow arrow in (e) indicates the equivalence of the triangulation in (e) and (f) by a local shift of the vertices. The curved green arrows in (h) represents long-range SWAPs.}
\label{fig:double_the_size_TV}
\end{figure*}

In this section we prove \textbf{Theorem 2} in the context of Turaev-Viro codes. We demonstrate how to implement
$n$ Dehn twists along any cycle of a genus-$g$ surface using constant-depth local circuits and permutations. Specifically, we show
that in the code subspace, $\mathcal{D}_{n \omega} = \prod_{i=1}^{k} \mathcal{P}_{\sigma_i} \mathcal{LU}_i$, where the $\mathcal{LU}_i$
are local quantum circuits with depth independent of $n$, code distance $d$ and system size, while $k = \mathcal{O}(\log_2 n)$.

We start by considering a torus with its length in the $\beta$-cycle being $n$ times as much as its length along the
$\alpha$-cycle, as illustrated in Fig.~\ref{fig:multi_Dehn_twists_string_net} (where $n=2$ in this example).  After
applying the F-moves [Fig.~\ref{fig:multi_Dehn_twists_string_net}(b)] similar to the case in Fig.~\ref{fig:Dehn-twist-string-nets}
and permutation of qubits along the $\beta$-cycle [Fig.~\ref{fig:multi_Dehn_twists_string_net}(b)], we achieve a Dehn twist
$\mathcal{D}_{n \alpha}$ in a single shot, corresponding to the following transformation: $\mathcal{D}_{n\alpha} :W^a_\beta  \longmapsto W^a_{n\alpha+\beta}$
[with the $n=2$ case illustrated in Fig.~\ref{fig:multi_Dehn_twists_string_net}(c)].  By changing the opposite aspect ratio of the torus,
we can also achieve the Dehn twist $\mathcal{D}_{n\beta}$ in a single shot.   This again suggests that by increasing the
number of qubits participating in the topological state by $n$ times and with fixed code distance $d$, the time complexity of implementing a particular
logical gate sequence can be decreased by $n$ times, even in the case of the more computationally powerful non-abelian code.

Now to prove the theorem, we also need to be able to freely adjust the aspect ratio of the torus,
as in the case of abelian stabilizer codes in Sec.~\ref{sec:multi_dehn_twist_aspect_ratio_ZN}.
To do this, we apply a plaquette-dividing protocol to grow the torus along a particular direction,
allowing us to adjust its aspect ratio, as shown in Fig.~\ref{fig:double_the_size_TV}. Here, by a sequence of
1-3 Pachner moves and F-moves, we see that we can double the size of the code by a constant depth local unitary, followed by a permutation
on the qubits.
By repeating this procedure, one is able to stretch the torus along a particular direction by $n$ times with
$\log_2{n}$ steps of the above procedure.  Therefore, one gains an exponential speedup when merging $n$ Dehn twists together,
as compared with applying them sequentially.

It is straightforward to generalize the above procedure to Dehn twists along any of the $3g-1$ loops on the genus-$g$ surface.

Finally, we note that, unlike the situation in abelian stabilizer codes where multiple Dehn twists can be applied in a single shot
by increasing the range of the local quantum circuit (second part of \textbf{Theorem 2}) (Sec.~\ref{sec:increase_range}), we have not found such a
general property for the Turaev-Viro codes.  The key difference is that, in the abelian stabilizer codes, it is always possible to
create an arbitrarily long-range stabilizer in just a single shot (2 small steps) with long-range entangling gates (e.g.~CNOT). In contrast,
in the non-abelian codes, we create a plaquette involving an edge with length $\mathcal{O}(n)$ by applying
$\mathcal{O}(n)$ steps of F-moves.

\section{Arbitrary MCG elements: Proof of Theorem 3}\label{sec:theorem3}

So far we have demonstrated how one can implement all elementary Dehn twists and braids by a constant depth local
circuit followed by a permutation. Here we consider how to implement arbitrary elements of the MCG, proving \textbf{Theorem~3}.

Specifically, let $\zeta \in \text{MCG}(\Sigma_{g,p})$ be an arbitrary element of the mapping class group of $\Sigma_{g,p}$,
and $\mathcal{V}_{\zeta}$ be its representation on the code subspace. $\zeta$ has a presentation in terms of $k$ Dehn twists and braids,
as the latter generate the mapping class group.
According to \textbf{Theorem \ref{theorem1}}, each Dehn twist and braid can be implemented by a constant depth local quantum circuit $\mathcal{LU}_i$
followed by a permutation $\mathcal{P}_{\sigma_i}$. Therefore,
\begin{equation}\label{Rk1}
  \mathcal{V}_\zeta \otimes I \ket{\Psi} = (\mathcal{V}_\zeta\ket{\Phi}) \otimes \ket{\Pi}_a = \prod_{i=1}^{k} \mathcal{P}_{\sigma_i}\mathcal{LU}_i\ket{\Phi}\otimes \ket{\Pi}_a.
\end{equation}
Recall that $\ket{\Phi}$ is the many-qubit topological state, and $\ket{\Pi}_a$ is a product state on ancilla qubits.

Now the idea is that since the $\mathcal{P}_{\sigma_i}$'s are just relabelings of the qubits, they can be deferred
until after all of the circuits $\mathcal{LU}_i$.

To get an intuition first, we set the general case aside and look at the braiding protocol described in Section \ref{sec:braiding_protocol}.
Note that the braiding protocol was also in the form of $\mathcal{B}$$=$$\prod _{i=1}^{12} \mathcal{P}_{\sigma_i}\mathcal{LU}_i$.
After performing $\mathcal{LU}_1$, just before doing the long range permutations shown in Fig.~\ref{fig:braiding_circuits}(d), one can
see that the code is actually in square lattice form and just stretched in some parts and squeezed in other parts. So one can
perform the circuit $\mathcal{LU}_2$ on the same lattice, skipping the long range permutations of Fig.~\ref{fig:braiding_circuits}(d).
The price to pay is that in the deformed square lattice, some plaquettes are elongated by a factor of $2$, so if the $\mathcal{LU}_2$ gates
involve qubits that are one lattice constant apart, now the range will increase to $2$. By using the same idea in each step, one can
defer all permutations until the end and perform all of them at once.

Now we return to the general case. By inserting appropriate factors of $\mathcal{P}_{\sigma_i} \mathcal{P}_{\sigma_i}^{-1}$, we can turn the right hand side of Eq.~\ref{Rk1} into
\begin{widetext}
\begin{equation}
  (\prod_{i=1}^{k}\mathcal{P}_{\sigma_i}) \prod_{i=1}^{k} (\mathcal{P}_{\sigma_1}^{-1}\mathcal{P}_{\sigma_2}^{-1}\cdots
\mathcal{P}_{\sigma_{i-1}}^{-1} \mathcal{LU}_i \mathcal{P}_{\sigma_{i-1}}\cdots \mathcal{P}_{\sigma_2} \mathcal{P}_{\sigma_1}) \ket{\Phi}\ket{\Pi}.
\end{equation}
\end{widetext}
Let $\tilde{\sigma}_i = \prod_{j=1}^{i-1}\sigma_j$. Then Eq.~\eqref{Rk1} takes the simple form
\begin{align}\label{Rk2}
 \non \mathcal{V}_\zeta \otimes I \ket{\Psi}&=(\mathcal{V}_\zeta\ket{\Phi}) \otimes \ket{\Pi}_a \\
  &= \mathcal{P}_{\tilde \sigma_{k+1}}  \prod_{i=1}^{k} \mathcal{P}_{\tilde \sigma_i}^{-1} \mathcal{LU}_i \mathcal{P}_{\tilde \sigma_i} \ket{\Phi}\otimes\ket{\Pi}_a.
\end{align}
One can interpret $\widetilde{\mathcal{LU}_i} = \mathcal{P}_{\tilde \sigma_i}^{-1} \mathcal{LU}_i \mathcal{P}_{\tilde \sigma_i}$ as
$\mathcal{LU}_i$ but performed over relabeled qubits. However, locality of $\mathcal{LU}_i$ alone, does not guarantee
locality of $\widetilde{\mathcal{LU}_i}$ since the permutations are non-local operations. Nevertheless, as discussed in more detail in
the subsequent section, the permutations that we use for the Dehn twist and braid protocols have a special structure (referred
to in the subsequent section as a connectivity-preserving isomorphism),
which keep local gates local, although they may increase the range of the gates by a constant factor.
Thus $\widetilde{\mathcal{LU}_i}$ would still be a local quantum circuit.

Locality of each term in the product, together with the fact that $k$ does not depend on code distance, ensures that the
whole product in Eq.~\eqref{Rk2} is a local constant depth circuit, with depth independent of code distance and system size,
which we denote by $\mathcal{LU}_\zeta$. Then,
\begin{equation}
  \mathcal{V}_\zeta \otimes I \ket{\Psi}=(\mathcal{V}_\zeta\ket{\Phi}) \otimes \ket{\Pi}_a = \mathcal{P}_{\tilde \sigma_{k+1}}  \mathcal{LU}_\zeta \ket{\Phi}\otimes \ket{\Pi}_a.
\end{equation}

Since the permutations appearing in \textbf{Theorem \ref{theorem1}} change distance, the range of gates in $\mathcal{LU}_i$ and $\widetilde{\mathcal{LU}_i}$
differ by a constant factor (independent of code distance). Let us consider a typical permutation $\sigma$ which we use to implement one of braids or Dehn twists.
If $D(m,n)$ denotes the distance between two qubits labeled by $m$ and $n$, then,
\begin{equation}\label{c_ratio}
  \max_{m,n}\frac{D(m,n)}{D(\sigma(m),\sigma(n))}=c_\sigma
\end{equation}
is independent of code distance and system size since otherwise $\sigma$ would turn a local operator into a non-local one.

Let $c$ be the maximum $c_\sigma$,  and $\rho$ be the maximum gate range used in the Dehn twists and braids
that compose $\zeta$. According to Eq. \ref{c_ratio} the range of gates in
$\mathcal{P}^{-1}_{\sigma_i} \mathcal{LU}_i \mathcal{P}_{\sigma_i}$ would be $c\,\rho$ at most. This in turn means the
range of gates in $\widetilde{\mathcal{LU}_i}$ would be at most $c^i\,\rho$. And since $i$ can be $k$ at most, the range
of gates in $\mathcal{LU}_\zeta$ is $r=\mathcal{O}(c^k)$, which completes the proof of \textbf{Theorem 3}.

\section{Topological protection, fault tolerance and experimental platforms}\label{sec:fault_tolerance}

\begin{figure*}
  \includegraphics[width=1.8\columnwidth]{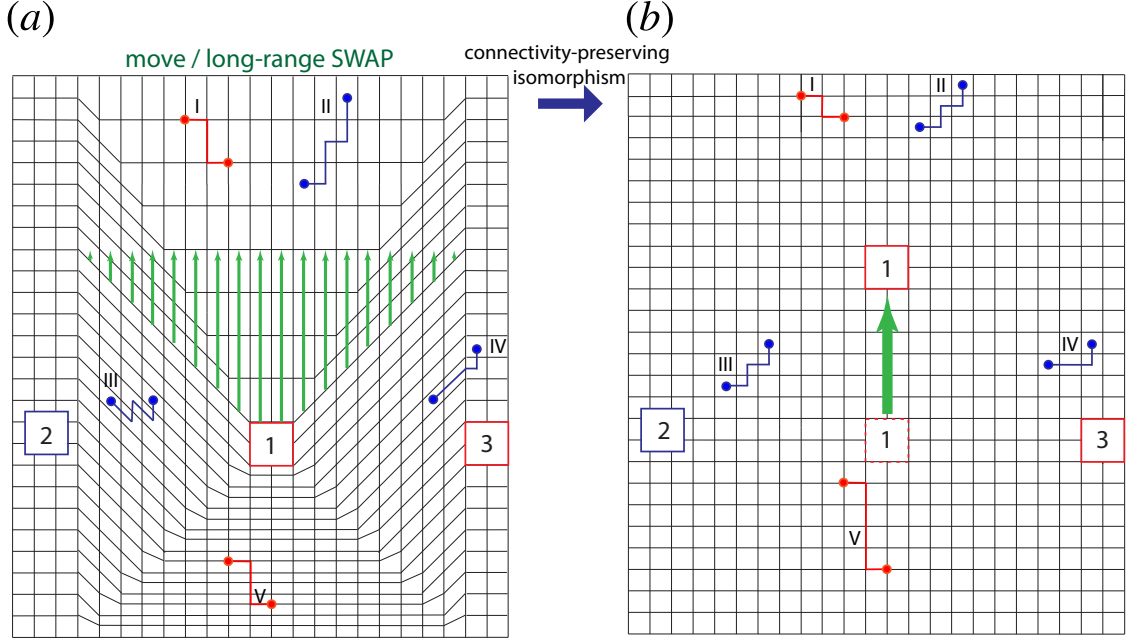}
  \caption{Propagation of pre-existing errors under the qubit permutation $\mathcal{P}_\sigma$ in the braiding scheme from Fig.~\ref{fig:braiding_circuits}.  Five error strings before and after the permutation are illustrated.}
  \label{fig:error_string_map}
\end{figure*}

In this section, we argue that logical operations consisting of the 
$\mathcal{LU}$ and $\mathcal{P}_\sigma$ operations, which compose the Dehn twists and braids,
are topologically protected and fault-tolerant logical operations, assuming appropriate error models. 

\subsection{Intrinsic protection and error propagation}
A major consideration of fault tolerance is the extent to which logical gates propagate errors. 
For a given state with a pre-existing local error on site $j$, i.e., $E_j \ket{\psi}$,
one can apply a logical gate $U$ to it.  Even if the logical gate $U$ is perfect, i.e. it does not generate new errors,
the pre-existing error may propagate under $U$, which can be seen from the following identity:
\be  
   U E_j \ket{\psi} = U E_j U^\dag (U\ket{\psi}).
\ee
This means that the ideal target state $U\ket{\psi}$ suffers from the propagated error $U E_j U^\dag$.

We first consider the constant depth local quantum circuit $\mathcal{LU}$ implementing the local geometry deformation discussed in the previous sections, which consists only of a finite
sequence of geometrically local gates supported on a region of radius $\mathcal{O}(1)$. These are also referred to 
as \textit{locality-preserving unitaries} \cite{Beverland:2016bi, Bravyi:2013dx}. Any logical gate that can be performed in terms of a
locality-preserving unitary is fault tolerant because the propagation of pre-existing errors is restricted by causality:
a single-site error can only spread to an error supported within a Lieb-Robinson `light cone' \cite{Lieb_Robinson} with radius $\mathcal{O}(1)$. 
Therefore constant depth $\mathcal{LU}$, which composes the first part of the Dehn twist and braid protocols presented in this paper, 
also preserves the locality of errors, and are hence intrinsically fault-tolerant. 

Now we consider the second part of our proposed logical gates, the qubit permutation $\mathcal{P}_\sigma$. 
While the Eastin-Knill theorem rules out the possibility of a universal transveral logical gate set, it was commented
in Refs.~\cite{Eastin:2009cj, Zeng:2011gs} that qubit permutation could be a possible loop-hole to circumvent the theorem \footnote{Qubit 
permutation was also extensively studied in depth by Ref.~\cite{Zeng:2011gs} in the context of stabilizer codes, despite the fact 
that it was realized that universal gate sets cannot be achieved even with the addition of such permutations.}.  
It was also pointed out in Ref.~\cite{Eastin:2009cj} that such qubit permutations, taken in isolation, would be fault-tolerant. 

Below we analyze the topological protection and fault-tolerance property of the long-range permutations $\mathcal{P}_{\sigma}$ that we use in some detail,
and also specify the experimental platforms with which fault-tolerance of such qubit permutations are expected. 

We note that in this paper, we use a specific class of permutations $\mathcal{P}_\sigma$ that we can refer to as
a \textit{connectivity-preserving isomorphism} (CPI). While the CPI can permute qubits over long distances, it preserves
the local connectivity of the underlying lattice structure of the codes / Hamiltonian. More concretely, for a pair of neighboring vertices $v_1$ and $v_2$ 
in the original lattice (triangulation) $\Lambda$, and the permuted vertices $\sigma(v_1),  \sigma(v_2)$,  
the edge $e[v_1, v_2]$ connecting the original vertices $v_1$ and $v_2$ is exactly transformed to the edge connecting the new vertices, i.e.,  
\be
\mathcal{P}_\sigma: e[v_1, v_2] \longmapsto e[\sigma(v_1), \sigma(v_2)].
\ee 
Moreover, the new edge $e[\sigma(v_1), \sigma(v_2)]$ also remains local, in the sense that it has length of $\mathcal{O}(1)$. 
More specifically, $\mathcal{P}_{\sigma}$ performs an isomorphism from the original lattice $\Lambda$ to a new deformed 
lattice $\Lambda'$ with completely the same connectivity. 
Thus we see that CPI is clearly a more general type of transformation compared to a locality preserving unitary $\mathcal{LU}$. 
As we will describe in more detail, a CPI is intrinsically protected, because the many-qubit quantum state
remains in the ground-state subspace of a local Hamiltonian; in other words, the many-body quantum state remains in the
code subspace of a geometrically local topological QECC. A generic permutation does not have this property, as it typically 
moves the quantum state out of the ground state subspace of a geometrically local Hamiltonian and thus necessarily generates
errors (anyons). 
On the other hand, the qubit permutation considered in our paper -- the CPI -- is still more general than 
the type of qubit permutations that were considered in Ref.~\cite{Zeng:2011gs}, which are automorphisms, 
i.e., isomorphisms that maps the code space back to itself, namely  $AUT : \mathcal{H}_{\Lambda} \mapsto  \mathcal{H}_{\Lambda} $. 
In our case, $\mathcal{P}_{\sigma}$ maps the code space on a triangulation $\Lambda$ to that of a different triangulation, 
$\Lambda'$: $\mathcal{P}_{\sigma} : \mathcal{H}_{\Lambda} \rightarrow \mathcal{H}_{\Lambda'}$. The $\mathcal{LU}$ is then needed
to map back to the original code space $\mathcal{H}_{\Lambda}$. 

To analyze the topological protection of a CPI, we first consider propagation of pre-existing errors under a perfect (error-free) CPI. 
Let us consider an error string (the two end points of the string corresponding to anyons) with a length $l$ much smaller
than the code distance $\textit{l} \ll d$, and which has support on sites $\{ j_1, j_2, j_3, ..., j_n \}$, as illustrated in Fig.~\ref{fig:error_string_map}. 
Our CPI permutation $\mathcal{P}_\sigma$ maps the string onto the new sites  $\{ \sigma(j_1), \sigma(j_2), \sigma(j_3), ..., \sigma(j_n) \}$, 
which is a deformed error string with a length of the same order as before , i.e., $\mathcal{O}(\textit{l})$$\ll$$d$. 
One can compare  the configurations in Fig.~\ref{fig:error_string_map}(a) and (b), and see that the error string I and II gets 
squeezed, III and IV gets deformed, and V gets stretched.  Therefore, although our connectivity-preserving isomorphism 
does not preserve the location of errors, it only changes the length of the error string by a constant factor (independent of code distance)
and hence preserves the characteristic length of the error strings. It is also worth emphasizing that the above properties suggest that the history of errors (anyons) can be completely tracked by the software (with the assistance of syndrome measurement) for the purpose of decoding in active error corrections. We note that the decoder for our scheme 
will differ from the standard ones, and thus requires further development in future works. 

To summarize, the combined ($\mathcal{LU}$ and $\mathcal{P}_\sigma$) constant-depth logical gate (denoted by $U$)  maps a local operator $O$ with support in a region $\mathcal{R}$ to another local operator $O'=U^\dag O U$
with support in a region $\mathcal{R}'$, such that the area ratio between $\mathcal{R'}$ and $\mathcal{R}$ is bounded by an $\mathcal{O}(1)$ constant factor $c$, similar to the property of a locality-preserving unitary, namely,
\be
\textsf{supp}(U^\dag O U) \le c \  \textsf{supp}(O).
\ee
Therefore, the constant-depth logical gate is naturally topologically protected and can be made fault-tolerant. Note that the difference from the definition of the locality-preserving unitary in Ref.~\cite{Beverland:2016bi} is that here the regions $\mathcal{R}$ and $\mathcal{R}'$ do not have to be in the vicinity of each other.

\subsection{Additional error generation for the qubit permutation taking into account the hardware implementation}
Now we further consider additional errors that can be generated during the process of the qubit permutation. 
Just as with any discussion of fault-tolerance, we have to assume a particular reasonable noise model. 
In our case, the permutations will yield fault-tolerant operations if errors in any site-to-site
permutation process,  i.e.  $j \mapsto \sigma(j)\equiv j'$, are independent. We consider two schemes 
and their experimental platforms for implementing the permutations: (1) moving qubits; (2) long-range SWAP gates utilizing ancillas. 

\textit{(1) Moving qubits.} For certain experimental systems, one can directly move the qubits to desired positions. 
Ion traps are currently the most promising platform for this purpose, since high-fidelity fast shuttling of individual 
ion qubits has been realized experimentally  \cite{Bowler:2012fr,  Walther:2012iz, Wright:2013df}. Shuttling of ions 
has also been  proposed and demonstrated for a scalable quantum computation architecture \cite{Home:jr, Lekitsch:2015ua, Kaufmann:2017kc}, 
including fault-tolerant quantum computation with surface codes on a 2D qubit array \cite{Lekitsch:2015ua}.

In practice, the shuttling process may suffer from noise, the major one being heating of ions \cite{Wright:2013df}, which is independent 
for individual ions. That is, there is no correlated noise between individual shuttling processes. One can model the shuttling process 
of a single qubit by the standard Pauli error channel (for simplicity here we only consider qubits, as opposed to qudits):
 \begin{align}\label{Pauli_channel}
 \mathcal{E}_{j'} =\{ \sqrt{1-p^x_{j'}-p^y_{j'}-p^z_{j'}} I, \sqrt{p^x_{j'}} X_{j'}, 
 \sqrt{p^y_{j'}} Y_{j'}, \sqrt{p^z_{j'}} Z_{j'} \} ,
 \end{align}
where $p^x_{j'}$, $p^y_{j'}$ and $p^z_{j'}$ are error probabilities of the Pauli-$X$, $Y$ and $Z$ errors to occur at site $j'$.  For uncorrelated noise, the 
joint probability of a Pauli-$i$ error ($i=X,Y,Z$) occurring  at site $j_1'$ and a Pauli-$k$ error occurring at site $j_2'$ is $p^{i}_{j_{1'}} p^{k}_{j_{2'}}$. 
Therefore, for a length-$n$ error string of the form $\bigotimes_{j'=1}^{n} X_{j'}$, the probability of the error is $\prod_{j'=1}^{n} p^{x}_{j'}$. 
This result can also be easily generalized to the case where the operators on each site of the error string are of different types. For 
an error string of $O(d/2)$ length, the error becomes uncorrectable and a logical error occurs. However, such an error string 
is exponentially suppressed in this uncorrelated noise model, which implies the existence of an error threshold.

\textit{(2) Long-range SWAP}.  For experimental systems with long-range connectivity, one 
can implement the permutation $\mathcal{P}_{\sigma}$ using long-range SWAP operations. 

Here we focus on the simplest implementation: we use two sets of qubits, a data register
labeled by the set \{$j$\} and storing the topological state  $\ket{\Psi(\{j\})}$, and a 
temporary register labeled by the set \{$j'$\} storing  a product state of zeros, i.e., $\bigotimes_{j'} \ket{0}_{j'}$.  
We can take the two sets of qubits to form a superlattice structure in real space, such that corresponding qubits 
in the two registers are nearest neighbors if $j=j'$.  We then SWAP all the information in the data 
register \{$j$\} to the temporary register \{$j'$\}, according to the map $j'=\sigma(j)$.
We can then just continue the computation in the new register \{$j'$\}, which now becomes the data register, 
and will later SWAP the data back to original data register \{$j$\} if another permutation is needed.  The 
qubits in the temporary  register can also serve as ancilla qubits for syndrome measurement during the 
active error correction so extra resources may not be required.     

Assuming the errors that occur on individual SWAP operations between the pair of sites $j$ and $j'$ 
are independent (uncorrelated), the errors that occur during this SWAP process can again be captured 
by the Pauli noise channel [Eq.~\eqref{Pauli_channel}] on each site. Therefore, the noise property 
will be similar to the case of moving qubits, and the operation is thus fault-tolerant. 

A more complicated implementation will be making an additional SWAP of the information in the temporary register 
$\{j'\}$ back to the original data register $\{j\}$.  This two-step SWAP procedure makes the error analysis slightly
more complicated, but does not change the essence of the fault tolerance. 

Some experimental platforms that can be used for such long-range SWAP operations are listed below. 

\textbf{(I)} \textit{Long-range connectivity in ion traps} mediated by motional (phonon) modes of 
ions \cite{Linke:2017bz}. All the long-range SWAPs can be performed in parallel and have uncorrelated noise 
if there is a separate phonon mode mediating each individual SWAP.  
 
\textbf{(II)}~\textit{Modular architecture of 3D superconducting cavities} \cite{CampagneIbarcq:2017wq, Kurpiers:2017ub, Axline:2017uq, Chou:2018vz}. 
The quantum information is stored in microwave cavity photons. This architecture has reconfigurable long-range connectivity 
between cavity nodes, routed by microwave circulators and superconducting cables \cite{CampagneIbarcq:2017wq}.
One possible scheme is through direct quantum state transfer between remote cavity nodes in a network, which is equivalent to a 
long-range SWAP \cite{Axline:2017uq, Kurpiers:2017ub}. The noise is uncorrelated if different cables are used for individual 
SWAP processes. An alternative scheme is through remote entanglement generation and teleportation \cite{CampagneIbarcq:2017wq, Chou:2018vz}, 
which also has uncorrelated noise for individual teleportation channels. 

\textbf{(III)} 
~\textit{Circuit QED with cavity buses} \cite{Blais:2007hh, Schoelkopf:2008vi}. Here, long-range interaction between 
superconducting qubits or semiconductor spin qubits can be mediated by cavity array serving as quantum buses \cite{Majer:2007em, Helmer:2009de, houck2012, Zhu:2013cm, Fitzpatrick:2016ws, Naik:2017bo, Zhu:2018aa}.  

\textbf{(IV)} 
\textit{Rydberg atoms}. Here, long-range gates can be realized via Rydberg-blockade mechanisms \cite{Saffman:2010ky, Comparat:2010cb, Maller:2015is,   Pichler:2016ec}.

\begin{figure*}
  \includegraphics[width=1.8\columnwidth]{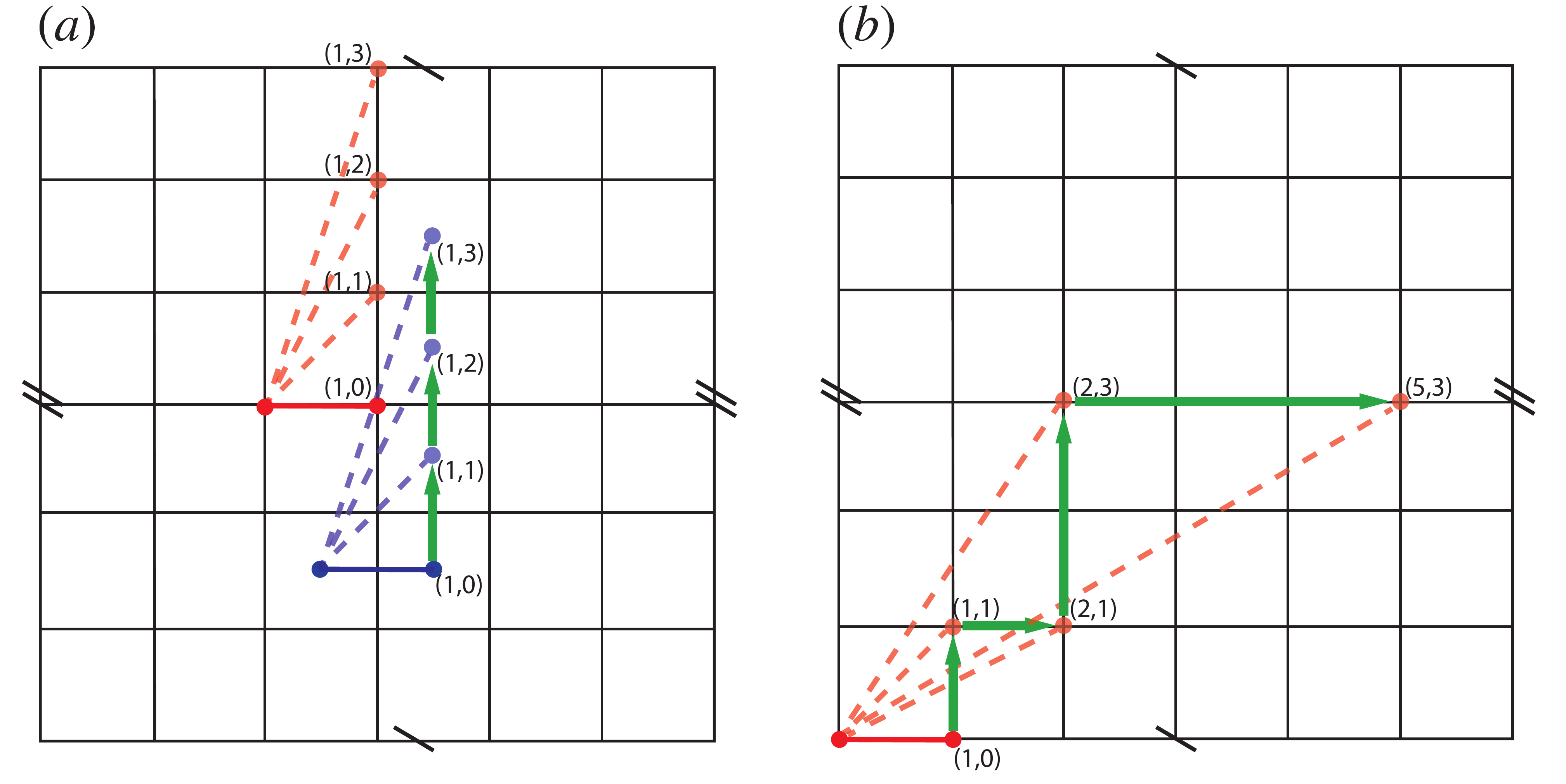}
  \caption{Illustration of the propagation of error strings after a successive application of the instantaneous logical gates (Dehn twists) on a torus (for the $\mathbb{Z}_N$ toric code example). (a) A sequence of Dehn twists along the same $\alpha$-cycle (vertical) leads to a linear growing of error strings. (b) An alternating sequence of Dehn twists along both cycles leads to an exponential growing of error strings. }
  \label{fig:error-stretching}
\end{figure*}

\subsection{Additional time overhead to achieve full fault tolerance}
So far we have argued that our protocols for logical gates are topologically protected and can be made fault-tolerant because they do not spread errors by more than a $\mathcal{O}(1)$ factor. 
Fault-tolerant quantum error correction requires syndrome measurements, decoders to deduce the error strings from the syndrome measurements, 
and finally an error recovery procedure. 


Although the logical gate applied here is constant-depth itself, dramatically differing from the situation of conventional braiding,  extra time overhead might be introduced due to the error correction and decoding processes after the application of the gate in order to achieve full fault tolerance.
 In the absence of measurement error,  the error syndrome can be decoded immediately after the application of the gate, since the error string is only changed by an $\mathcal{O}(1)$ constant factor independent of the code distance $d$, and hence remains correctable. 

However, in the presence of faulty measurement, one is expected to perform $\mathcal{O}(d)$ rounds of measurements to decode the error
syndrome, similar to the case of surface code error correction.  In the context of transversal gates or constant-depth local quantum circuits \cite{Bravyi:2013dx},
the error string either does not change or is stretched at most linearly, and one hence expects that only $\mathcal{O}(d)$ rounds of measurements
are needed for every $\mathcal{O}(d)$ times of the application of such logical gates. The overall time overhead is hence $\mathcal{O}(1)$ in these
situations.  In the context of our logical gates, which are geometrically non-local, the overall time overhead depends on the detailed property
of the logical circuits of the corresponding quantum algorithms. We note that our protocols do not introduce plaquette or vertex operators
whose measurement outcomes are unknown (in contrast to, for example, lattice surgery methods for surface code). In other words, the eigenvalues
of all new plaquette and vertex operators in our protocols are predetermined from the values of previously known plaquette and vertex operators.
Thus, while not studied explicitly here, we expect that the space-time paths of the error syndromes can be fault-tolerantly decoded.  

In the cases when the error string is stretched linearly with time,  the overall time complexity is still  $\mathcal{O}(1)$. We can illustrate this situation with the $\mathbb{Z}_N$ toric code suppported on a torus as shown in Fig.~\ref{fig:error-stretching}(a).  Here, we apply a sequence of Dehn twists along the $\alpha$-cycle of the torus, i.e., $(\mathcal{D}_\alpha)^n$. We use $(l_x, l_y)$ to denote the length of error string. We can see that for both types of error strings ($e$ and $m$), the error string starting at the $(1,0)$ configuration grows with the following sequence $(1,0) \rightarrow  (1,1) \rightarrow   (1,2) \rightarrow   (1,3) \cdots \rightarrow   (1,n-1) \rightarrow   (1,n)$. We note that, in Fig.~\ref{fig:error-stretching}(a), we make the illustration in a way that the position of the left anyon is fixed, which occurs in the situation that the cut of the Dehn twist goes through the left anyon. In the more general situations, the left anyon will also be permuted, while the relative motion of the right anyon and hence length of the anyon string remains the same as illustrated in the figure. As one can easily conclude,  the error string is stretched linearly, similar to the situation of constant-depth local quantum circuit.  Therefore, $\mathcal{O}(d)$ rounds of measurements are needed for applying every $\mathcal{O}(d)$ Dehn twists $\mathcal{D}_\alpha$.  We note that the same conclusion applies exactly to the more general family of Turaev-Viro codes.  

In the worst case scenario, when certain sequence of braids are repetitively applied in the same region and stretch the error string exponentially with time which leads to it growing to the code distance $d$ in $\log d$ time steps. Therefore, we expect that $\mathcal{O}(d)$ rounds of syndrome measurements are needed for every $\mathcal{O}(\log d)$  logical gates, giving rise to an overall $\mathcal{O}(d/ \log d)$ time overhead.  This type of situation can be illustrated with Fig.~\ref{fig:error-stretching}(b).  Here, we apply an alternating sequence of Dehn twists along both cycles of a torus, i.e., $\mathcal{D}_\alpha\mathcal{D}_\beta\mathcal{D}_\alpha\mathcal{D}_\beta \cdots$.  For an error string starting with the $(1,0)$ configuration, we end up with the following sequence of string growing: $(1,0) \rightarrow  (1,1) \rightarrow   (2,1) \rightarrow   (2,3) \rightarrow (5, 3) \rightarrow (5, 8)  \cdots (l_x, l_y) \rightarrow (l_x+l_y, l_y) \rightarrow (l_x+l_y, l_x+2l_y) \cdots$. The property of adding the length component along both directions at the next round leads to an exponential growing the of the length of the error string.  

Finally we remark that, since the logical gates in our cases have bounded support in space, i.e., around a single handle or pair of anyons, the logical gates applied in separate regions may  not stretch the error strings consecutively. Therefore, for certain class of sequential logical circuits, we expect that the additional average time overhead for the  decoding is still $\mathcal{O}(1)$. 

Regarding the error recovery protocols, we note that these protocols are trivial for Abelian surface codes, 
because any logical errors can be stored and corrected in software by ``updating the Pauli frame'' of the computation. However for the 
non-Abelian Turaev-Viro codes, the error recovery schemes are more sophisticated \cite{Fibonacci_error_correction, Feng:2018, Dauphinais:2017bz} 
and need further exploration.

\section{Discussion}\label{sec:discussion}

We have demonstrated that braids and Dehn twists, which are the generators of MCG$(\Sigma_{g,p})$, can be achieved by a constant depth local unitary circuit, 
followed by a permutation on qubits. By utilizing long-range SWAP operations and physical ancilla qubits, the permutations can also be achieved through a constant depth
quantum circuit. 
These results thus imply that for topological codes with local interactions (local syndromes) in Euclidean space (excluding the class of hyperbolic codes), the space-time overhead for implementing a single logical operation on encoded qubits can be made optimal, which scales as $\mathcal{O}(d^2)$. 
Other proposed protocols for realizing 
universal fault-tolerant gate sets, such as those which use magic state distillation, code switching, 
or other measurement-based schemes, require a space-time cost for a single logical gate that scales as $\mathcal{O}(d^3)$,
including a time overhead that necessarily diverges with code distance when classical computational resources are included. 
We note that this estimate of the overhead cost is simply for implementing logical gates on logical qubits encoded using QECCs with code distance $d$.
To estimate the total space-time overhead for \it fault-tolerant \rm computation, we should also include the additional space-time overhead
required for the error correction protocols, which include the syndrome measurements, decoder, and error recovery procedures. We estimate that, in the presence of measurement errors, an additional $\mathcal{O}(d/\log d)$ overhead is introduced in the worst case in order to decode the space-time paths of the error syndromes. We leave a more detailed
study of this to future work. 

In the previous section we discussed how the long-range permutations are natural to implement in a number of different 
experimental platforms. From a broader perspective, our scheme demonstrates at a fundamental level 
the significant advantage of long-range connectivity in quantum architectures for implementing fault-tolerant quantum 
computation. In addition, our study essentially provides a vision to bridge ideas from quantum communication, such 
as robust quantum state transfer and teleportation, and ideas from fault-tolerant quantum computation.

With respect to systems with purely local interactions between physical qubits, our results can still be of practical relevance. 
Long-range SWAP operations can be implemented through local operations together with a long-range entangled Bell pair.
Thus we can envision a quantum architecture where during the quantum computation, long-range Bell pairs are continuously being
created, perhaps with entirely local operations, in parallel but independently of the quantum computation. The logical gates
then utilize these long-range Bell pairs as a resource as part of the computation. This approach also shares the spirit with distributed quantum computation \cite{Cabrillo:1999co, Bose:1999di}.

From a conceptual perspective, our results imply a new view on braiding and Dehn twists. Conventionally, braiding and Dehn 
twists are considered in terms of \it adiabatic\rm \  processes. In terms of ground states of gapped Hamiltonians, braiding 
and Dehn twists are achieved by adiabatic evolution with a local Hamiltonian, which takes a time that diverges with system size or distance between 
anyons. This is manifested in active error correction approaches to quantum computation by requiring that braids of holes, twist defects, and
non-Abelian anyons to require either (a) a local unitary quantum circuit whose depth grows with code distance, or (b) a sequence of measurements, where fault-tolerance
requires that the measurements be performed $d$ times \footnote{ In certain cases in three-dimensional codes, fault-tolerance with a single round 
of measurements may be possible, but the classical processing to determine the measurement outcome still grows with the code distance.}.  Our 
results, on the other hand, show that braiding and Dehn twists fundamentally need not be thought of as an adiabatic process. 

Rather, the essence of Dehn twists can in some sense be thought of as actually an appropriate permutation on the qubits, which
effectively implements the appropriate large diffeomorphism (diffeomorphism not continuously connected to the identity) on the space. 
The constant depth local unitary circuit can be thought of as a `trivial' diffeomorphism that, while it changes the geometry of the space,
can be thought of in the continuum limit as being continuously connected to the identity operation. However this trivial operation is required
because we need to return the system after the permutation back to the original subspace. 

Our braid protocols that utilize the movement of punctures (e.g.~anyons) over large distances of the order of the code distance can instead be interpreted
as follows. Rather than considering moving the anyon through the many-body state by applying a string operator of length $d$, we leave the anyon where it is
and we `grow' the state to one side of the anyon by a factor of two and `shrink' it to the other side by a factor of two, through a type of 
local entanglement renormalization protocol. This allows us to effectively change the distance between anyons by a factor of 2; if the anyons are separated
by a distance of order $d$, this effectively allows motion of the anyon by $d$ steps in essentially a single shot. A remarkable 
consequence of this is that the number of steps required to bring two anyons a distance $\ell$ apart to the same location goes like $\log \ell$,
because our protocol only allows motion by a distance that is bounded by the distance to the nearest anyon. 

Let us now compare our results to those of Ref.~\onlinecite{Zhu:2017tr}, where it was shown that certain mapping class group elements can be realized through
a finite sequence of transveral SWAP operations in a multi-layer topological state with appropriate boundary conditions and defects. 
The result of Ref.~\onlinecite{Zhu:2017tr} can, in our context, be stated as follows. Certain mapping class group operations $\gamma \in \text{MCG}(\Sigma_{g,p})$ 
that are of finite order (i.e. $\gamma^k = 1$ for some $k$, called torsion elements) can be implemented purely in terms of a 
permutation on qubits. Furthermore, the structure of the permutation is such that it is an automorphism of the lattice (it keeps the lattice exactly invariant). 
This allows us to consider folding the system into a multilayer system, (`quantum origami') such that the permutation reduces to SWAP operations between layers. 
However, braids and Dehn twists are fundamentally different, as they have infinite order. The permutations they require are not automorphisms
of the lattice, but rather more general connectivity-preserving isomorphisms. It is intriguing that we now have two classes of mapping 
class group operations that can be achieved with constant time overhead: (1) braids and Dehn twists, and (2) torsion elements. 

From the point of view of mathematics, our result demonstrates how mapping class group elements can be achieved as a map on the triangulation of a surface. 
Braids and Dehn twists correspond to permutations of vertices of the triangulation, followed by a finite sequence of Pachner moves to recover the 
original triangulation. Crucially, the length of the sequence of Pachner moves is independent of the length of any non-contractible cycle and number of vertices
in the triangulation. 

We have established our results for all non-chiral topologically ordered states, which can be captured by Turaev-Viro-Barrett-Westbury TQFTs. These
include all of the Kitaev quantum double models and Levin-Wen string-net models as special cases. While we have not explicitly considered
higher dimensions, we expect analogous results to straightforwardly apply in higher dimensions as well, as the TVBW TQFTs have natural 
generalizations to higher dimensions in terms of higher categories, such as the Crane-Yetter-Walker-Wang TQFTs \cite{crane1993, walker2006, walker2012}.

\section{Acknowledgements}

We thank Michael Freedman, John Preskill, Matthew Hastings, Jeongwan Haah,  Mohammad Hafezi,  Zhenghan Wang, and Stephen Jordan for helpful discussions. A.L. and M.B. were  supported by 
NSF CAREER (DMR-1753240) and JQI-PFC-UMD.  G.Z. was supported by NSF CAREER (DMR-1753240),  JQI-PFC-UMD, ARO-MURI and YIP-ONR.

\bibliographystyle{apsrev4-1.bst}

%

\end{document}